# A molecule with half-Möbius topology

**Authors:** Igor Rončević[1,2#]*, Fabian Paschke[3#], Yueze Gao[2], Leonard-Alexander Lieske[3], Lene A. Gödde[2], Stefano Barison[4,5], Samuele Piccinelli[4,6], Alberto Baiardi[4], Ivano Tavernelli[4], Jascha Repp[7], Florian Albrecht[3], Harry L. Anderson[2] and Leo Gross[3]*

**Affiliations:**

[1] Department of Chemistry, The University of Manchester, Oxford Road, Manchester, United Kingdom.

[2] Department of Chemistry, Oxford University, Chemistry Research Laboratory, Oxford, United Kingdom.

[3] IBM Research Europe – Zurich, Rüschlikon, Switzerland.

[4] IBM Quantum, IBM Research Europe – Zurich, Rüschlikon, Switzerland.

[5] Institute for Theoretical Physics, ETH Zürich, CH-8093 Zürich, Switzerland

[6] Institute of Physics, École Polytechnique Fédérale de Lausanne (EPFL), Lausanne, Switzerland.

[7] Institute of Experimental and Applied Physics and Halle-Berlin-Regensburg Cluster of Excellence CCE, University of Regensburg, Regensburg, Germany.

[#] Equally contributing first authors

* Corresponding authors. Email: igor.roncevic@manchester.ac.uk; lgr@zurich.ibm.com

**Abstract:** Stereoisomers of $C_{13}Cl_2$ exhibiting helical orbitals around a ring of carbon atoms were synthesized by atom manipulation on NaCl surfaces. We resolved the enantiomeric geometries of the singlet states by atomic force microscopy and mapped their helical orbital densities by scanning tunnelling microscopy. A π-orbital basis of the helical, non-planar singlets that twists by 90° in one circulation is consistent with a half-Möbius topology. In such a topology, the π-orbital basis changes sign with respect to two circumnavigations and is periodic with respect to four circumnavigations. A quasiparticle on a ring with this boundary condition could be interpreted as carrying a Berry phase of $\pi/2$. We demonstrate reversible switching of the topology, between the two singlets of oppositely threaded half-Möbius topology, and the planar, topologically trivial, triplet state. Multireference calculations, including large-scale sample-based ab initio calculations executed on quantum hardware, reveal that the switching is associated with a helical pseudo Jahn-Teller effect.

**One-Sentence Summary:** We demonstrate a molecular topology in which the π-orbital basis twists by 90° in one circumnavigation of the molecule.

## Main Text:

Topologically nontrivial molecules are rare, and chemists have long theorized about molecules with geometries resembling a Möbius strip (*1*). In π-conjugated rings of this type, the *p*-orbitals form a helical basis that can give rise to a Möbius topology, with profound implications on stability, coherent transport (*2, 3*), and aromaticity (*4, 5*). The realization of molecules with Möbius geometries (*6-13*) has spurred theoretical investigations of Möbius aromaticity and helical orbitals *(2, 3, 14-17*).

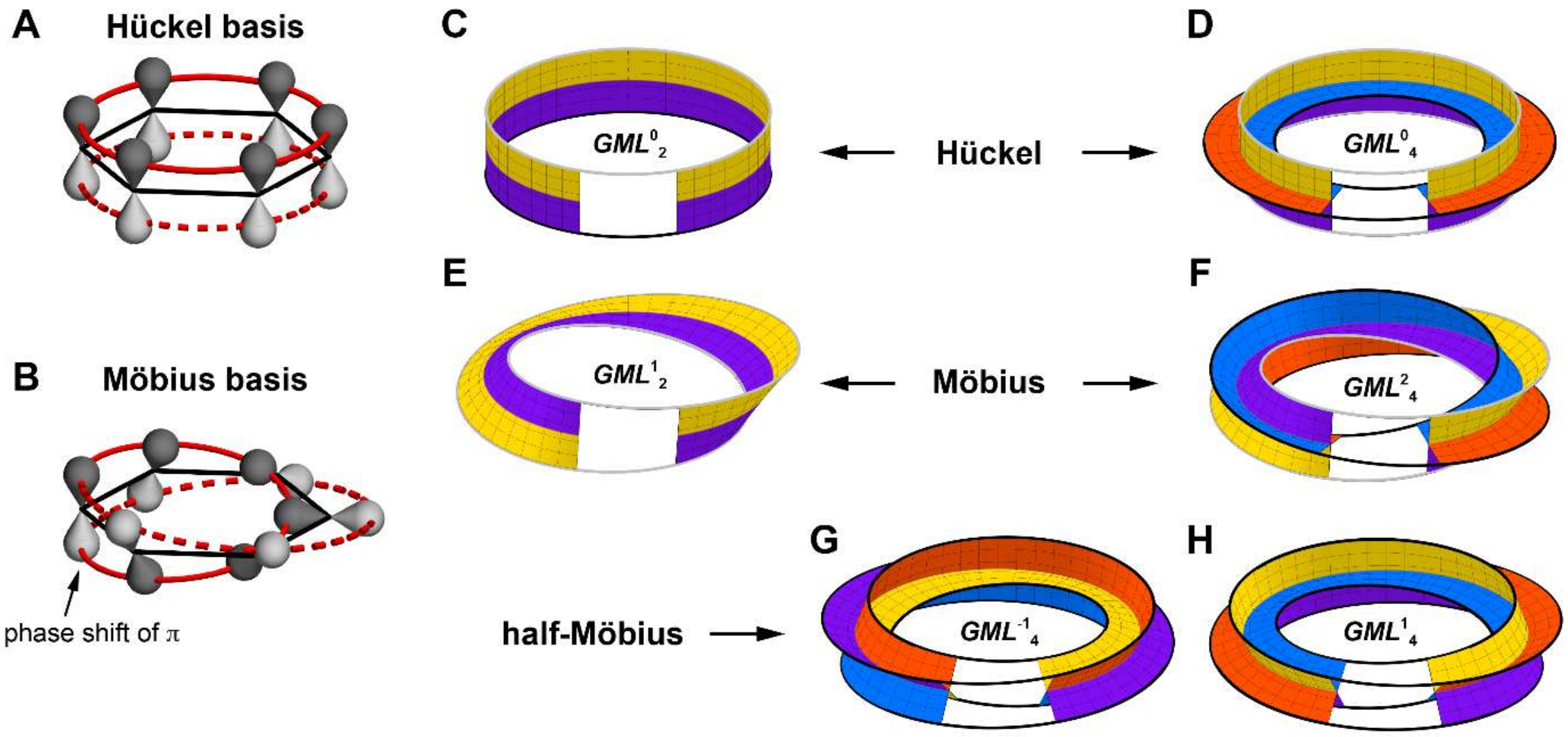

**Figure 1. Orbital Bases and Generalized Möbius-Listing (*GML*) bodies. (A, B)** Schematic Hückel and Möbius orbital basis, respectively (*8*). Orbital bases can be represented as *GML* bodies. In a $GML^{n}_{m}$ body with polygonal cross section, *m* denotes an *m*-fold symmetry of the cross section and *n/m* is the fraction of a full 360° twist upon one circulation. A positive (negative) *n* denotes a right-hand (left-hand) thread. (**C, D**) *GML* bodies with two- and four-fold cross section with no twists, corresponding to Hückel bases. (**E, F**) *GML* bodies with a twist of 180° upon one circulation of the ring, corresponding to Möbius bases. (**G, H**) *GML* bodies with four-fold cross section with a twist of 90° upon one circulation of the ring. Here, we demonstrate the molecular implementation of an orbital basis with a 90° twist corresponding to a $GML^{1}_{4}$ topology (G, H), that we term half-Möbius. In (C-H), colors are used to facilitate the observation of the twisting. Segments in the front are omitted to reveal the cross sections.

Hückel and Möbius topologies can be represented as generalized Möbius-Listing (*GML*) bodies (*18*), see Fig. 1. The π-orbital basis of a trivial Hückel aromatic molecule (Fig. 1A) shows no twist. In contrast, the π-orbital basis of a Möbius aromatic molecule twists by 180° in one circulation, which implies a shift in the phase of the π-orbital basis by π (*2, 4, 5, 8, 14*), see Fig. 1B. A trivial Hückel topology corresponds to a $GML^{0}_{2}$ body, see Fig. 1C. An example is benzene, in which the π-orbital basis is out-of-plane (of the atoms) and shows no twist. Cyclocarbons (*19-22*), rings of sp-hybridized two-coordinate carbon atoms, have two non-twisted separated π-systems, one in-plane (red-blue in Fig. 1D) and the other out-of-plane (yellow-purple), corresponding to a $GML^{0}_{4}$ body (Fig. 1D) also representing a Hückel topology.

In contrast, a Möbius strip corresponds to a $GML^{1}_{2}$ body (Fig. 1E). Möbius hydrocarbons have been reported, with famous examples realized by Ajami et al. (*7*) and Segawa et al. (*12*). In these realized Möbius molecules the geometry corresponds to the Möbius topology, e.g., by forming a twisted carbon nanobelt of 6-membered carbon rings (*12*). Even Möbius molecules with geometrical triple twists of three times 180° have been realized, which correspond to $GML^{3}_{2}$ bodies (*11, 13*). A $GML^{2}_{4}$ body (Fig. 1F) also represents a Möbius topology, or more precisely two orthogonal Möbius $GML^{1}_{2}$ bodies, one colored yellow-purple and one colored blue-red in Fig. 1F. Hypothetically, such a topology might be formed in a helically twisted cyclocarbon, but we are not aware of a realization of this topology in a molecule. Möbius topologies are chiral and two enantiomers can be distinguished ($GML^{\pm 1}_{2}$ and $GML^{\pm 2}_{4}$). Both $GML^{1}_{2}$ and $GML^{2}_{4}$ bodies must be circumnavigated twice along an edge to return to the starting point, but $GML^{1}_{2}$ has one edge, whereas $GML^{2}_{4}$ has two edges.

The topology of $GML^{\pm 1}_{4}$ bodies, which only have a half-twist, that is, a twist by 90° in one circulation (Fig. 1, G and H), we term half-Möbius topology. A $GML^{\pm 1}_{4}$ body has only one edge and needs to be circumnavigated four times along its edge to return to the starting point, compared to one time in $GML^{0}_{2}$ and $GML^{0}_{4}$ bodies with no twist and Hückel topology, or two times in $GML^{\pm 1}_{2}$ and $GML^{\pm 2}_{4}$ bodies of Möbius topology. In this work, we present the on-surface synthesis of a molecule with a π-system that corresponds to a half-Möbius $GML^{\pm 1}_{4}$ topology and can be switched in handedness (between $GML^{1}_{4}$ and $GML^{-1}_{4}$) and between half-Möbius and Hückel ($GML^{0}_{4}$) topology. Finally, we discuss the implications of a molecule with half-Möbius topology.

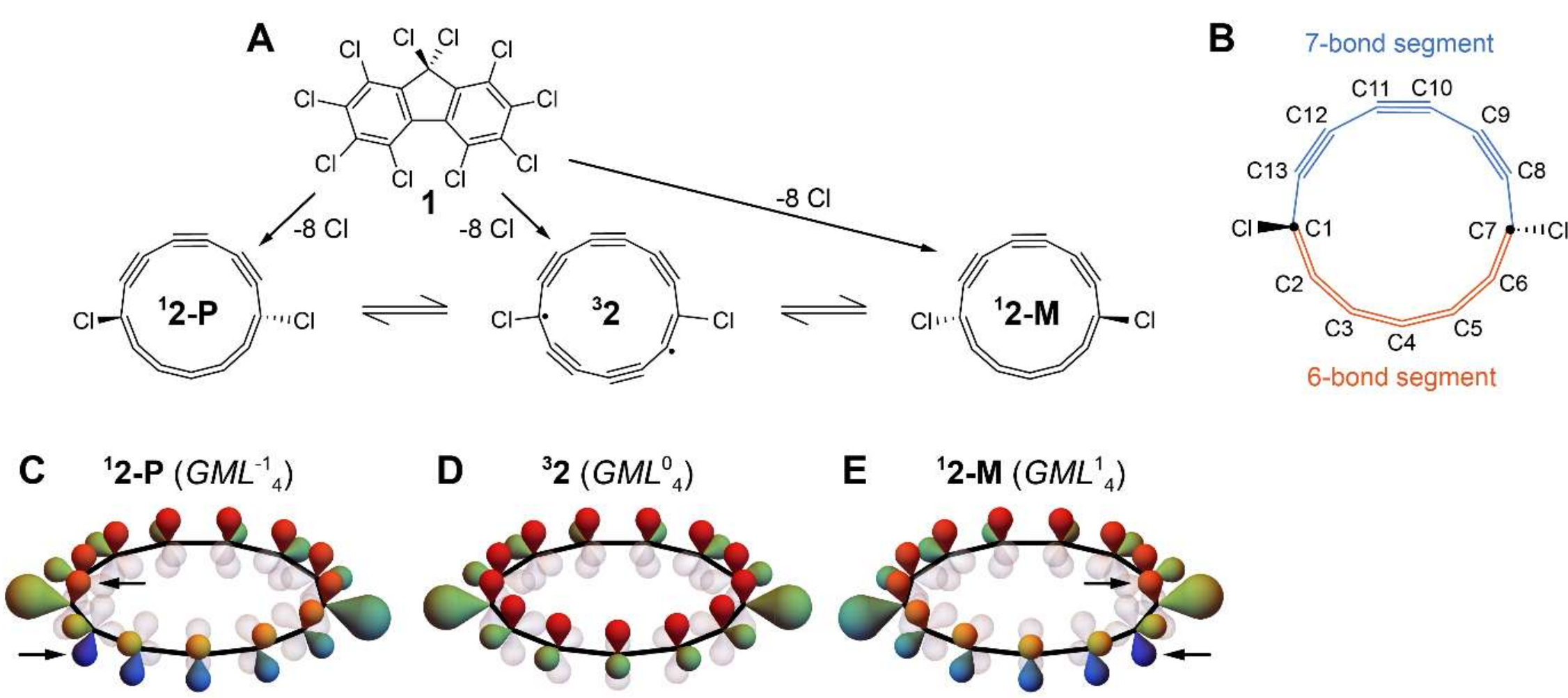

**Figure 2. On-surface synthesis and orbital basis of $C_{13}Cl_2$.** (**A**) Reaction Scheme. By tip-induced dissociation of eight chlorines from **1**, the product **2** was obtained. By atom manipulation we induced transitions between three states of **2**, that are, two non-planar, singlet stereoisomers $^{1}$**2-P** and $^{1}$**2-M**, and a planar triplet state $^{3}$**2**. (**B**) Lewis structure of $^{1}$**2-P**. (**C–E**) Orbital basis of $^{1}$**2-P**, $^{3}$**2** and $^{1}$**2-M**, corresponding to $GML^{-1}_{4}$, $GML^{0}_{4}$, and $GML^{1}_{4}$ bodies (Fig. 1, G, D and H, respectively). Colors

indicate *p*-orbital orientations – with only one colored lobe per *p*-orbital for clarity. Neighboring *p*-orbital lobes with similar colors have significant overlap. The orbital bases of **$^1$2-P** (C) and **$^1$2-M** (E) acquire a phase shift of π after two circulations (following the sequence of rainbow colors), indicated by the black arrows.

The molecule under study, **2** (Fig. 2A), consists of a ring of thirteen carbon atoms, labelled C1 to C13 in the closed-shell Lewis structure in Fig. 2B, and two chlorine atoms attached to C1 and C7. Therefore, eleven of the C atoms, (C2–C6 and C8–C13) are two-coordinate (sp-hybridized), whereas two (C1 and C7) are three-coordinate ($sp^2$-hybridized), because of the Cl heteroatoms. The odd number of C atoms in the ring results in two sp-hybridized segments of different length. The segment including C8 to C13 has an odd number of bonds (7-bond segment, blue in Fig. 2B), whereas the segment including C2 to C6 has an even number of bonds (6-bond segment, orange in Fig. 2B).

Molecule **2** is related to odd-*N* cyclo[*N*]carbons (*21, 22*), but with two heteroatoms added. Similar molecules, that is, cyclic $C_{11}H_2$ and $C_{13}H_2$, have been proposed and discussed in terms of possible Möbius aromaticity by Martín-Santamaría and Rzepa (*23*), albeit with a short two-bond segment. Here, we realized cyclic $C_{13}Cl_2$ molecules and found that a similar length of the odd-bond and even-bond segment was beneficial for stabilizing a singlet ground state with helical *π*-orbitals.

## **Theory**

In $sp^2$(-hybridized) systems such as benzene or the previously reported Möbius hydrocarbons, each carbon atom has three nearest neighbors. The remaining *p*-orbital must be orthogonal to the plane defined by these nearest neighbors, making the choice of the π-basis *determined by geometry*. In sp systems such as cyclocarbons, every carbon has only two nearest neighbors and thus two orthogonal *p*-orbitals, the orientation of which is not fixed, allowing different orbital bases for a given geometry (*24*) and rendering sp systems uniquely suited for engineering unconventional electronic topologies (*25*). In systems with both sp and $sp^2$ carbons such as $C_{13}Cl_2$, both helical and non-helical π-bases of the sp carbons connecting the $sp^2$ carbons are allowed by geometry, and the realization of a trivial or nontrivial topology is then *determined by electronic structure*. Electronic and geometric structure are mutually inter-dependent and a change in electronic structure can affect the geometry by a helical Jahn-Teller distortion, as we show in the following paragraphs. The $sp^2$-bonded heteroatoms can be viewed as boundary conditions determining the local orientation of the π-system, and as handles for controlling and observing the presence, extent and handedness of helicality. Although the global π-conjugation is partly interrupted at the $sp^2$ carbons, it can persist through one *p*-orbital on the $sp^2$ center, and through hyperconjugation (*26, 27*), which is captured as next-nearest-neighbor coupling in a tight-binding approach (**figs. S1-S3**). Hyperconjugation, or more generally σ-π mixing, is important for stabilizing a Möbius topology (*28*), as in a monocyclic pure π-system the Möbius and Hückel topologies can at best be equal in energy (*4, 8*).

The two segments of compound **2** relate to linear carbon chains (*2, 3, 14-17*). Garner et al. (*2, 3*) predicted helical orbitals for twisted linear carbon chains and showed that their torsion profiles are remarkably different for even and odd chains. An even-bond cumulenic chain has its energy minimum at a twist angle $\varphi = 90°$ and its π-system has Möbius topology, based on a co-arctate Möbius orbital basis. In contrast, an odd-bond cumulenic chain has its energy minimum at a twist angle $\varphi = 0°$ and its π-system has Hückel topology (*2*). The helicality of orbitals was shown to increase with increasing twist angle and to be larger for cumulenic compared to polyynic chains (*17*). In **2**, an even- and an odd-bond chain are connected and bent into a ring. Additionally, their π-systems couple via the *p* orbitals ($p_z$ in the case of $\varphi = 0°$) of the shared $sp^2$ carbons, and via hyperconjugation. As a result of this coupling, the equilibrium value of the tilt angles of the Cl–C bonds with respect to the overall ring structure will result from competing strain from both segments. The twist angle $\varphi$ is the sum of the out-of-plane tilt angles $\theta$ of the Cl–C bonds (**Scheme S1** and **fig. S4**). In the absence of such competing effects (such as in $C_{12}Cl_2$ or $C_{14}Cl_2$, see **Table S3**, we expect a planar ring and a twist angle of $\varphi = 0°$, i.e., the minimization of bond angle strain.

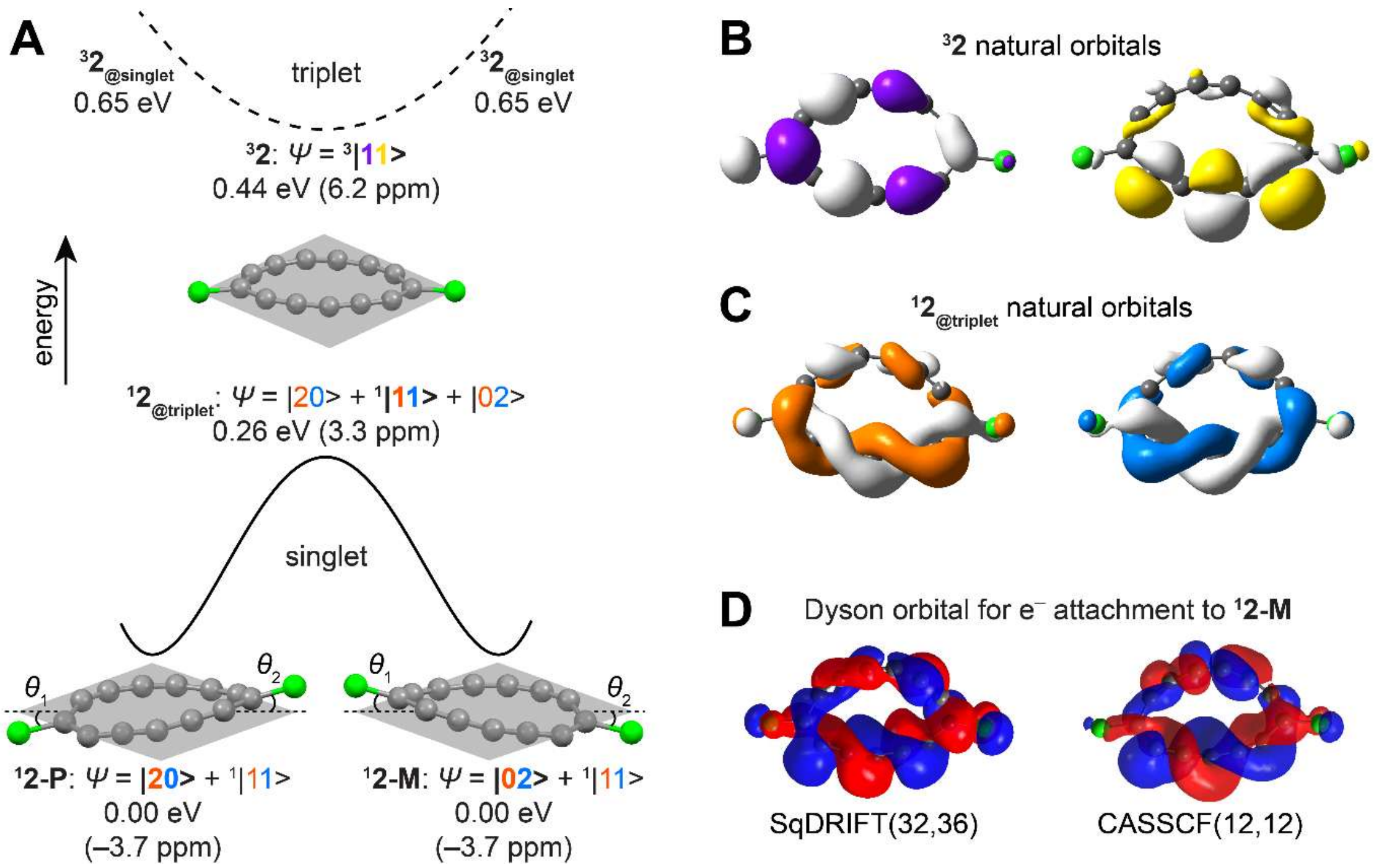

**Figure 3. Multireference calculations of $C_{13}Cl_2$.** (**A**) Energy diagramin the relaxed geometries of the singlet $^{1}$**2**, and the triplet $^{3}$**2**, adsorbed on NaCl (not shown). C atoms gray, Cl atoms green, tilt angles $\theta$ are indicated. NICS values in ppm for the compounds in the gas phase. $^{1}$**2$_{@triplet}$** ($^{3}$**2$_{@singlet}$**) is the singlet (triplet) in the geometry of $^{3}$**2** ($^{1}$**2**)**.** Important configurations contributing to the respective states $\psi$ are labelled by occupations of the corresponding natural frontier orbitals, with the colors referring to the natural orbitals shown in (B) and (C) and **fig. S6**, with the dominant configuration for each geometry shown in bold. (**B, C**) Natural frontier orbitals of $^{3}$**2** and $^{1}$**2**, respectively, computed at the optimized geometry of the triplet in the gas phase. (See also **figs. S1** and **S4-S9** and **Tables S1** and **S2**.) (**D**) Dyson orbital for electron attachment to $^{1}$**2-M** in the on-surface optimized geometry. Orbital on the left obtained from SqDRIFT-calculated wave functions in (33,36) and (32,36) active spaces using 72-qubit circuits, see also **fig. S10**. Orbital on the right obtained by complete active space self consistent field (CASSCF) calculations with (13,12) and (12,12) active spaces.

Our multireference calculations on NaCl and in the gas phase (**fig. S5**), predict as ground state for **2** a non-planar singlet $^{1}\mathbf{2}$ of chiral geometry, with out-of-plane tilts $\theta$ of the Cl-bonds by about 12°, corresponding to a twist angle $\varphi$ of about 24° (Fig. 3A and **figs. S4-S9** and associated text). For the triplet $^{3}\mathbf{2}$, a nearly planar geometry, $\varphi \approx 1°$ (see **Table S1**), with non-twisted molecular orbitals is predicted, which is higher in energy than the singlet $^{1}\mathbf{2}$ by 0.44 eV on NaCl. The planar equilibrium geometry of the triplet $^{3}\mathbf{2}$ and the near-complete separation of its orbitals into in-plane and out-of-plane π-systems, see Fig. 3, A and B, indicate its trivial Hückel topology and an orbital basis corresponding to a $GML^{0}_{4}$ body (Fig. 2D).

The out-of-plane, chiral, geometric distortion in the singlet $^{1}\mathbf{2}$ gives rise to finite C–Cl tilt angles $\theta$ and twisted molecular orbitals (Fig. 3, A and C). The geometric distortion does not originate from simple bond-angle strain and suggests a change of the π-system topology. In the triplet-optimized geometry of $^{3}\mathbf{2}$, our calculations find that the singlet state, that is $^{1}\mathbf{2}_{\mathbf{@triplet}}$, remains the electronic ground state. This nearly planar singlet $^{1}\mathbf{2}_{\mathbf{@triplet}}$ is a diradicaloid (*29, 30*), that is, a multireference state with contributions from both open-shell and closed-shell configurations, with mostly open-shell ($^{1}|11\rangle$) character (see Fig. 3A). Its frontier orbitals (Fig. 3C) are approximately mirror symmetric to each other and show pronounced twists of opposite helicity in the six-bond segment. In the seven-bond segment, these orbitals show very similar out-of-plane densities. Structural relaxation of the singlet geometry from planarity lowers the energy of one of the two closed-shell configurations (e.g., $|20\rangle$), making it the dominant contribution to the wavefunction, while increasing the energy of the other closed-shell (e.g., $|02\rangle$) and the open-shell configuration $^{1}|11\rangle$. This symmetry breaking is also accompanied by anti-aromaticity relief, see the NICS values shown in Fig. 3A. The symmetry-breaking ($C_{\mathrm{s}} \rightarrow C_{1}$) is a helical analogue of the pseudo-Jahn-Teller effect in cyclobutadiene ($D_{4\mathrm{h}} \rightarrow D_{2\mathrm{h}}$), (*31, 32*) and thus can be considered a helical pseudo-Jahn-Teller effect (*31, 33, 34*). According to our calculations it lowers the energy by 0.26 eV. Much like its planar analogue in cyclobutadiene, the helical pseudo-Jahn-Teller effect lifts the degeneracy of the frontier orbitals of the singlet, increasing the closed-shell character of **2**, with one of the formerly degenerate closed-shell configurations becoming dominant.

The shapes of the frontier orbitals of $^{1}\mathbf{2}$ showing pronounced twisting in the six-bond segment, but not in the seven-bond segment, see Fig. 3C, indicate that the distortion is caused by a helical electronic structure of the six-bond segment, suggesting a description based on an orbital basis with a twist within the six-bond segment, shown in Fig. 2, C and E. In such a basis the six-bond segment electronically couples the in-plane orbital of one $sp^2$ carbon to the out-of-plane orbital of the other $sp^2$-hybridized carbon, whereas the seven-bond segment couples in-plane with in-plane and out-of-plane with out-of-plane orbitals. The topology of this basis corresponds to a $GML^{1}_{4}$ body. The $GML^{1}_{4}$ topology of such an orbital basis differs from those of molecules with Hückel and Möbius π-systems. Separated in-plane and out-of-plane π-systems of a $GML^{0}_{4}$ topology mix into one π-system of $GML^{1}_{4}$ topology. The $GML^{1}_{4}$ topology implies a sign change, i.e., a phase shift of π of the orbital system, after two circulations, shown in Fig. 2C, in contrast to a sign change after one circulation for Möbius systems. We performed tight-binding calculations using such a basis and obtained frontier orbital densities that approximate those of the multireference calculations of $^{1}\mathbf{2}$ (**fig. S9**).

The finite twist angle of the relaxed singlet $^{1}\mathbf{2}$ results in chirality, meaning that we distinguish the singlet, enantiomers $^{1}$**2-P** and $^{1}$**2-M** with positive (clockwise) and negative (anticlockwise) twist angle $\varphi$ relative to the six-bond segment (see **Scheme S1**), respectively. The thread of the orbital basis is reflected in the molecular geometry, specifically in out-of-plane positions of the Cl heteroatoms defining the structural twist angle $\varphi$ (**Scheme S1** and **fig. S4**), and carbon atoms of the ring. The thread of the orbital basis is opposite to the structural twist of the six-bond segment $\varphi$. The basis twists left-handed ($GML^{-1}_{4}$) in $^{1}$**2-P**, see Fig. 2C, and right-handed ($GML^{1}_{4}$) in $^{1}$**2-M**, see Fig. 2E. For a finite twist angle $\varphi$ such orbital basis twist by $\mp(90° - \varphi)$ in the six-bond segment and by $\mp\varphi$ in the seven-bond segment, resulting in an overall twist of the orbital basis by –90° or +90° in one circumnavigation, see Fig. 2, C and E, respectively.

We term the π-system topology related to a $GML^{1}_{4}$ orbital basis, half-Möbius. The observables of such topology in **2** include non-zero tilt angles $\theta$ of the Cl–C bonds and those resulting from the twisted molecular orbitals. In contrast, a Hückel topology of the π-system would imply a planar geometry of **2** and non-twisted molecular orbitals that are separated into in-plane and out-of-plane orbital systems.

Observables of systems with strong multireference character can strongly depend on the number of correlated electrons (*35*). In the case of **2**, experimentally accessible observables are Dyson orbitals corresponding to molecular charge transitions that can be probed by STM (see below). To validate our findings, we performed ab initio calculations of the energies (see **fig. S10**) and the electron attachment Dyson orbital for $^{1}$**2** in an active space including 32 electrons, which contains all 24 carbon π electrons and four electrons per chlorine. These calculations were done using SqDRIFT, an approximate full configuration interaction quantum computing algorithm (*36*). Using current quantum hardware, SqDRIFT enables large-scale quantum chemical calculations without resorting to brute force full configuration interaction. Our SqDRIFT calculations (see Fig. 3D left and **figs. S10**) found no substantial changes relative to the classical calculations using an active space of 12 correlated electrons used in the remainder of this work, suggesting that the (12,12) active space is sufficient to describe the electronic structure of **2**. Implementing the quantum algorithm for the system considered here on a quantum processor required 72 qubits, rendering it one of the largest sample-based calculations to date (*37*), and exceeding the capability of exact classical simulations.

## Experiment

We generated isomers of **2** (see Fig. 2A) from the precursor $C_{13}Cl_{10}$ (**1**) on bilayer (two atomic layers) NaCl on Au(111) (*22*), and investigated their structural and electronic properties by atomic force microscopy (AFM) and scanning tunneling microscopy (STM). As we show in the following, we identified and characterized the molecule in its non-planar, chiral singlet state geometries $^{1}$**2-M** and $^{1}$**2-P** and in its planar triplet state geometry $^{3}$**2**. Reversible switching of the molecular topology (*38*) was achieved by atom manipulation, between all three configurations. Moreover, we imaged the helical orbital density of the lowest unoccupied molecular orbital (LUMO) of $^{1}$**2-M**, visualizing its twisted helical orbital density, in agreement with theoretical results for a $GML^{1}_{4}$ topology.

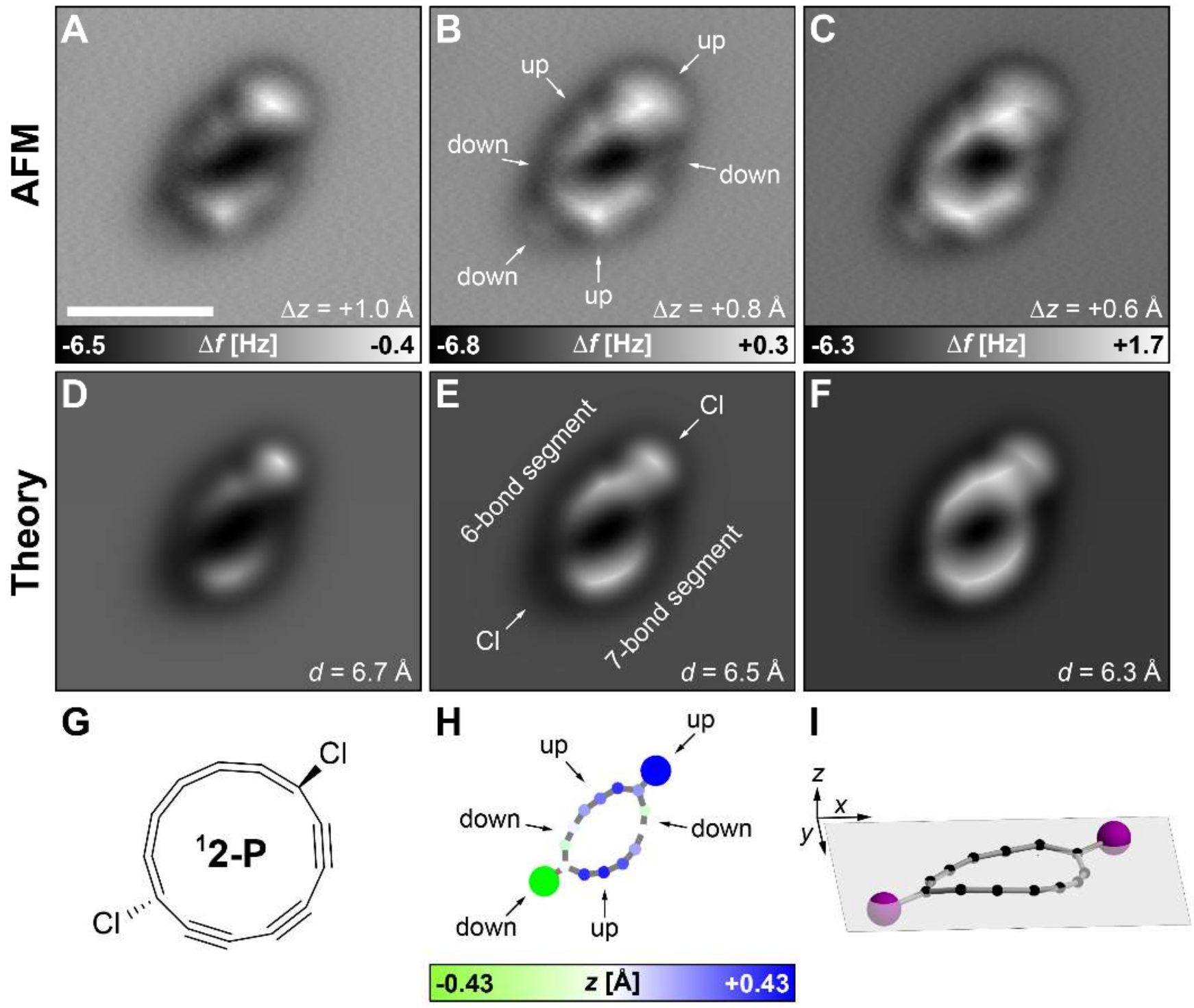

**Figure 4**. **Characterization of the singlet $^1$2-P**. (**A–C**) Constant-height CO-tip AFM data on defect-free bilayer NaCl on Au(111) at different tip-height offsets $\Delta z$. (**A**) AFM-far; (**B**) AFM-intermediate, with indicated out-of-plane distortions deduced from the AFM data ("down": toward the surface, "up": away from the surface); (**C**) AFM-close. Parameters: Setpoint $V = 100$ mV, $I = 0.9$ pA, tip-height offsets $\Delta z$ from the setpoint indicated. (**D–F**) AFM simulations based on the relaxed geometry of **$^1$2-P** on NaCl using the Probe-Particle Model (*39*); tip-molecule distances $d$ and (E) molecular moieties indicated. (**G**) Lewis structure. (**H**) Relaxed geometry of **$^1$2-P** adsorbed on NaCl, with relative out-of-plane displacements of the atoms indicated by color. (**I**) Corresponding 3D-representation with a half-transparent plane, parallel to the NaCl surface, to visualize the out-of-plane distortion of the molecule. (See **figs. S13** to **S17** for experimental determination and calculations of the adsorption site and geometry of **$^1$2-P**, **$^1$2-M** and **$^3$2**).

With the tip placed above the precursor molecule **1**, eight Cl atoms were dissociated by applying voltage pulses of $V = 4.5$ V to 5 V and tunneling currents $I$ on the order of 1 pA. We obtained different isomers of $C_{13}Cl_2$ (see **Scheme S2** for an expanded reaction scheme). All regioisomers of $C_{13}Cl_2$ other than **2** displayed planar geometries, indicating triplet ground states (**figs. S11** and **S12**). Figure 4, A to C, show CO-tip AFM data (*40*) of **2** adsorbed on defect-free bilayer NaCl, that is, without defects or adsorbates in the vicinity of the molecule. AFM with a CO tip provides information about the molecular geometry (*41*) and bond order (*39*). The AFM contrast of the molecule shown in Fig. 4, A to C, suggested a chiral geometry and a pronounced out-of-plane topography of the molecule including out-of-plane distortions of the molecular ring. Figure 4, D to F, show results obtained with AFM simulations (*39*) based on the geometries of the multireference calculations of **$^1$2-P** on NaCl (Fig. 4, H and I and **fig. S8**). The chiral molecular geometry reflected the effect of the chiral electronic structure on the geometry, due to the helical pseudo-Jahn-Teller effect (see above).

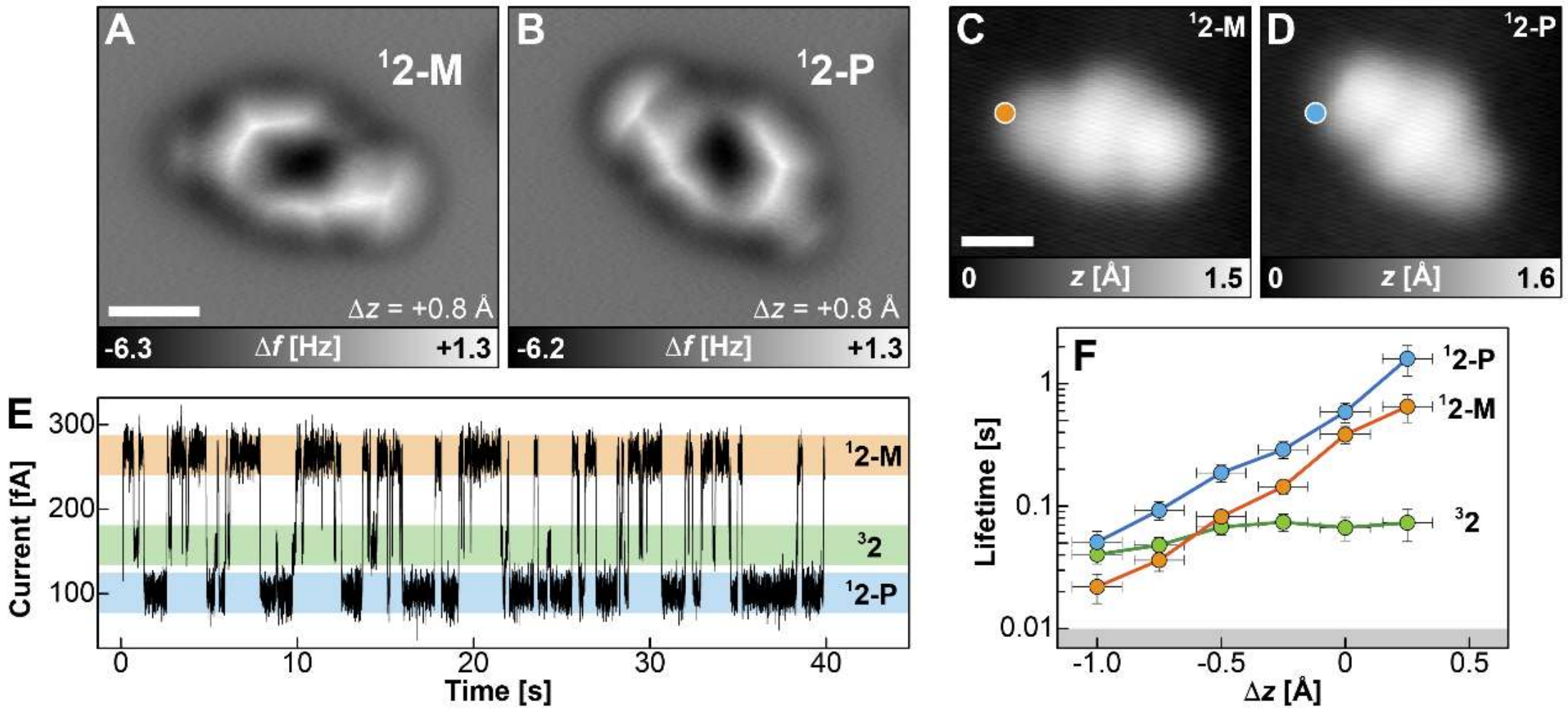

**Figure 5. Induced transitions.** (**A, B**) Constant-height AFM data (at $V$ = 0 V) of **$^1$2-M** and **$^1$2-P**, respectively. (**C, D**) Corresponding STM data (at $V$ = 150 mV, $I$ = 0.4 pA). (**E**) $I(t)$ data at $V$ = 250 mV and constant tip height, setpoint $V$ = 400 mV, $I$ = 0.2 pA, tip-height offset $\Delta z$ = 0 Å. Background color bars indicate three observed current plateaus (low current: blue; intermediate current: green; high current: orange). (**F**) Average lifetimes of the three current plateaus at different tip-height offsets $\Delta z$ at $V$ = 250 mV. All data were obtained on the same individual molecule at the lateral position indicated in C and D. For $|V|$ < 200 mV, the molecule was only stable in the high and low current states, corresponding to **$^1$2-M** and **$^1$2-P**, respectively. The plateau at intermediate current corresponds to a state that was not stable on defect-free NaCl, that we assign as **$^3$2**. (For AFM and STM data at different $V$ see **figs. S15** and **S18**, and for additional $I(t)$ data as a function of $\Delta z$ and $V$ see **figs. S19** to **S24**.)

When tunneling was induced at increased sample voltages $|V|$ > 210 mV, we observed sudden changes in the contrast corresponding to switching events (for STM images at such conditions see **fig. S18**). AFM and STM images below the switching threshold, that is, at $|V|$ < 210 mV, obtained after switching events, revealed the molecule with either of two mirrored contrasts (Fig. 5, A to D). The AFM data suggested that the handedness of **$^1$2** was switched between the two enantiomers **$^1$2-M** and **$^1$2-P**. The AFM contrast, the adsorption position and the orientation of **$^1$2-M** and **$^1$2-P** all agreed with theory (see **fig. S14**).

We obtained $I(t)$ data with the tip located above the molecule (Fig. 5E). Switching events could be observed as steps between three different current plateaus. The high- and the low-current plateau in Fig. 5E corresponded to the molecular configurations **$^1$2-M** and **$^1$2-P**, imaged by AFM in Fig. 5, A and B, and by STM in Fig. 5, C and D, respectively. The currents being high and low resulted mainly from different adsorption orientations of the molecule with respect to the (fixed) tip position during the $I(t)$ measurement. The states were assigned to the currents by comparison to the related STM and AFM images (Fig. 5, C and D, and **fig. S18**), and by imaging the molecule after induced transitions in the stable **$^1$2-M** and **$^1$2-P** states. Note that for the spectroscopy we deliberately choose a tip position, at which the three states could be distinguished by their different currents. Figure 5F shows the extracted lifetimes (average times the states were occupied) for different tip-height offsets $\Delta z$ (and thus different tunneling currents *I*). The state related to the plateau of intermediate currents (green in Fig. 5E) exhibits a relatively short lifetime even at large $\Delta z$, and its lifetime depended less on $\Delta z$ (see **fig. S21**) and also less on *V* (see **fig. S22**) as compared to transitions out of $^{1}$**2-M** and $^{1}$**2-P**. The plateau at intermediate currents we assigned to the triplet state, $^{3}$**2** (see **figs. S15** to **S24**).

Analysis of the *I*(*t*) data at different tip-height offsets $\Delta z$ (and different *V*, see **fig. S22**) indicates switching induced by tunneling electrons for transitions out of $^{1}$**2-M** and $^{1}$**2-P** (Fig. 5F and **fig. S21**). The voltage threshold of 0.2 V in the experiment agrees reasonably well with the calculated energy difference between $^{1}$**2** and $^{3}$**2** on the order of few 0.1 eV (Fig. 3). The lifetime of $^{3}$**2** depended much less on the current and decreased only by a factor of two when increasing *I* by a factor of ten from 0.1 pA to 1 pA (**fig. S21**). This observation suggested that the transitions from $^{3}$**2** were not predominantly triggered by tunnelling electrons, but might be assisted by inelastic excitations. At voltages below the switching threshold, the molecule was not stable in $^{3}$**2** on defect-free NaCl. We assume that $^{3}$**2** decayed with an intrinsic lifetime on the order of 0.1 s to $^{1}$**2**$_{\mathbf{@triplet}}$ (the value at the largest tip height probed in Fig. 5F provides a rough estimate). The consecutive Jahn-Teller distortion from to $^{1}$**2**$_{\mathbf{@triplet}}$ to either $^{1}$**2-M** or $^{1}$**2-P** (Fig. 3) likely happened on a much faster time scale. At increased currents and voltages, the intrinsic lifetime of $^{3}$**2** was only moderately reduced which we relate to inelastic tunnelling events, the electric field, or both effects that might reduce the potential barrier for transitions.

However, the triplet state $^{3}$**2** became the state with the longest lifetime for voltages well above the switching threshold ($V > 280$ mV, at currents *I* on the order of 1 pA, see **fig. S22**). We attributed this stability to the lifetime of $^{3}$**2** being less reduced by tunneling electrons, compared to the lifetimes of $^{1}$**2-M** and $^{1}$**2-P**. AFM (and STM) images, in which the molecule was most of the time in the $^{3}$**2** state were obtained at parameters at which the molecule was switching faster than the AFM (STM) bandwidth, see **fig. S15** (**S18**). They revealed a mirror symmetry and an image that results from switching between different adsorption sites and orientations, which agreed with a superposition of the two lowest energy adsorption sites of $^{3}$**2** on NaCl (see **fig. S15**).

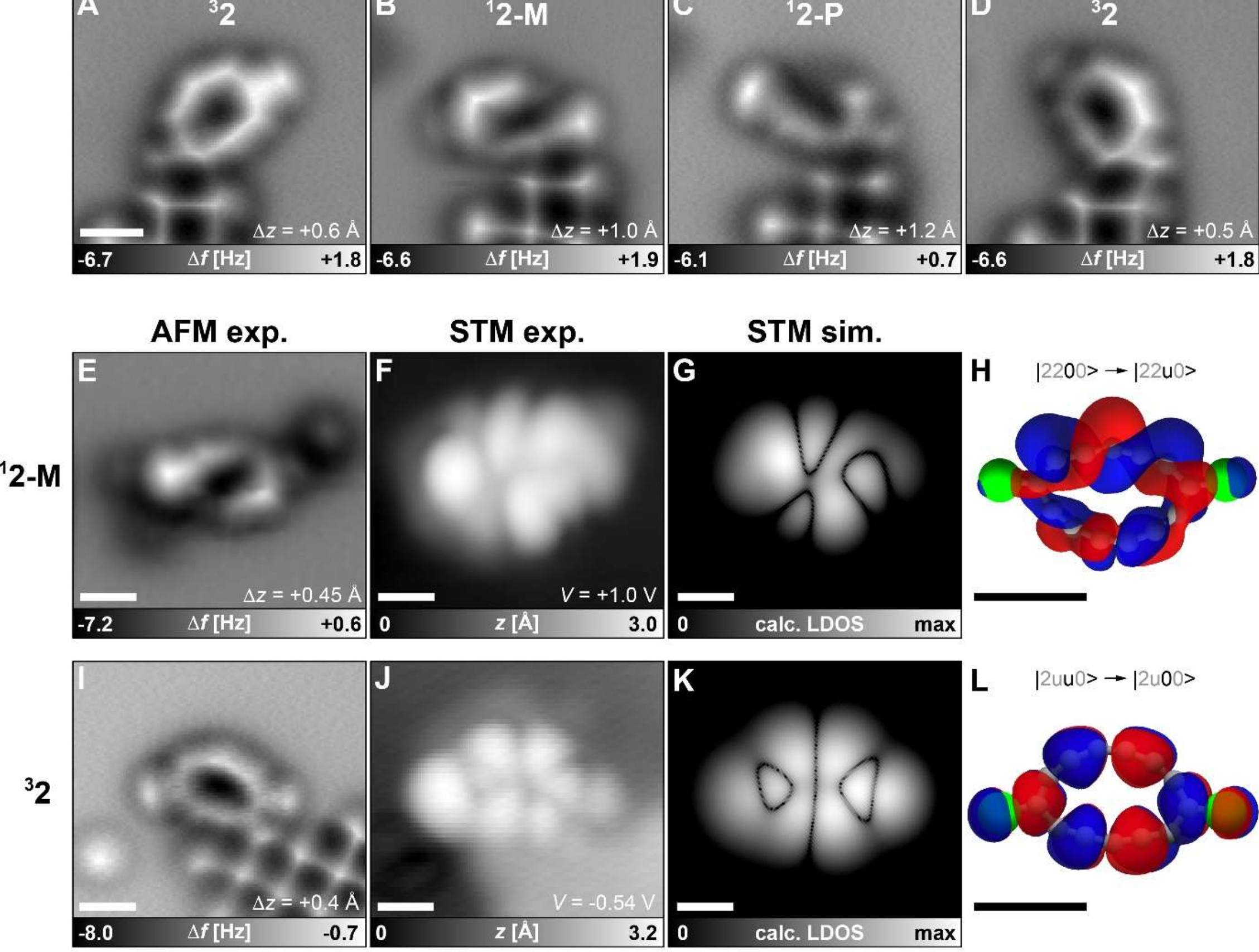

**Figure 6**. **Stabilized geometries and imaging of helical orbital density**. (**A–D**) AFM data of **2** adsorbed near Cl adatoms on bilayer NaCl on Au(111). By increasing the voltage ($V > 250$ mV during STM imaging) the molecule was manipulated between different adsorption sites that stabilized different states, that is, (A) $^{3}$**2**, (B) $^{1}$**2-M**, (C) $^{1}$**2-P**, and (D) $^{3}$**2**. Setpoint: $V = 200$ mV, $I = 0.6$ pA. (**E**) AFM data (setpoint: $V = 200$ mV, $I = 0.6$ pA) of $^{1}$**2-M** adsorbed on bilayer NaCl, next to three Cl adatoms. At this site the $^{1}$**2-M** state was sufficiently stable for obtaining an (**F**) STM image of the orbital density at the negative ion resonance ($V = 1.0$ V, $I = 0.4$ pA). (**G**) Simulated STM orbital density image (*43*) based on (**H**) the Dyson orbital of the neutral to anionic transition of $^{1}$**2-M** on NaCl. (**I**) AFM data (setpoint: $V = 200$ mV, $I = 1.2$ pA) of $^{3}$**2** adsorbed on bilayer NaCl next to a 3$^{rd}$ layer NaCl island. At this site the $^{3}$**2** state was sufficiently stable for obtaining an (**J**) STM image of the orbital density at the positive ion resonance ($V = –540$ mV, $I = 0.5$ pA). (**K**) Simulated STM orbital density image based on (**L**) the Dyson orbital of the neutral to cationic transition of $^{3}$**2** on NaCl. For details of the calculated Dyson orbitals see **fig. S9** and for STM simulations see **figs. S25** and **S26**. Scale bars 5 Å.

Near defects such as adsorbates or third-layer NaCl island edges, the potential energy landscape was modified and different molecular states could become stable (*22, 44, 45*). Fields from uncompensated ionic charges in the vicinity ($Cl^{–}$ and $Na^{+}$ in the edge and $Cl^{–}$ adatoms) could stabilize different states (*46*). We observed adsorption sites that stabilized a more planar geometry of **2**, see Fig. 6, A and D. Calculations showed that the triplet $^{3}$**2** was adsorbed with a nearly planar geometry, but the molecular plane was tilted with respect to the surface (**fig. S15**). The tilt gave rise to different brightness of the Cl atoms in Fig. 6, A and D (**figs. S16** and **S17**). Based on the good agreement between experiment and theory (Fig. 3), we assigned the observed planar geometry to $^{3}$**2**. In contrast to $^{1}$**2**, the carbon ring of $^{3}$**2** was planar, resulting in an AFM contrast that showed little variation of brightness above the ring.

Note that for the adsorption sites shown in Fig. 5 and in Fig. 6, A to D, increased voltages $|V| > 250$ mV led to switching between different configurations and adsorption sites (**fig. S18**). However, at other adsorption sites, such as near Cl adatoms, configurations of **2** were additionally stabilized. At the adsorption site shown in Fig. 6E, next to three Cl adatoms, $^{1}$**2-M** was stable enough for obtaining STM data at its negative ion resonance (NIR) at $V = 1.0$ V without switching or displacing it. The STM orbital density image (Fig. 6F) reflected the LUMO density of $^{1}$**2-M** (*47*). Comparison to a simulated STM image (Fig. 6G) (*43*), derived from the corresponding calculated Dyson orbital of $^{1}$**2-M** (Fig. 6H and **figs. S9** and **S25**) showed an excellent match. The helical shape of the orbital implied a twisted orbital basis and that the π orbitals of $^{1}$**2** were not separated into in-plane and out-of-plane systems. In contrast, for $^{3}$**2** (Fig. 6I) orbital images (Fig. 6J and **fig. S16**) showed a non-twisted, out-of-plane orbital density, consistent with a $GML^{0}_{4}$ topology and matching the simulated STM image (Fig. 6K) based on the Dyson orbital for electron detachment from $^{3}$**2** (Fig. 6L and **figs. S9** and **S26**).

## **Discussion**

The change in topology between the triplet and singlet and the observed switching may be rationalized by looking at the occupation numbers of the in-plane and out-of-plane systems. In the triplet state there are 13 electrons in the out-of-plane system and 11 in the in-plane system, which matches with the numbers of *p*-orbitals contributing to these systems in a Hückel basis. This is consistent with the triplet remaining planar. The energy of the singlet can usually be lowered by having closed-shell contributions. But a single-reference, closed-shell singlet state with odd occupation numbers of the two π-systems in Hückel topology is not possible. However, by mixing the in-plane and out-of-plane systems of a Hückel $GML^{0}_{4}$ basis in a $GML^{1}_{4}$ orbital basis, a closed-shell configuration with 24 electrons in the mixed π-system can be achieved. The singlet $^{1}$**2**$_{\text{@triplet}}$ is a multireference state of electronic configurations with opposite helicity of orbitals, which will lead to a helical Jahn-Teller distortion, stabilizing one of these helical configurations. In contrast, the triplet state in the non-planar geometry of $^{1}$**2-P** or $^{1}$**2-M**, that is, $^{3}$**2**$_{\text{@singlet}}$, is Jahn-Teller inactive, because the Jahn-Teller split former HOMO and LUMO of opposite helicity are occupied with one same-spin electron each.

A Möbius $GML^{1}_{2}$ topology entails the boundary conditions of a sign change for one complete circumnavigation and a periodic boundary condition for two circumnavigations. This boundary condition is analogous to the one of an electron under a perpendicular magnetic field equal to ½ of the Dirac magnetic flux quantum (*48-50*). Moreover, this boundary condition associated with the Möbius topology results in a real-space Berry phase of π (*51-54*), with the Berry phase being the phase difference acquired over the course of a cyclic evolution in parameter space (*51*).

The half-Möbius $GML^{1}_{4}$ topology entails a boundary condition with a wavefunction's sign change, but only after *two* complete circumnavigations. This entails a Berry phase of π for *two* complete circumnavigations, which can nominally be written as a Berry phase of π/2;

however, interference effects become observable only after two full circumnavigations. Although a Berry phase of $\pi/2$ would suggest broken time reversal symmetry, the implications of the present boundary condition remains to be clarified (see **Note S2**). Along the same lines, the analogy to the Möbius $GML^{1}_{2}$ topology (*52-54*) suggests that the boundary condition of the half-Möbius $GML^{1}_{4}$ topology is analogous to a quantum ring threaded by a magnetic field of ¼ of the Dirac magnetic flux quantum.

## Conclusion

The AFM contrasts, determined adsorption sites and orbital density images provide evidence for a molecule being observed and switched between its two ground state singlet enantiomers $^{1}$**2-M** and $^{1}$**2-P** of half-Möbius $GML^{\pm 1}_{4}$ topologies, respectively, and its planar triplet state $^{3}$**2** of trivial Hückel $GML^{0}_{4}$ topology, see **fig. S27**. The singlets feature helical orbitals, resolved by STM. Multireference calculations revealed that the chiral geometry of the singlets results from a helical pseudo Jahn-Teller effect, consistent with a twist of the helical basis in the six-bond segment of the molecule.

A Möbius topology (with a phase shift of $\pi$ in one circulation) can give rise to projected orbital angular momentum eigenvalues $L_z$ of integer and half-integer values and pseudo-double-valued eigenfunctions (*49*). Thus, we conclude that the molecules $^{1}$**2-M** and $^{1}$**2-P** realized in this work, which feature a half-Möbius topology and thus periodicity only after multiples of $8\pi$ circumnavigations, can give rise to projected orbital angular momentum eigenvalues $L_z$ of integer, half-integer and quarter-integer values. The wave function of such systems can be characterized as pseudo-quadruple-valued (see **Note S1**). Consequences of the unusual nominal Berry phase of $\pi/2$ (see **Note S2**) remain to be explored (*55*).

A strong response to an external magnetic field as noted for Möbius topologies (*52*) can be expected. Currents in such molecules might result in large magnetic fields, due to their twisted molecular frontier orbitals (*56*). The different orbital character of the (helical) singlet and the (non-helical) triplet orbitals results in relatively large spin-orbit coupling (*57*) (2.3 $cm^{-1}$ at the singlet geometry, see ***Multireference calculations***) indicating an avenue for accessing attractive phenomena such as spin-momentum locking (*58*).

The implementation of the non-trivial topology realized here is not structurally encoded, but imposed by monovalent substituents, representing small perturbations. This enables switching between trivial and non-trivial topologies and even different chiralities on demand, offering the exploration of quasiparticle properties rooted in topology. This exploration could be done by leveraging advantages provided by quantum hardware, which enables fast and scalable modeling of topologically non-trivial systems challenging for standard electronic structure methods. Building on the half-Möbius topology, more complex molecules and molecular networks with braiding of connectivity and topology that might even be switchable, can be envisioned.

**Reference list**

(*1*) I. M. James, ed., *History of topology* (Elsevier, 1999).

(*2*) M. H. Garner, R. Hoffmann, S. Rettrup, G. C. Solomon, Coarctate and Möbius: The helical orbitals of allene and other cumulenes, *ACS Cent. Sci.* **4**, 688–700 (2018).

(*3*) M. H. Garner, W. Bro-Jørgensen, G. C. Solomon, Three distinct torsion profiles of electronic transmission through linear carbon wires, *J. Phys. Chem. C* **124**, 18968–18982 (2020).

(*4*) E. Heilbronner, Hückel molecular orbitals of Möbius-type conformations of annulenes, *Tetrahedron Letters* **5**, 1923–1928 (1964).

(*5*) H. E. Zimmerman, Möbius-Hückel concept in organic chemistry. Application of organic molecules and reactions," *Accounts Chem. Res.* **4**, 272–280 (1971).

(*6*) M. Mauksch, V. Gogonea, H. Jiao, P. v. R. Schleyer, Monocyclic $(CH)_9^+$ — A Heilbronner Möbius aromatic system revealed, *Angew. Chem. Int. Ed.* **37**, 2395–2397 (1998).

(*7*) D. Ajami, O. Oeckler, A. Simon, R. Herges, Synthesis of a Möbius aromatic hydrocarbon, *Nature* **426**, 819–821 (2003).

(*8*) R. Herges, Topology in chemistry: designing Möbius molecules, *Chem. Rev.* **106**, 4820–4842 (2006).

(*9*) Y. Tanaka, S. Saito, S. Mori, N. Aratani, H. Shinokubo, N. Shibata, Y. Higuchi, Z. S. Yoon, K. S. Kim, S. B. Noh, *et al.*, Metalation of expanded porphyrins: a chemical trigger used to produce molecular twisting and Möbius aromaticity, *Angew. Chem. Int. Ed.* **47**, 681–696 (2008).

(*10*) Z. S. Yoon, A. Osuka, D. Kim, Möbius aromaticity and antiaromaticity in expanded porphyrins, *Nat. Chem.* **1**, 113–122 (2009).

(*11*) G. R. Schaller, F. Topic, K. Rissanen, Y. Okamoto, J. Shen, R. Herges, Design and synthesis of the first triply twisted Möbius annulene, *Nat. Chem.* **6**, 608–613 (2014).

(*12*) Y. Segawa, T. Watanabe, K. Yamanoue, M. Kuwayama, K. Watanabe, J. Pirillo, Y. Hijikata, K. Itami, Synthesis of a Möbius carbon nanobelt, *Nat. Synth.* **1**, 535–541 (2022).

(*13*) W. Fan, T. M. Fukunaga, S. Wu, Y. Han, Q. Zhou, J. Wang, Z. Li, X. Hou, H. Wei, Y. Ni, H. Isobe, J. Wu, Synthesis and chiral resolution of a triply twisted Möbius carbon nanobelt, *Nat. Synth.* **2**, 880–887 (2023).

(*14*) H. S. Rzepa, Möbius aromaticity and delocalization, *Chem. Rev.* **105**, 3697–3715 (2005).

(*15*) C. H. Hendon, D. Tiana, A. T. Murray, D. R. Carbery, A. Walsh, Helical frontier orbitals of conjugated linear molecules, *Chem. Sci.* **4**, 4278–4284 (2013).

(*16*) P. W. Thulstrup, S. V. Hoffmann, B. K. Hansen, J. Spanget-Larsen, Unique interplay between electronic states and dihedral angle for the molecular rotor diphenyldiacetylene, *Phys. Chem. Chem. Phys.* **13**, 16168–16174 (2011).

(*17*) W. Bro-Jørgensen, M. H. Garner, G. C. Solomon, Quantification of the helicality of helical molecular orbitals, *J. Phys. Chem. A*, **125**, 8107–8115 (2021).

(*18*) I. Tavkhelidze, *Proceedings of Ukrainian Mathematical Congress, S* (2011), vol. 2, pp. 177–190.

(*19*) K. Kaiser, L. M. Scriven, F. Schulz, P. Gawel, L. Gross, and H. L. Anderson, An sp-hybridized molecular carbon allotrope, cyclo[18]carbon, *Science*, **365**, 1299–1301 (2019).

(*20*) L. Sun, W. Zheng, W. Gao, F. Kang, M. Zhao, and W. Xu, On-surface synthesis of aromatic cyclo[10]carbon and cyclo[14]carbon, *Nature*, **623**, 972–976 (2023).

(*21*) G. V. Baryshnikov, R. R. Valiev, L. I. Valiulina, A. E. Kurtsevich, T. Kurtén, D. Sundholm, M. Pittelkow, J. Zhang, H. Ågren, Odd-number cyclo[n]carbons sustaining alternating aromaticity, *J. Phys. Chem. A,* **126**, 2445–2452 (2022).

(*22*) F. Albrecht, I. Roncevic, Y. Gao, F. Paschke, A. Baiardi, I. Tavernelli, S. Mishra, H. L. Anderson, L. Gross, The odd-number cyclo[13]carbon and its dimer, cyclo[26]carbon, *Science,* **384**, 677–682 (2024).

(*23*) S. Martin-Santamaria, H. S. Rzepa, Double aromaticity and anti-aromaticity in small carbon rings, *Chem. Commun.* **16** 1503–1504 (2000).

(*24*) H. Fischer, H. Kollmar, Zur Invarianz in der LCAO MO Theorie. *Theoret. Chim. Acta,* **12**, 344–348 (1968).

(*25*) I. V. Alabugin, B. Gold, "Two functional groups in one package": using both alkyne π-bonds in cascade transformations, *J. Org. Chem.* **78**, 7777–7784 (2013).

(*26*) I. V. Alabugin, G. dos Passos Gomes, M. A. Abdo, Hyperconjugation, *WIREs Comput. Mol. Sci.* **9**, e1389 (2019).

(*27*) J. I.-C. Wu, P. v. R. Schleyer, Hyperconjugation in hydrocarbons: not just a "mild sort of conjugation", *Pure and Applied Chemistry* **85**, 921–940 (2013).

(*28*) C. S. Wannere, H. S. Rzepa, B. C. Rinderspacher, A. Paul, C. S. Allan, H. F. Schaefer III, P. v. R. Schleyer, The geometry and electronic topology of higher-order charged Möbius annulenes, *J. Phys. Chem. A* **113**, 11619–11629 (2009).

(*29*) M. Abe, Diradicals, *Chem. Rev.* **113**, 7011-7088 (2013).

(*30*) T. Stuyver, B. Chen, T. Zeng, P. Geerlings, F. De Proft, R. Hoffmann, Do diradicals behave like radicals? *Chem. Rev.* **119**, 11291–11351 (2019).

(*31*) B. Bersuker, Jahn–Teller and Pseudo-Jahn–Teller effects: from particular features to general tools in exploring molecular and solid state properties, *Chem. Rev.* **121**, 1463–1512 (2020).

(*32*) E. Monino, M. Boggio-Pasqua, A. Scemama, D. Jacquemin, P.-F. Loos, Reference energies for cyclobutadiene: Automerization and excited states, *J. Phys. Chem. A*, **126**, 4664–4679 (2022).

(*33*) M. H. Garner, R. Laplaza, C. Corminboeuf, Helical versus linear Jahn–Teller distortions in allene and spiropentadiene radical cations, *Phys. Chem. Chem. Phys.* **24**, 26134–26143 (2022).

(*34*) S. Hirotsu, Jahn-Teller induced phase transition in $CsCuCl_3$: structural phase transition with helical atomic displacements, *J. Phys. C: Solid State Phys.* **10**, 967 (1977).

(*35*) D. Jacob, J. Fernández-Rossier, Theory of intermolecular exchange in coupled spin-1/2 nanographenes, *Phys. Rev. B* **106**, 205405 (2022).

(*36*) S. Piccinelli, A. Baiardi, M. Rossmannek, A. C. Vazquez, F. Tacchino, S. Mensa, E. Altamura, A. Alavi, M. Motta, J. Robledo-Moreno, *et al*. Quantum chemistry with provable convergence via randomized sample-based quantum diagonalization, arXiv:2508.02578 (2025)

(*37*) T. Weaving, A. Mingare, A. Ralli, P. V. Coveney, Towards Compact Wavefunctions from Quantum-Selected Configuration Interaction, *arXiv:2509.02525* (2025).

(*38*) M. Stepien, B. Szyszko, L. Latos-Grazzynnski, Three-level topology switching in a molecular Möbius band, *J. Am. Chem Soc.* **132**, 3140–3152 (2010).

(*39*) P. Hapala, G. Kichin, C. Wagner, F. S. Tautz, R. Temirov, P. Jelínek, The mechanism of high-resolution STM/AFM imaging with functionalized tips, *Phys. Rev. B* **90**, 085421 (2014).

(*40*) L. Gross, F. Mohn, N. Moll, P. Liljeroth, G. Meyer, The Chemical Structure of a Molecule Resolved by Atomic Force Microscopy, *Science* **325**, 1110–1114 (2009).

(*41*) B. Schuler, W. Liu, A. Tkatchenko, N. Moll, G. Meyer, A. Mistry, D. Fox, L. Gross, Adsorption geometry determination of single molecules by atomic force microscopy, *Phys. Rev. Lett.* **111**, 106103 (2013).

(*42*) L. Gross, F. Mohn, N. Moll, B. Schuler, A. Criado, E. Guitián, D. Peña, A. Gourdon, G. Meyer, Bond-order discrimination by atomic force microscopy, *Science* **337**, 1326–1329 (2012).

(*43*) F. Paschke, L.-A. Lieske, F. Albrecht, C. J. Chen, J. Repp, L. Gross, Distance and voltage dependence of orbital density imaging using a CO-functionalized tip in scanning tunneling microscopy, *ACS nano* **19**, 2641–2650 (2025).

(*44*) Y. Gao, F. Albrecht, I. Roncevic, I. Ettedgui, P. Kumar, L. M. Scriven, K. E. Christensen, S. Mishra, L. Righetti, M. Rossmannek, I. Tavernelli, H. L. Anderson, L. Gross, On-surface synthesis of a doubly anti-aromatic carbon allotrope, *Nature* **623**, 977–981 (2023).

(*45*) S. Mishra, M. Vilas-Varela, L.-A. Lieske, R. Ortiz, S. Fatayer, I. Roncevic, F. Albrecht, T. Frederiksen, D. Peña, L. Gross, Bistability between π-diradical open-shell and closed-shell states in indeno [1, 2-a] fluorene, *Nat. Chem.* **16**, 755–761 (2024).

(*46*) G. dos Passos Gomes, I. V. Alabugin, Drawing catalytic power from charge separation: stereoelectronic and zwitterionic assistance in the Au (I)-catalyzed Bergman cyclization, *J. Am. Chem. Soc.* **139**, 3406–3416 (2017).

(*47*) J. Repp, G. Meyer, S. M. Stojkovic, A. Gourdon, C. Joachim, Molecules on Insulating Films: Scanning-Tunneling Microscopy Imaging of Individual Molecular Orbitals, *Phys. Rev. Lett.* **94**, 026803 (2005).

(*48*) Y. Anusooya-Pati, Z. Soos, A. Painelli, Symmetry crossover and excitation thresholds at the neutral-ionic transition of the modified Hubbard model, *Phys. Rev. B* **63**, 205118 (2001).

(*49*) E. Miliordos, Particle in a Möbius wire and half-integer orbital angular momentum, *Phys. Rev. A* **83**, 062107 (2011).

(*50*) L. Muechler, J. Maciejko, T. Neupert, R. Car, Möbius molecules and fragile Mott insulators, *Phys. Rev. B* **90**, 245142 (2014).

(*51*) M. V. Berry, Quantal phase factors accompanying adiabatic changes, *Proceedings of the Royal Society of London. A. Mathematical and Physical Sciences* **392**, 45–57 (1984).

(*52*) Z.-L. Guo, Z. Gong, H. Dong, C. Sun, Möbius graphene strip as a topological insulator, *Phys. Rev. B* **80**, 195310 (2009).

(*53*) M. A. Davidovich, E. V. Anda, J. R. Iglesias, G. Chiappe, Bohm-Aharonov and Kondo effects on tunneling currents in a mesoscopic ring, *Phys. Rev. B* **55**, R7335 (1997).

(*54*) J. Wang, S. Valligatla, Y. Yin, L. Schwarz, M. Medina-Sánchez, S. Baunack, C. H. Lee, R. Thomale, S. Li, V. M. Fomin, *et al.*, Experimental observation of Berry phases in optical Möbius-strip microcavities, *Nat. Photonics* **17**, 120–125 (2023).

(*55*) L. Levy, G. Dolan, J. Dunsmuir, H. Bouchiat, Magnetization of mesoscopic copper rings: Evidence for persistent currents, *Phys. Rev. Lett.* **64**, 2074 (1990).

(*56*) W. Bro-Jørgensen, S. P. Sauer, G. C. Solomon, M. H. Garner, Substantial Magnetic Fields Arising from Ballistic Ring Currents in Single-Molecule Junctions. *JACS Au*, **5**, 4073–4085 (2025).

(*57*) P. Baronas, R. Komskis, E. Tankeleviciute, P. Adomenas, O. Adomeniene, S. Jursenas, Helical molecular orbitals to induce spin–orbit coupling in oligoyne-bridged bifluorenes, *J. Phys. Chem. Lett.* **12**, 6827–6833 (2021).

(*58*) K. Gotlieb, C.-Y. Lin, M. Serbyn, W. Zhang, C. L. Smallwood, C. Jozwiak, H. Eisaki, Z. Hussain, A. Vishwanath, A. Lanzara, Revealing hidden spin-momentum locking in a high-temperature cuprate superconductor, *Science* **362**, 1271–1275 (2018).

(*59*) Zenodo: 10.5281/zenodo.15495263

(60) F. J. Giessibl, High-speed force sensor for force microscopy and profilometry utilizing a quartz tuning fork, *Appl. Phys. Lett.* **73**, 3956–3958 (1998).

(*61*) T. R. Albrecht, P. Grütter, D. Horne, D. Rugar, Frequency modulation detection using high-Q cantilevers for enhanced force microscope sensitivity, *J. Appl. Phys.* **69**, 668–673 (1991).

(*62*) I. Roncevic, F. J. Leslie, M. Rossmannek, I. Tavernelli, L. Gross, H. L. Anderson, Aromaticity reversal induced by vibrations in cyclo[16]carbon, *J. Am. Chem. Soc.* **145**, 26962–26972 (2023).

(*63*) G. Li Manni, I. Fdez. Galvan, A. Alavi, F. Aleotti, F. Aquilante, J. Autschbach, D. Avagliano, A. Baiardi, J. J. Bao, S. Battaglia, *et al.*, The OpenMolcas Web: A community-driven approach to advancing computational chemistry, *J. Chem. Theory Comput.* **19**, 6933–6991 (2023).

(*64*) Y. Nishimoto, Analytic gradients for restricted active space second-order perturbation theory (RASPT2), *J. Chem. Phys.* **154**, 194103 (2021).

(*65*) P.-Å. Malmqvist, B. O. Roos, The CASSCF state interaction method, *Chem. Phys. Lett.* **155**, 189–194 (1989).

(*66*) K. Aidas, C. Angeli, K. L. Bak, V. Bakken, R. Bast, L. Boman, O. Christiansen, R. Cimiraglia, S. Coriani, P. Dahle, *et al.*, The Dalton quantum chemistry program system, *WIREs Comput. Mol. Sci.* **4**, 269–284 (2014).

(*67*) Y. Hong, J. Oh, Y. M. Sung, Y. Tanaka, A. Osuka, D. Kim, The extension of Baird's rule to twisted heteroannulenes: Aromaticity reversal of singly and doubly twisted molecular systems in the lowest triplet state. *Angew. Chem. Int. Ed.* **129**, 2978–2982 (2017).

(*68*) J.-D. Chai, M. Head-Gordon, Long-range corrected hybrid density functionals with damped atom-atom dispersion corrections, *Phys. Chem. Chem. Phys.* **10**, 6615–6620 (2008).

(*69*) M. J. Frisch, G. W. Trucks, H. B. Schlegel, G. E. Scuseria, M. A. Robb, J. R. Cheeseman, G. Scalmani, V. Barone, G. A. Petersson, H. Nakatsuji, *et al*. Gaussian 16 Revision C. 02. 2016; Gaussian Inc. *Wallingford CT* (2016).

(*70*) S. Arulmozhiraja, T. Ohno, CCSD calculations on C14, C18, and C22 carbon clusters, *J. Chem. Phys.* **128**, 114301 (2008).

(*71*) J. J. Eriksen. The shape of full configuration interaction to come. *J. Phys. Chem. Lett.* **12**, 418–432 (2021).

(*72*) E. Campbell, Random Compiler for Fast Hamiltonian Simulation. *Phys. Rev. Lett*., **123**, 070503 (2019).

(*73*) S. Barison, J. R. Moreno, M. Motta, Quantum-centric computation of molecular excited states with extended sample-based quantum diagonalization, *Quantum Sci. Technol.* **10**, 25034 (2025).

(*74*) P. Jordan, E. Wigner, Über das Paulische Äquivalenzverbot, *Zeitschrift für Physik*, **47**, 631–651 (1928).

(*75*) A. Miessen, P. J. Ollitrault, F. Tacchino, and I. Tavernelli, Quantum algorithms for quantum dynamics, *Nat. Comput. Sci.* **3**, 25–37 (2023).

(*76*) M. Motta, K. J. Sung, K. B. Whaley, M. Head-Gordon, J. Shee, Bridging physical intuition and hardware efficiency for correlated electronic states: the local unitary cluster Jastrow ansatz for electronic structure. *Chem. Sci*. **14**, 11213–11227 (2023).

(*77*) N. Pavlicek, A. Mistry, Z. Majzik, N. Moll, G. Meyer, D. J. Fox, L. Gross, Synthesis and characterization of triangulene, *Nat. Nano.* **12**, 308–311 (2017).

(*78*) L. Gross, F. Mohn, P. Liljeroth, J. Repp, F. J. Giessibl, G. Meyer, Measuring the Charge State of an Adatom with Noncontact Atomic Force Microscopy, *Science*, **324**, 1428–1431 (2009).

(*79*) L. L. Patera, S. Fatayer, J. Repp, L. Gross, Probing Molecular Properties at Atomic Length Scale Using Charge-State Control, *Chem. Rev.* **125**, 5830–5847 (2025).

(*80*) B. Stipe, M. Rezaei, W. Ho, S. Gao, M. Persson, B. Lundqvist, Single-molecule dissociation by tunneling electrons, *Phys. Rev. Lett.* **78**, 4410–4413 (1997).

**Funding**: This work was funded by European Research Council grant no. 885606, ARO-MAT (H.L.A. and Y.G.); European Community Horizon 2020, grant project 101019310 CycloCarbonCatenane (Y.G. and H.L.A.); UKRI Horizon Europe Guarantee MSCA Postdoctoral Fellowship ElDelPath, EP/X030075/1 (I.R. and H.L.A.); European Research Council Synergy grant no. 951519, MolDAM (F.P., F.A., L.-A.L., J.R. and L.Gr.); University of Manchester (I.R). I.R. acknowledges the assistance given by Research IT and the use of the Computational Shared Facility at The University of Manchester.

**Acknowledgements**: The authors thank Diego Peña (University of Santiago de Compostela), Shantanu Mishra (Chalmers University of Technology), Lisanne Sellies (IBM Research), Titus Neupert (University of Zurich), Lukas Muechler (Pennsylvania State University), Sieglinde Pfaendler (IBM Research) and Alessandro Curioni (IBM Research) for discussions.

**Author contributions**: Conceptualization: I.R. and L.Gr.; Synthesis of the precursors: Y.G.; On-surface synthesis and STM and AFM measurements: L.-A.L., F.A. and L.Gr.; Theoretical analysis and computational simulations: F.P., L.Gö. and I.R.; quantum computations: S.B., S.P., A.B. and I.T. ; Writing: First draft I.R. and L.Gr. All authors commented on the manuscript and discussed the results. Competing interests: The authors declare that they have no competing interests.

**Data and materials availability:** All experimental data are reported in the main text and supplementary materials. Additional data, i.e., results of calculations, are deposited at (*59*) Zenodo: 10.5281/zenodo.15495263

**Supplementary Materials**

Materials and Methods

Supplementary Text

Supplementary Schemes S1 and S2

Supplementary Figures S1 to S27

Supplementary Tables S1 to S3

Supplementary Notes 1 and 2

References (*60-80*)

The Supplementary Materials contain information about the experimental methods, Supplementary Schemes S1 and S2, computational details and additional computational results (**figs. S1** to **S10** and Tables S1 to S3), additional experimental results and analyses (**figs. S11** to **S27**), Supplementary Notes 1 and 2, and references (*60*) to (*80*).

Supplementary Material

# A molecule with half-Möbius topology

**Authors:** Igor Rončević[1,2#]*, Fabian Paschke[3#], Yueze Gao[2], Leonard-Alexander Lieske[3], Lene A. Gödde[2], Stefano Barison[4,5], Samuele Piccinelli[4,6], Alberto Baiardi[4], Ivano Tavernelli[4], Jascha Repp[7], Florian Albrecht[3], Harry L. Anderson[2] and Leo Gross[3]*

**Affiliations:**

[1] Department of Chemistry, The University of Manchester, Oxford Road, Manchester, United Kingdom.

[2] Department of Chemistry, Oxford University, Chemistry Research Laboratory, Oxford, United Kingdom.

[3] IBM Research Europe – Zurich, Rüschlikon, Switzerland.

[4] IBM Quantum, IBM Research Europe – Zurich, Rüschlikon, Switzerland.

[5] Institute for Theoretical Physics, ETH Zürich, CH-8093 Zürich, Switzerland

[6] Institute of Physics, École Polytechnique Fédérale de Lausanne (EPFL), Lausanne, Switzerland

[7] Institute of Experimental and Applied Physics and Halle-Berlin-Regensburg Cluster of Excellence CCE, University of Regensburg, Regensburg, Germany.

[#] Equally contributing first authors

* Corresponding authors. Email: igor.roncevic@manchester.ac.uk; lgr@zurich.ibm.com

## 1. STM and AFM methods

The on-surface characterization and reactions were performed in a home-built combined STM/AFM, operated at a temperature $T = 5$ K in ultra-high vacuum. The decachlorofluorene precursor **1** (*22*) was thermally sublimed onto a cold ($T < 10$ K) Au(111) surface partially covered with NaCl bilayer (two atomic layer thick) islands. AFM measurements were performed in non-contact mode with a qPlus sensor (*60*). The sensor was operated in frequency-modulation mode (*61*) with the oscillation amplitude $A$ kept constant at $A = 0.5$ Å. All data were recorded on molecules adsorbed on bilayer NaCl with CO functionalized tips (*40*). STM images were recorded at constant current and AFM images at constant height. The STM-controlled setpoint for constant-height AFM images provides the tip height at the top center of the image at which position the tip-height offset $\Delta z$ was added. For spectroscopy, the STM setpoint corresponds to the lateral tip position at which the spectra had been obtained. Positive tip-height offsets $\Delta z$ correspond to an increase in tip-sample distance with respect to the setpoint. With AFM-far, we denote an imaging height just at the onset of atomic resolution. With AFM-close, we denote imaging at $\Delta z$ further reduced by about few 0.1 Å, with respect to AFM-far. AFM images were acquired at $V = 0$ V, if not stated otherwise (in **fig. S15A**).

The bias voltage $V$ was applied to the sample with respect to the tip. Dehalogenation of **1** and skeletal rearrangement towards a carbon ring was induced by applying voltage pulses with typically $V = 4.5$ V for a few 100 ms and at constant height with the tip above the molecule and the tip being retracted by 6 to 8 Å from a setpoint of $I = 1$ pA and $V = 0.2$ V, resulting in currents $I$ on the order of few pA at $V = 4.5$ V (*22*).

## 2. Supplementary Schemes

**Scheme S1**. **Nomenclature of $^{1}$2-P and $^{1}$2-M**. The out-of-plane locations of the Cl heteroatoms give rise to a helical twist. The Cl-C bonds are tilted out of the molecular plane by $\theta_1$ and $\theta_2$ (see **fig. S4**). In the **P** enantiomer, the rotation of the second C-Cl bond with respect to the first, upon moving through the shorter segment of the ring, in **2** the six-bond segment, is clockwise by $\varphi = \theta_1 + \theta_2$, whereas in the **M** enantiomer it is counterclockwise by $\varphi$. Positive (negative) angles correspond to clockwise (counterclockwise) rotations. The sign of the tilt angles $\theta_1$ and $\theta_2$ is defined with respect to movement form the shorter segment to the longer segment. With these definitions, positive (negative) tilt angle $\varphi$ and twist angles $\theta_1$, $\theta_2$ are observed in $^{1}$**2-P** ($^{1}$**2-M**). The diagrams in the top row show a view parallel to the shorter segment and the resulting twist angles $\varphi$. Note that the winding of the half-Möbius orbital basis is left-handed (anticlockwise) in $^{1}$**2-P** and in hypothetical $^{1}$**4-P**, see **Scheme S2**, because in these the shorter segment has an even number of bonds, but right-handed (clockwise) in hypothetical $^{1}$**3-P**, because in it the shorter segment has an odd number of bonds.

**Scheme S2. Expanded reaction scheme, showing other isomers of $C_{13}Cl_2$.** From **1** we generated **2**, **3** and **4** by tip-induced dissociation of 8 Cl atoms. As discussed in the main text, **2** was observed in the singlets $^{1}$**2-M** and $^{1}$**2-P** and the triplet $^{3}$**2**. For **3**, we only observed the planar triplet $^{3}$**3** (see **fig. S11**). For **4**, we only observed the planar triplet $^{3}$**4** (see **fig. S12**).

## 3. Computational details

### *3.1. Tight-binding calculations*

Following the approach of Garner and Hoffman (*2*), a tight-binding nearest-neighbor Hamiltonian for linear carbon chains with *N* sp-hybridized atoms can be written as:

$$\hat{H}_{nn} = -\sum_{\langle i,j \rangle}^{N} \sum_{\langle k,l \rangle}^{2} t_{kl} c_{ik}^{\dagger} c_{jl} \quad (1)$$

in which $\langle i,j \rangle$ denotes summation over neighboring atoms *i* and *j*, and $\langle k,l \rangle$ denotes summation over two orthogonal $p$-orbitals (green and purple in **fig. S1**), which are coupled by $t_{kl}$. In the case that *i* or *j* is $sp^2$-hybridized, the sum only runs to 1. The $t_{kl}$ coupling depends on the angle between the orbitals *k* and *l* on neighboring atoms *i* and *j*:

$$t_{kl} = t_0 \cos \phi_{ijkl} \quad (2)$$

In a non-helical topology, we can assign these orbitals as $p_z$, (e.g., when $k, l = 1$), and $p_{xy}$ $(k, l = 2)$, which gives trivial solutions:

$$k = l \colon \phi_{ij11} = \phi_{ij22} = 0;\ t_{kl} = t_0 \quad (3)$$

$$k \neq l \colon \phi_{ij12} = \phi_{ij21} = \frac{\pi}{2};\ t_{kl} = 0 \quad (4)$$

In a helical basis shown in **fig. S1** an angle of $2\theta$ is accumulated over *N* atoms, which gives:

$$\phi_{ij11} = \phi_{ij22} = \frac{2\theta}{N-1} \quad (5)$$

$$\phi_{ij12} = \phi_{ij21} = \frac{\pi}{2} - \frac{2\theta}{N-1} \quad (6)$$

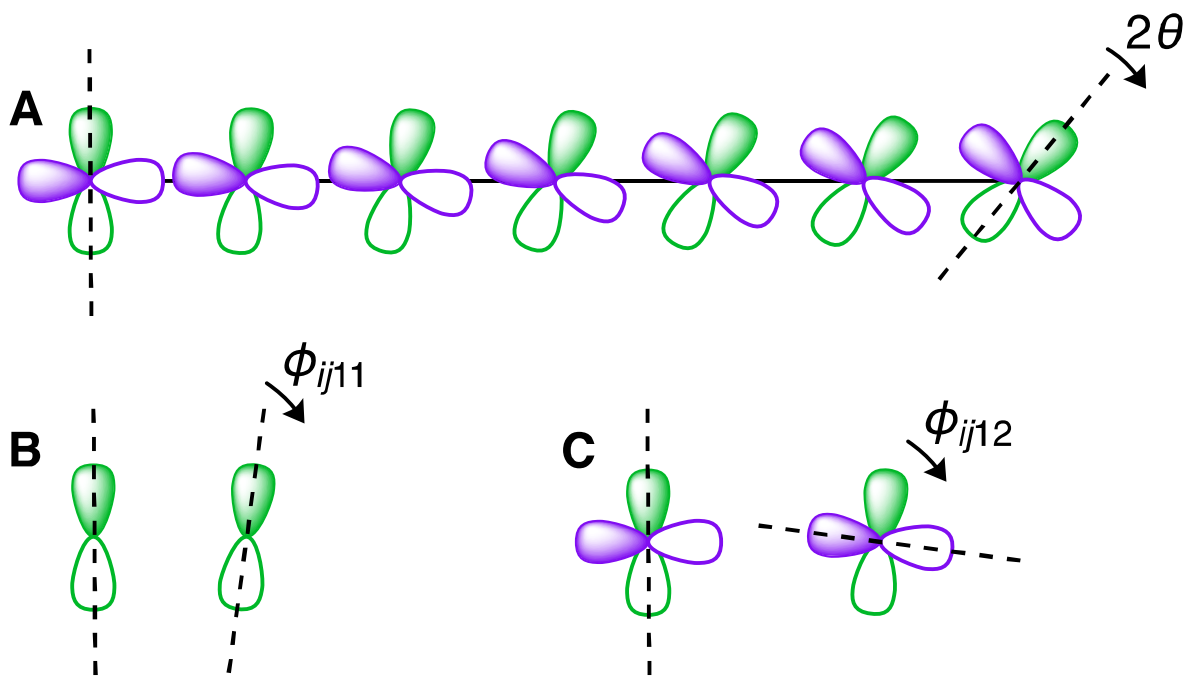

**Figure S1. The helical orbital basis for linear carbon chains.** (**A**) An angle of $2\theta$ is accumulated over the chain length. (**B**) Angle between neighboring orbitals with $k = l$. (**C**) Overlap between neighboring orbitals with $k \neq l$.

Single-bond terminated carbon chains with an odd number of bonds tend to undergo a Peierls distortion towards a polyynic structure with bond length alternation (BLA), whereas double-bond terminated carbon chains with an even number of bonds tend to be cumulenic with all bonds equal in length. In closed-shell $C_{13}Cl_2$ as drawn in Fig. 2B there is a seven-bond segment ($N = 8$ including two sp$^2$-hybridized carbons), which we can expect to have BLA $\neq$ 0, and a six-bond segment ($N = 7$), in which we set BLA = 0. To evaluate the effect of BLA in the seven-bond segment (labelled 7bs), we assume it increases or decreases $t_0$ by a percentage $\delta$ (*62*):

$$t_{0,(7\mathrm{bs})} = t_0\left(1+(-1)^i\delta\right) \tag{9}$$

$$t_{0,(6\mathrm{bs})} = t_0 \tag{10}$$

To extend the conjugation through the whole ring, we add next-nearest-neighbor couplings between the sp-hybridized atoms (6 and 8; 13 and 2) on opposing sides of sp$^2$-hybridized atoms 7 and 1 (**fig. S2** and Fig. 2B in main text):

$$\hat{H}_{nnn} = -\sum\nolimits_{\langle k,l\rangle}^{2} t_{6,8kl} c_{6k}^{\dagger} c_{8l} - \sum\nolimits_{\langle k,l\rangle}^{2} t_{13,2k}\, c_{13k}^{\dagger} c_{2l} \tag{11}$$

where the next-nearest-neighbor coupling values $t_{6,8kl}$ and $t_{13,2k}$ depend on $\theta$ analogously to Eq. 2:

$$t_{6,8kl} = t_{nnn} \cos \phi_{6,8kl} \tag{12}$$

$$t_{13,2kl} = t_{nnn} \cos \phi_{13,2kl} \tag{13}$$

These approximations yield the Hamiltonian:

$$\hat{H} = \hat{H}_{nn}(t_0, \delta, \theta) + \hat{H}_{nnn}(t_{nnn}, \theta) \tag{14}$$

which depends on nearest-neighbor and next-nearest-neighbor couplings $t_0$ and $t_{nnn}$, the proportion of BLA $\delta$ in the seven-bond segment, and the tilt angle $\theta$ (**figs. S2A, S4**). Finally, out-of-plane bending of the sp$^2$-hybridized carbons can be added as:

$$\hat{H} = \hat{H}_{nn}(t_0, \delta, \theta) + \hat{H}_{nnn}(t_{nnn}, \theta) + 2\frac{1}{2}k_{\mathrm{bend}} \sin^2\theta \tag{15}$$

where the final harmonic term is multiplied by two as bending occurs at atoms C1 and C7 (**fig. S2A**).

In the case of $\theta = 0$, we assume a Hückel $GML^0{}_4$ basis, with separate in-plane and out-of-plane π-systems (**fig. S2B**), and construct the tight-binding Hamiltonian using Eqs. 3, 4. For a finite value of $\theta$ we choose a half-Möbius $GML^1{}_4$ basis (**fig. S2C**). Similarly to the STM data and the obtained Dyson orbitals (see main text), this basis has a large ($90° - 2\theta$) twist in the six-bond segment, and and a small $2\theta$ twist in the seven-bond segment, and it is described by Eqs 5, 6.

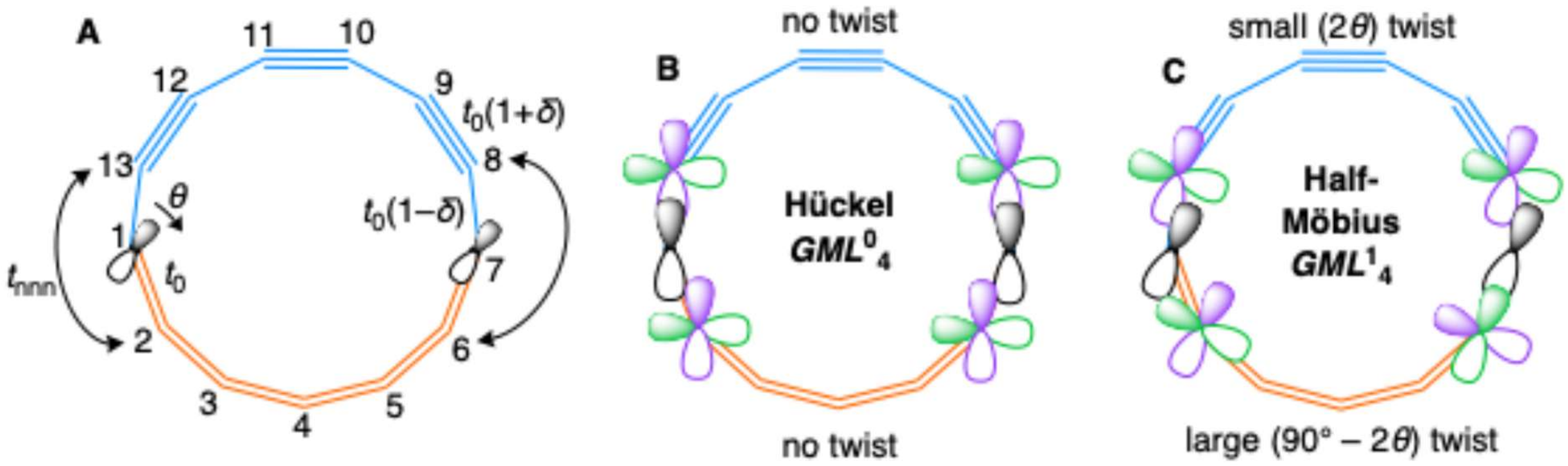

**Figure S2. Tight-binding description (A) and possible bases (B, C) of $C_{13}Cl_2$.** (**A**) Atom labelling and parameters of the tight-binding Hamiltonian for $C_{13}Cl_2$. The lone *p*-orbitals of $sp^2$-hybridized atoms 1 and 7 are shown. (**B, C**) Possible bases in which the tight-binding Hamiltonian can be evaluated.

To explore the variation of the energy with $\theta$, we need to choose values of $t_0, t_{\mathrm{nnn}}, \delta$, and $k_{\mathrm{bend}}$. Results are shown in **fig. S3**. Working in units of $t_0$, we use $t_{\mathrm{nnn}} = 0.1t_0$ and use $\delta = 0.0$ (no BLA, left column of **fig. S3**), $\delta = 0.1$ (moderate BLA, middle column of **fig. S3**) and $\delta = 0.2$ (strong BLA, right column of **fig. S3**). To explore bending, we choose $k_{\mathrm{bend}} = 0$ (first row of **fig. S3**), $k_{\mathrm{bend}} = t_0/2$ (middle row of **fig. S3**), and $k_{\mathrm{bend}} = t_0$ (third row of **fig. S3**).

In the case of the singlet (which has the twelve lowest MOs doubly occupied), we find that the combination of $\delta = 0.1$ and $k_{\mathrm{bend}} = t_0$, shown in **fig. S3H**, gives a good agreement with multireference calculations (**table S1** and shaded region of **fig. S3**).

For all explored values, the triplet (in which one electron is promoted from the HOMO to the LUMO, i.e. it is indistinguishable from the open-shell singlet) has an energy minimum at $\theta = 0$, in agreement with our multireference calculations (**table S1**).

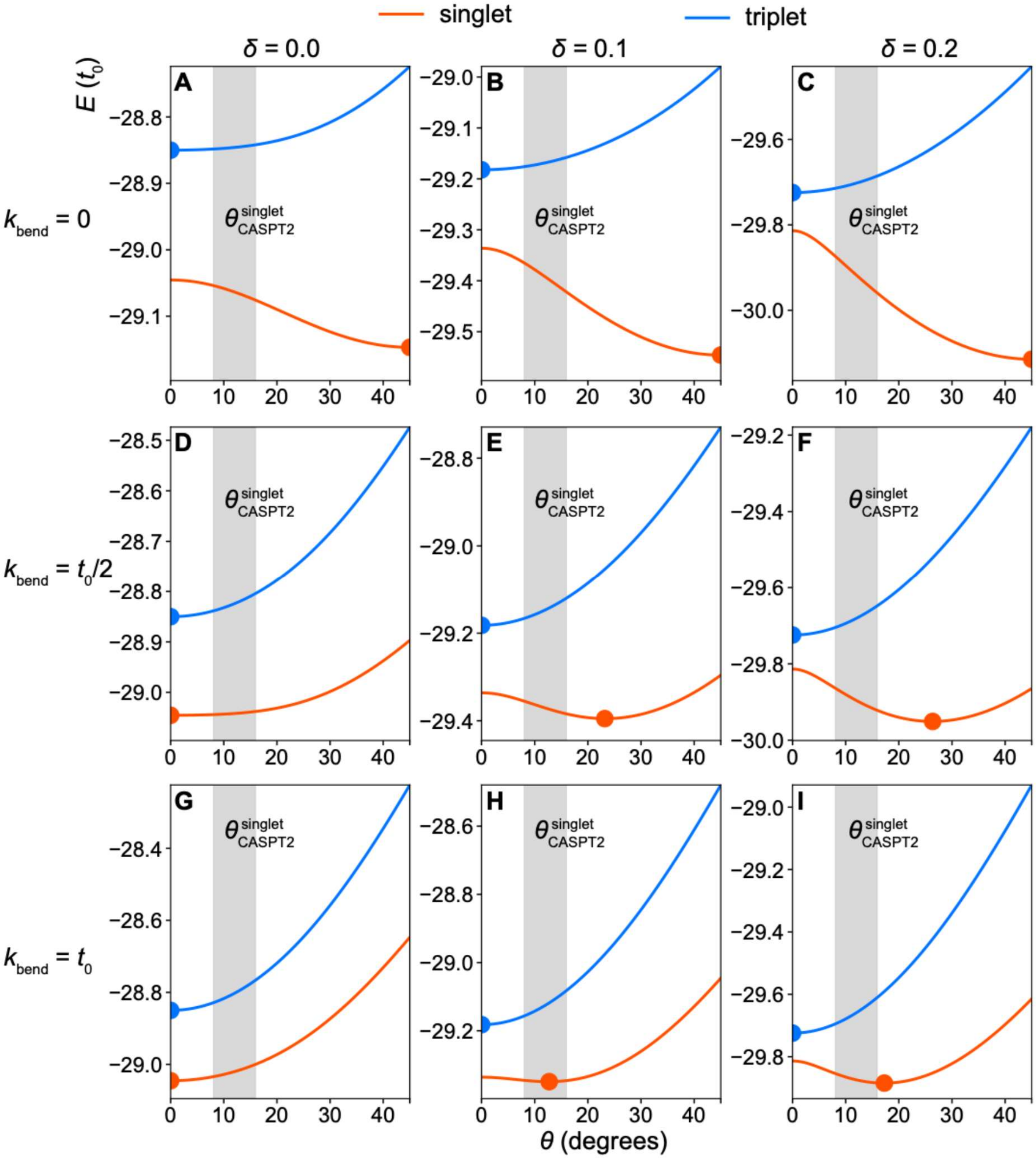

**Figure S3. Tight-binding calculations of $C_{13}Cl_2$.** Energies $E$ in units of $t_0$ for different tilt angles $\theta$ obtained using Eq. 15. Singlet energies (orange) are the sum of twelve lowest MO energies multiplied by two. Triplet energies (blue) were obtained by promoting one spinless electron from the HOMO to the LUMO. As spin is not considered here, the results for the triplet also describe an open-shell singlet. In all cases we assume $t_{nnn} = 0.1t_0$, For BLA we consider $\delta = 0$ (first column, **A**, **D**, **G** ), $\delta = 0.1$ (second column, **B**, **E**, **H** ), $\delta = 0.2$ (third column, **C**, **F**, **I**), and for bending $k_{\text{bend}} = 0$ (first row, **A–C**), $k_{\text{bend}} = \frac{1}{2}t_0$ (second row, **D–F**), $k_{\text{bend}} = t_0$ (third column, **G–I**). The CASPT2-obtained value of $\theta$ is ~0° for the triplet state and 8–15° for the singlet, corresponding to the calculated values on NaCl (shaded in gray). The respective equilibrium geometry predicted by tight-binding is highlighted as a dot.

### *3.2. Multireference calculations*

Geometry relaxations and Dyson orbital calculations were done using Molcas 24.02 (*63*). Geometries of $C_{13}Cl_2$ and $C_{13}H_2$ were relaxed in the singlet and triplet states at the CASPT2(12,12)/cc-pVDZ level of theory, that is 12 electrons in 12 orbitals. An IPEA shift of 0.25 a.u. was used. Analytical gradients (*64*) were computed in Cartesian coordinates, and relaxations were done using the quasi-Newton algorithm with a maximum step size of 0.05 a.u. until the gradient norm was below 0.01 a.u. and the energy in successive iterations oscillated by ~1 meV. $C_{13}H_2$ was only relaxed in the gas phase, while relaxations of $C_{13}Cl_2$ were done both in the gas phase and on-surface (see **fig. S5**). In the latter case, the surface was represented by a bilayer of point charges following the methodology established in our previous work (*22*). Geometric parameters of optimized $C_{13}Cl_2$ and $C_{13}H_2$ are compared in **Table S1**.

Dyson orbitals were computed by using the state interaction method (*65*) from MS-CASPT2 calculations with three roots for the –3, –2, –1, 0, +1, +2, and +3 charge states, done at the on-surface optimized singlet and triplet geometry of the neutral state. The ionization potentials and electron affinities obtained this way are shown in **Table S2**. The spin-orbit coupling matrix element between the singlet and triplet state was calculated at the CASSCF(12,12)/ANO-RCC-VTZP level of theory using the state-interaction method.

Optimized CASPT2 geometries and Dyson orbital cube files are available at (*59*) 10.5281/zenodo.15495263.

Nucleus-independent chemical shift (NICS) calculations were done using the development version of Dalton 2025 (*66*) at the CASSCF(12,12)/cc-pVDZ level of theory.

**Table S1.** Tilt ($\theta_1$ and $\theta_2$) and twist ($\varphi = \theta_1 + \theta_2$) angles of $C_{13}Cl_2$ and $C_{13}H_2$, estimated as an angle between the C–X (X = Cl or H) bond and the average plane defined by sp-hybridized atoms (see **fig. S4**), based on CASPT2(12,12)-optimized geometries.

| | $\theta_1$ | $\theta_2$ | $\varphi$ |
|---|---|---|---|
| $^1$**2** (NaCl) | 15.2 | 9.5 | 24.7 |
| $^1$**2** (gas) | 15.8 | 8.5 | 24.3 |
| singlet $C_{13}H_2$ (gas) | 19.9 | 20.4 | 40.2 |
| $^3$**2** (NaCl) | –1.6 | 3.6 | 2.0 |
| $^3$**2** (gas) | 0.2 | 1.0 | 1.2 |
| triplet $C_{13}H_2$ (gas) | 3.0 | 2.6 | 5.6 |

**Table S2.** First three ionization potentials (IP) and electron affinities (EAs) in eV, as calculated at the MS-CASPT2(12,x)/ANO-RCC-VTZP, x = 9–15, level of theory on the CASPT2(12,12)/cc-pVDZ optimized geometries of the neutral singlet and triplet.

| | $^1$**2** | $^3$**2** |
|---|---|---|
| IP3 | 35.5 | 34.7 |
| IP2 | 19.3 | 18.9 |
| IP1 | 7.8 | 7.2 |
| EA1 | 2.1 | 2.5 |
| EA2 | 0.1 | 0.9 |
| EA3 | –7.0 | –6.4 |

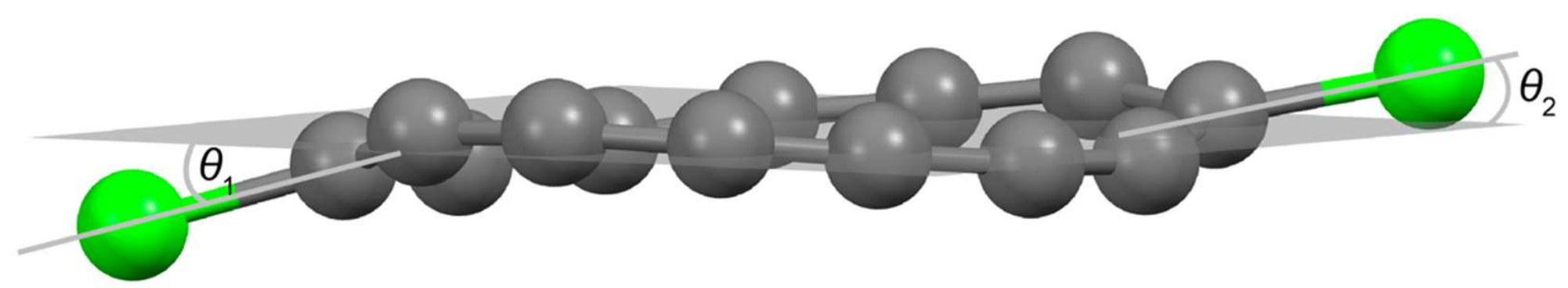

**Figure S4. Definition of tilt angles.** The plane is fit through all sp-hybridised atoms. The twist angle $\varphi$ corresponds to the sum of the tilt angles $\theta_1$ and $\theta_2$, i.e., $\varphi = \theta_1 + \theta_2$.

### *3.3. Aromaticity*

The change in topology from trivial to helical is also reflected in the aromaticity of **2**. The NICS trends are similar on the surface and in the gas phase (see **fig. S5**). As the weak interactions with the surface are not immediately relevant to our analysis, we shall focus on the gas-phase structures.

In the planar triplet minimum, **$^3$2** has roughly 13 electrons in the out-of-plane and 11 electrons in the in-plane π-system and the major contribution to its wavefunction is the $^3$|11> configuration. Such an electronic structure results in both diatropic and paratropic contributions to the induced current (*22, 62*) and results in overall anti-aromaticity, with $\mathrm{NICS(2)_{zz}}$ = 6.2 ppm. The planar singlet **$^1$2$_{@triplet}$**, which has helical natural orbitals and a dominant $^1$|11> contribution, is less anti-aromatic ($\mathrm{NICS(2)_{zz}}$ = 3.3 ppm) than **$^3$2**. This anti-aromaticity relief can be attributed to the contribution of closed-shell configurations |20> and |02>, as relaxing the singlet into its non-planar minima **$^1$2-M** or **$^1$2-P** (described as |20> and |02>, respectively) results in further anti-aromaticity relief, with $\mathrm{NICS(2)_{zz}}$ = –3.7 ppm. This $C_s \rightarrow C_1$ symmetry-breaking is highly analogous to the pseudo-Jahn–Teller $D_{4h} \rightarrow D_{2h}$ distortion in cyclobutadiene (*32*), which is also accompanied by anti-aromaticity relief.

A defining characteristic of the non-planar singlets **$^1$2-M** and **$^1$2-P** is the mixing of the in-plane and out-of-plane π-systems of the carbon ring into a single 24-electron helical π-system. Their weak aromaticity (or weak anti-aromaticity for the on-surface geometry) can be rationalized by noting that their orbital basis has the topology of a $GML^{\pm1}{}_4$ body. This means that they do not follow rules for Hückel aromaticity (valid for $GML^0{}_{2h}$ bodies, $h$ is an integer), which would imply strong anti-aromaticity for 24 (4*n*) electrons. They also do not closely follow the rules for conventional Möbius aromaticity (valid for $GML^h{}_{2h}$ bodies), as those would imply strong aromaticity for 24 electrons. Moreover, the non-planar triplet **$^3$2$_{@singlet}$** displays very similar NICS values to the singlet minima, suggesting that Baird's rule, which is valid both for Hückel and Möbius systems (*67*), may not apply here, possibly related to a change of topology. While further investigation of half-Möbius aromaticity is needed, our results suggest that both diatropic and paratropic contributions to the ring current are present, as the obtained NICS values appear to be highly sensitive to geometry (*62*), see **fig. S5**.

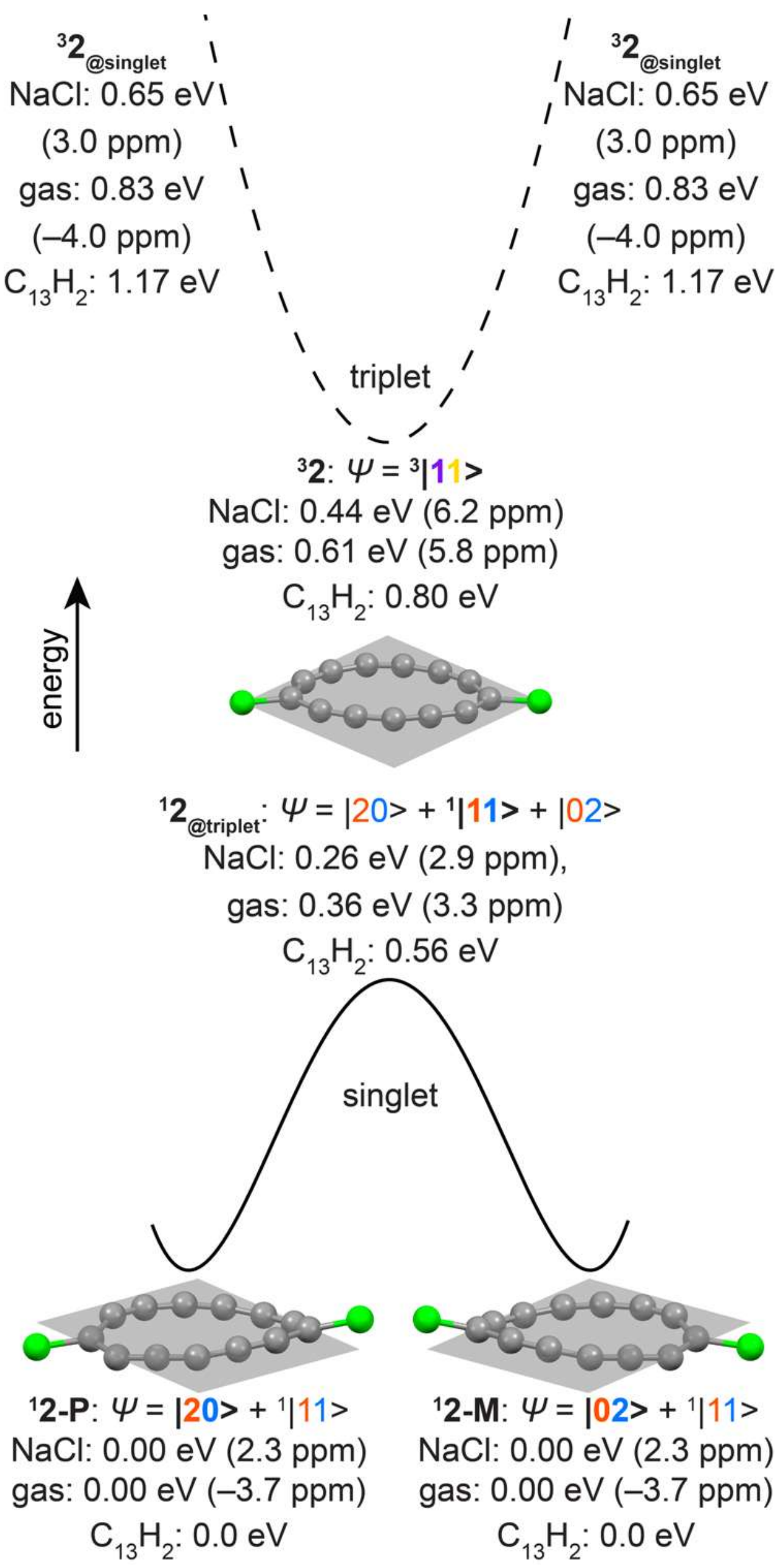

**Figure S5. Multireference calculations of $C_{13}Cl_2$ on the surface and in the gas phase, and $C_{13}H_2$ in the gas phase.** Energy diagram, with energies in eV, for the singlet $^{1}$**2** , and the triplet $^{3}$**2**. C atoms are shown in gray and Cl atoms in green. NICS(2)$_{zz}$ values in ppm for $C_{13}Cl_2$ gas-phase geometries are shown in parentheses. $^{1}$**2**$_{\textbf{@triplet}}$ is the singlet in the equilibrium geometry of $^{3}$**2**, and $^{3}$**2**$_{\textbf{@singlet}}$ is the triplet in the equilibrium geometry of the singlet. Multireference states are described by configurations that contribute significantly to their composition, with dominant contributions shown in bold.

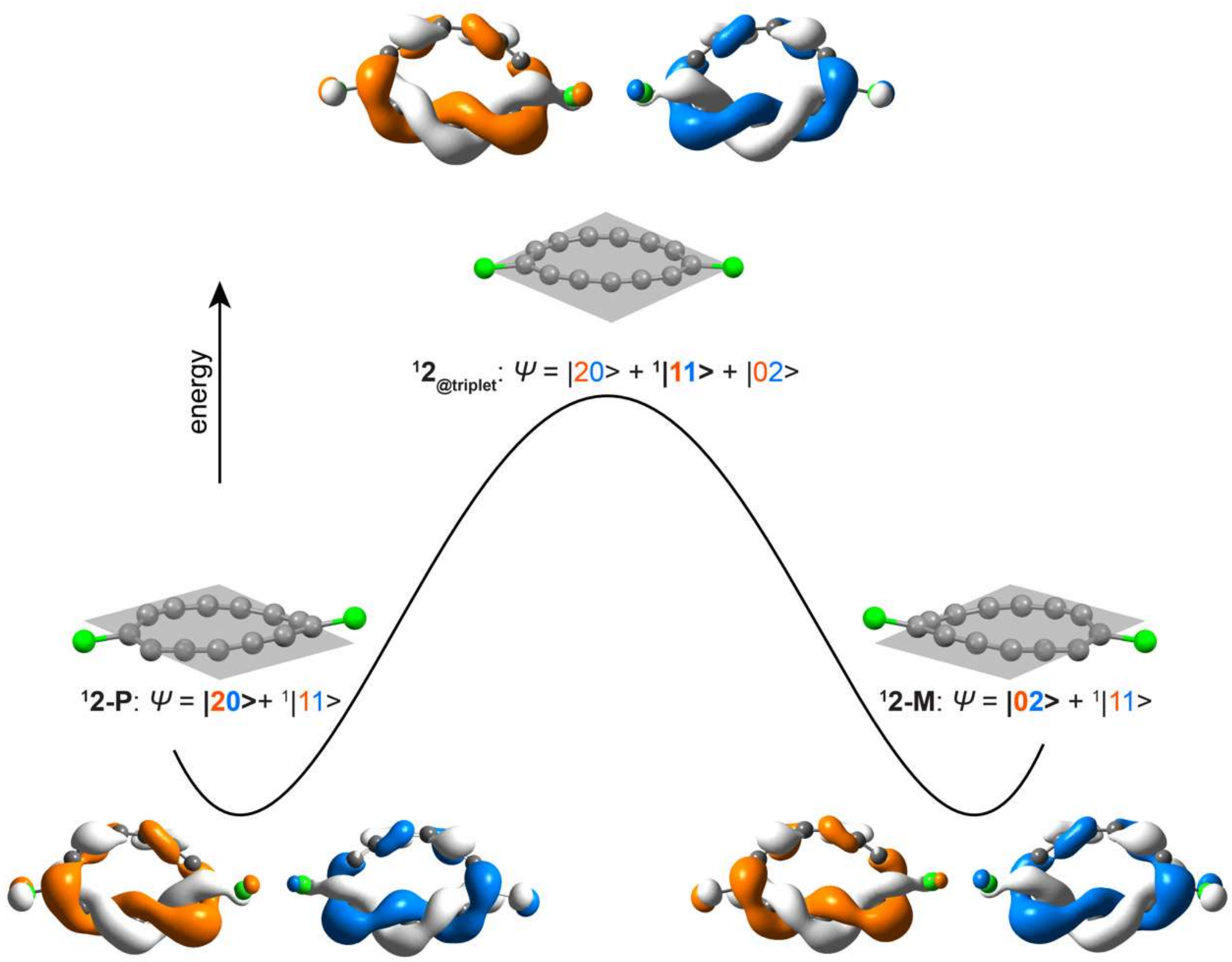

**Figure S6. Natural frontier orbitals of $^{1}$2 in different geometries.** Natural orbitals are shown for geometries calculated in the gas phase. $^{1}$**2**$_{\mathbf{@triplet}}$ denotes the singlet in the equilibrium geometry of $^{3}$**2**. Dominant contributions to the wavefunction are shown in bold. The natural occupations of the orange and blue orbital in $^{1}$**2**$_{\mathbf{@triplet}}$ (same as shown in Fig. 3C) are 0.97 and 1.02, respectively. As the triplet geometry is not perfectly planar, see **Table S1**, these two populations are not exactly equal, and the orange and blue orbitals are not exact mirror images of each other. For the relaxed gas phase geometry of the singlet $^{1}$**2-P** ($^{1}$**2-M**), the occupations of the orange and blue orbitals are 1.5 and 0.5 (0.5 and 1.5), respectively. The structures and natural orbitals of $^{1}$**2-P** are mirror symmetric with respect to those of $^{1}$**2-M**, e.g., the HOMO of $^{1}$**2-P** (colored orange) is mirror symmetric to the HOMO of $^{1}$**2-M** (colored blue). The natural occupations of the triplet, shown in Fig. 3B are 0.95 and 0.96 for the purple and yellow orbital, respectively.

### *3.4. DFT calculations of* $C_{12}Cl_2$*,* $C_{13}Cl_2$*, and* $C_{14}Cl_2$

The range-separated ωB97X-D functional (*68*) qualitatively reproduces the CASPT2-optimized singlet geometry of $C_{13}Cl_2$ well. We used the ωB97X-D/def2-SVP level of theory to compare equilibrium singlet geometries of $C_{12}Cl_2$ (two six-bond segments), $^{1}$**2** (i.e., $C_{13}Cl_2$, six- and seven-bond segment), and $C_{14}Cl_2$ (two seven-bond segments). DFT predicts that $^{1}$**2** is non-planar and that $C_{12}Cl_2$ and $C_{12}Cl_2$ singlets are planar, see **Table S3**. These calculations were done using Gaussian16 (*69*).

**Table S3.** Tilt ($\theta_1$ and $\theta_2$) and twist ($\varphi = \theta_1 + \theta_2$) angles, see **fig. S4** for definition, of optimized geometries of singlet $C_{13}Cl_2$, $^{1}$**2** ($C_{13}Cl_2$) and $C_{14}Cl_2$, estimated as an angle between the C–X (Cl or H) bond and the average plane defined by sp-hybridized atoms. In all three DFT-investiated systems, frequency calculations confirmed that the obtained geometries were minima.

| | $\theta_1$ | $\theta_2$ | $\varphi$ |
|---|---|---|---|
| $^{1}$**2** (gas, CASPT2 | 8.5 | 15.8 | 24.3 |
| $^{1}$**2** (gas, ωB97X-D) | 18.6 | 18.7 | 37.3 |
| $C_{12}Cl_2$ (gas, ωB97X-D) | 0.0 | 0.0 | 0.0 |
| $C_{14}Cl_2$ (gas, ωB97X-D) | 0.0 | 0.0 | 0.0 |

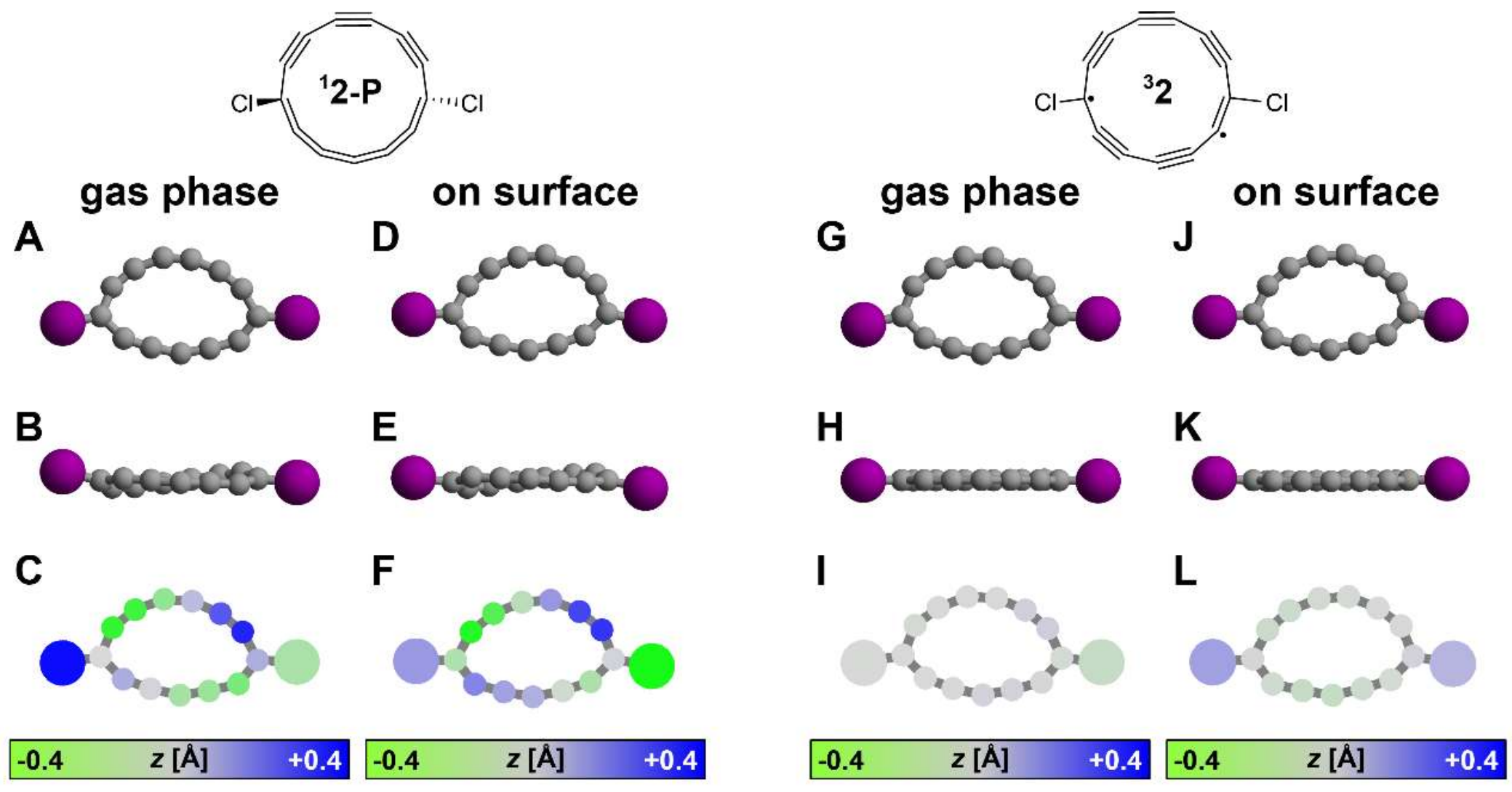

**Figure S7. CASPT2-optimized geometries.** CASPT2(12,12) calculated geometries of $^{1}$**2-P** and $^{3}$**2** in the gas phase and on NaCl. The bottom row shows the distortion with respect to the average plane of the nuclei of the molecule. Representations of the on-surface calculations that display also the NaCl substrate are shown in **figs. S14** and **S15**.

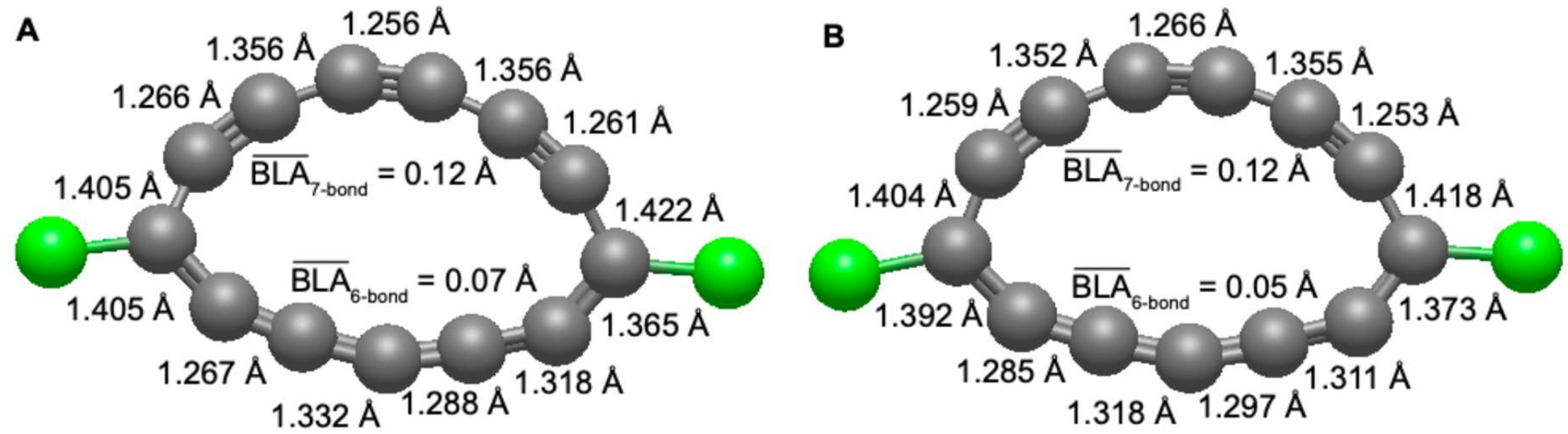

**Figure S8. Bond lengths in CASPT2-optimized on-surface geometries of $C_{13}C_2$ in the singlet (A) and triplet (B) state.** Average values of bond-length alternation (BLA) for the seven- and six-bond segments are shown (for the latter, the lower rightmost bond is not included in the average, because bond length is not alternating for this bond of the six-bond segment). For reference, the multireference results for cyclo[16] carbon predict a BLA value of ~0.12 Å (*44*), while coupled clusters results for cyclo[18]carbon predict a BLA of ~0.14 Å (*70*). Note that in the closed-shell Lewis structure of the singlet, see Fig. 2B, the seven-bond segment is drawn as a polyyne and the six-bond segment as a cumulene. The CASPT2-optimized geometry somewhat validates this structure, as the bond-length alternation in the seven-bond segment is similar to that in polyynes. However, we also find some bond-length alternation in the six-bond segment and irregular bond-length alternation through the seven-bond segment.

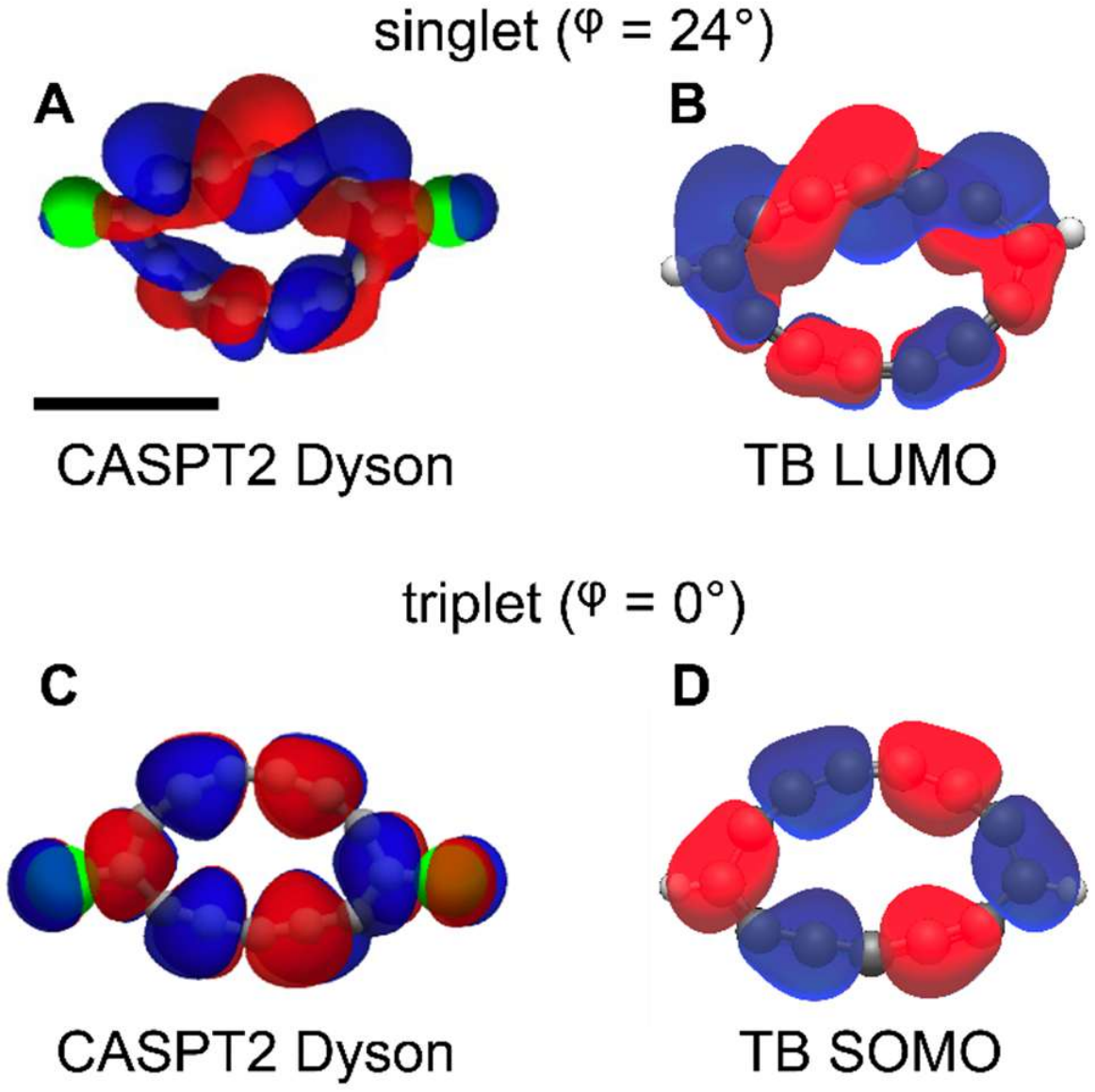

**Figure S9. Electron attachment to $^{1}$2 (A,B) and detachment from $^{3}$2 (C,D).** (**A**) Dyson orbital for electron attachment to the geometry-optimized $C_{13}Cl_2$ singlet calculated using CASPT2. (**B**) LUMO obtained using the tight-binding model and the half-Möbius topology $GML^{1}_{4}$, see **fig. S2C**. (**C**) Dyson orbital for electron detachment from the geometry optimized $C_{13}Cl_2$ triplet calculated using CASPT2. (**D**) The energetically lower SOMO obtained using the tight-binding model and the Hückel $GML^{0}_{4}$ topology, see **fig. S2B** (the higher SOMO is in-plane and therefore will result only in very weak contrast in STM (*22*)). Tight-binding parameters used are $t_{nnn} = 0.1t_0$, $\delta = 0.1$, (orbitals do not depend on $k_{\mathrm{bend}}$). The disagreement at sp$^2$-hybridized atoms 1 and 7 is explained by our tight-binding model not considering their substituents.

### *3.5. Quantum calculations*

The recently-proposed SqDRIFT quantum algorithm (*36*) is a quantum-computing counterpart of classical selected Configuration Interaction approaches (*71*). These approaches rely on the observation that, even in strongly-correlated systems, the exact full-configuration interaction wave function has a non-negligible support over an array of determinants, which can be written as $\{|\sigma_i^{\mathrm{SCI}}\rangle\}$, that is dramatically smaller than the full determinant space. SqDRIFT leverages the quantum computer as a sampling machine to construct the $\{|\sigma_i^{SCI}\rangle\}$ space.

Among the different sample-based quantum diagonalization variants, in SqDRIFT the quantum circuits used for the sampling phase are constructed as time-evolution circuits, i.e.

$$|\psi\rangle \; = \; e^{-iHt}|\psi_{\mathrm{o}}\rangle \; , \qquad (16)$$

where $|\psi_{\mathrm{o}}\rangle$ is a reference determinant (here taken as the lowest-energy closed-shell configuration), and $H$ is the electronic Hamiltonian, which we decompose as sum of elementary Fermionic operators $h_i$ as:

$$H \; = \; \sum_i c_i \, h_i \qquad (17)$$

The time-evolution operator is randomized through the qDRIFT protocol (*72*) as

$$e^{-iHt} \approx \prod_k e^{-ih_k t} \quad (18)$$

where the product includes only a subset of all the $h_k$ terms, sampled according to a probability distribution proportional to $|c_k|^2$. In practice, multiple randomized approximations are generated and executed on the quantum processors. This randomized compilation step enables a dramatic reduction of the circuit depth, making it feasible to utilise SqDRIFT on current quantum processors to solve complex quantum-chemical problems.

The parameters to be set in an SqDRIFT simulation are the following:

1. The number of randomized approximations of the time-evolution operator that are executed as quantum circuits. Each randomized approximation is constructed from Eq. 18, sampling a different subset of terms $h_k$ that are included in the product.
2. The number of Hamiltonian terms $h_k$ to be included in the time-evolution circuit for each randomization.
3. The number of samples to be collected from each circuit.
4. The dimension of the subspace in which the Hamiltonian is diagonalized.

Among the different sample-based quantum diagonalization approaches proposed in the literature, SqDRIFT is particularly appealing because it has been proven that sampling from the circuits defined above yields the $\{|\sigma_i^{\mathrm{SCI}}\rangle\}$ array with high probability (*36*). This is strictly true only when a large number of terms $h_k$ relative to the Hamiltonian norm $\sum_i \left|c_i^2\right|^2$ is included in each approximate time-evolution circuit. However, it has been empirically observed the relevant determinants $\{|\sigma_i^{\mathrm{SCI}}\rangle\}$ are already identified through sampling from relatively shallow circuits (*36*). In practice, the number of samples should be large enough to ensure that all configurations contributing to the ground-state wave function are included in the subsequent diagonalization step.

In addition to SqDRIFT, we also executed quantum calculations using the extended SqDRIFT (ext-SqDRIFT) (*73*) approach. Ext-SqDRIFT first executes an SqDRIFT calculation and diagonalizes the Hamiltonian in the resulting determinant space. A second subspace is then constructed including the determinants for which $|c_i| > \eta$ (with $\eta = 10^{-3}$ for the neutral species and $5 \cdot 10^{-4}$ for the anionic one) are retained. In an analogy with multireference CISD, this subspace is then extended by including all possible single (S) and double (D) excitations constructed starting from the existing ones. The Hamiltonian is then classically diagonalized in the resulting subspace.

To explore an active space beyond the reach of conventional approaches, we performed SqDRIFT calculations on an IBM quantum processor (see below for additional details on the quantum hardware) carried out on an active space of 32 electrons in 36 orbitals for the neutral molecular system, and 33 electrons in 36 orbitals for the anionic molecular system. Such an active space includes all carbon $\pi$ electrons and per chlorine four electrons that mix with the $\pi$-system of $C_{13}Cl_2$. The molecular orbitals were obtained from a smaller-scale CASSCF calculation, based on an active space including 12 electrons in 12 orbitals. The resulting Hamiltonian consists of a total of 222814 fermionic second-quantization operators.

To execute SqDRIFT on a quantum processor, the molecular Hamiltonian had to be mapped onto the quantum hardware. This was performed using Jordan-Wigner mapping (*74*), which resulted in 72-qubit circuits. For a system of this size, full implementation of Trotterized time evolution (*75*) would be impractical on current quantum hardware. Even a circuit specifically designed for near-term quantum devices, such as the Local Unitary Cluster Jastrow ansatz (*76*), would result in a final two-qubit circuit depth of 286, exceeding the capabilities of current hardware.

To sidestep these limitations, we employed the qDRIFT protocol, which randomizes the Hamiltonian terms to be implemented within the quantum circuit, effectively reducing the circuit depth. For each molecular system, several experiments are conducted using different numbers of sampled excitations and evolution times. Specifically:

- For each molecular system, circuits with 5, 10, and 15 fermionic excitations, constructed on top of the lowest-energy determinant obtained from the CASSCF orbitals, were considered;
- For each excitation count, circuits were prepared with three different evolution times, $t = 1,2,3$ (in atomic units);
- For each combination of excitation count and evolution time, 500 randomizations were executed.

This resulted in 1500 circuits per excitation count and a total of 4500 circuits per molecule. Each circuit was sampled 1024 times, yielding approximately 4.6 million samples per molecular system. The circuits were executed on *ibm_kingston*, a chip belonging to the second revision of IBM's Heron family of quantum processors.

The *ibm_kingston* quantum processor features 156 qubits, a median two-qubit gate error of $2.1 \cdot 10^{-3}$, and a median readout error of $8.5 \cdot 10^{-3}$. In the experiments, hardware noise was mitigated by discarding quantum samples that exhibited an incorrect particle number. Finally, the collected quantum samples were used to define the subspace in which the fermionic Hamiltonian was projected and diagonalized classically. Each Hamiltonian diagonalization was performed on a cluster composed of bare metal nodes consisting on four sockets each with a Intel Xeon Platinum 8260 (2.40 GHz) processor, for a total of 196 cores. The dimension of the subspace was adjusted by increasing the size of the batch of quantum samples considered. The reported calculations were obtained with batch sizes ranging from 500 to 5000 samples, leading to Hilbert subspaces including up to 20 million determinants. Using up to 30 cores of the aforementioned cluster, the largest diagonalizations required approximately 10 minutes. Results of the SqDRIFT calculations, showing the energy of $^{1}$**2** and its anion as a function of the subspace diagonalization dimension are presented in **fig. S10**

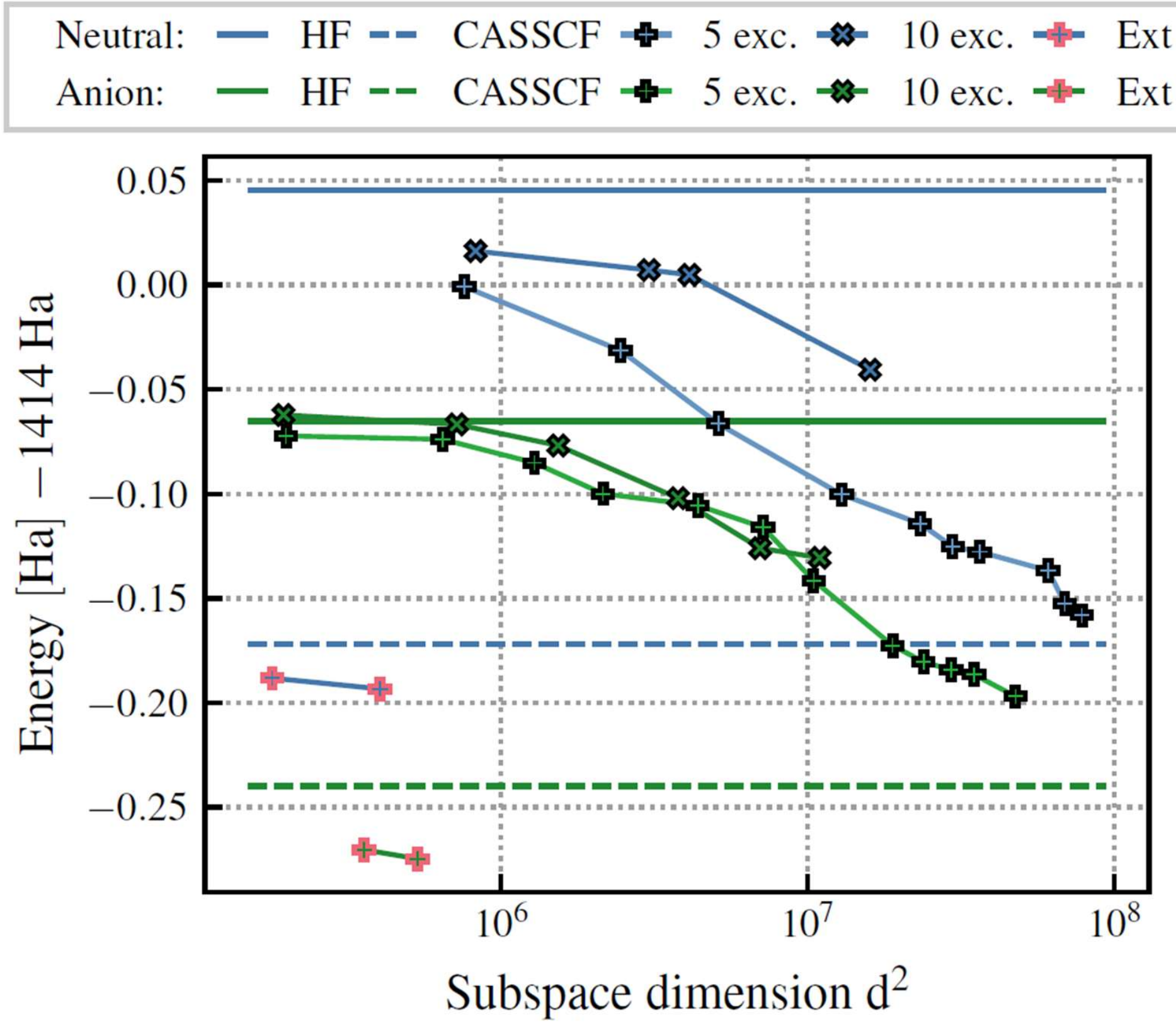

**Figure S10. Energy of the neutral and anion as a function of the subspace diagonalization dimension**. The Hartree-Fock (HF) energies serve as references and are reported in the solid lines at −38475.67 eV and −38478.69 eV for the neutral $^1$**2** (neu) and anion (ani) of **2**, respectively. The SqDRIFT energies obtained using 5 and 10 excitations (exc.) for the neutral $^1$**2** (shades of blue; min: −38481.24 eV for 5 exc.) and the anion of **2** (shades of green; min: −38482.29 eV for 5 exc.), respectively. Ext-SqDRIFT energies are also reported (red crosses; neu: −38482.192 eV, ani: −38484.405 eV).

As described above, SqDRIFT leverages the quantum processor solely as a sampling machine, and the subsequent diagonalization is executed classically. This means that classical representations of the neutral and charged (anionic) wave function are available, given as:

$$|\Psi_{\text{neutral}}\rangle = \sum_{i \in S_{\text{neutral}}} c_i^{\text{neutral}} |\sigma_i^{\text{neutral}}\rangle, \quad (19)$$

and

$$|\Psi_{\text{anionic}}\rangle = \sum_{i \in S_{\text{anionic}}} c_i^{\text{anionic}} |\sigma_i^{\text{anionic}}\rangle. \quad (20)$$

$S_{neutral}$ and $S_{anionic}$ are the subspaces sampled by SqDRIFT for the neutral-state and anionic-state wave functions, respectively. $\sigma_i^{neutral}$ and $\sigma_i^{anionic}$ are Slater determinants constructed from the molecular orbitals of the neutral and anionic species, respectively. Following our previous work (*22*), we expand the Dyson orbital in terms of the ground-state wave function molecular orbitals as:

$$\psi_{Dyson}(\boldsymbol{r}) = \sum_\mu d_\mu \phi_\mu(\boldsymbol{r}), \tag{21}$$

and calculate the expansion coefficient as:

$$d_\mu = \langle a_\mu^\dagger \Psi_{\text{neutral}} | \Psi_{\text{anion}} \rangle. \tag{22}$$

The coefficient is calculated, in practice, by expressing the correlated wave functions $|\Psi_{\text{neutral}}\rangle$ and $|\Psi_{\text{anionic}}\rangle$ according to their definition, obtaining:

$$d_\mu = \sum_{i,j} \left(C_i^{\text{neutral}}\right)^\star C_j^{\text{anionic}} \langle a_\mu^\dagger \sigma_i^{\text{neutral}} | \sigma_j^{\text{anionic}} \rangle. \tag{23}$$

The determinants $|a_\mu^\dagger \sigma_i^{\text{neutral}}\rangle$ and $|\sigma_j^{\text{anionic}}\rangle$ are expressed in terms of their respective CASSCF-optimised orbitals, which are however obtained from the same basis set. Hence, the overlaps appearing in the expression for $d_\mu$ can be calculated using standard strategies borrowed from non-orthogonal configuration interaction. Note that we accelerated the calculation of the Dyson orbital by neglecting, in the last equation, coefficients with a magnitude smaller, in absolute value, than 0.001 (this truncation has negligible impact on the Dyson orbital).

The thus obtained Dyson orbital for electron attachment to $^1$**2**, based on the SqDRIFT calculation, is shown in the main text in Fig. 3D, left panel. The SqDRIFT calculations used for both the neutral and anionic wave functions were run including 5 excitations per circuit. The subspace dimension $d^2$ used for the classical diagonalization is the largest amongst the ones reported in **fig. S10**. The Dyon orbital obtained by SqDRIFT agrees well with the the classical calculations using CASSCF in an active space of 12 correlated electrons shown in Fig. 3D, right panel (see also **fig. S9A** for the Dyson orbital calculated using CASPT2) indicating that the active space of 12 correlated electrons is sufficient to describe the system well.

## 4. Additional data and analyses

### *4.1. Different isomers of $C_{13}Cl_2$*

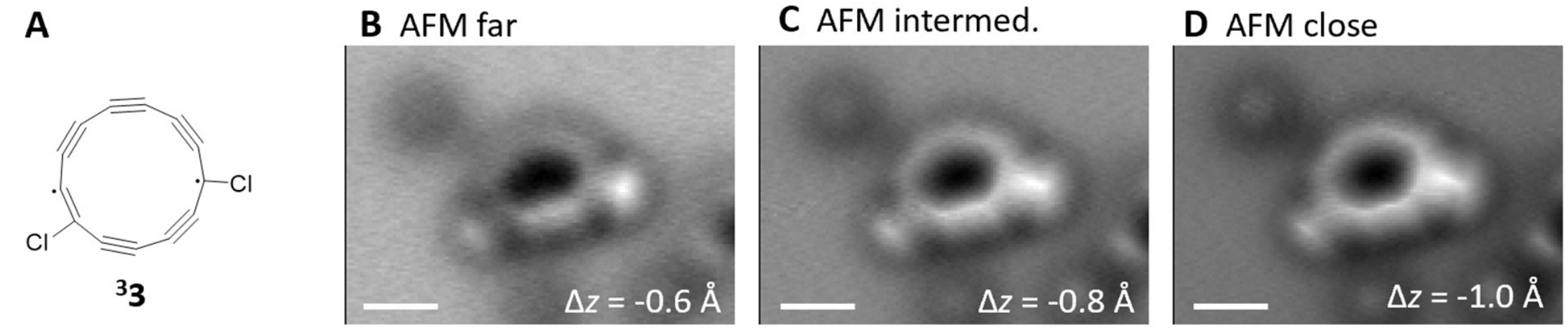

**Figure S11. $C_{13}Cl_2$ isomer 3**. (**A**) Resonance structure of $^3$**3**. (**B–D**) AFM data on monolayer NaCl (setpoint $I$ = 0.5 pA, $V$ = 0.2 V; $\Delta z$ as indicated). The AFM data indicates a planar ring geometry and thus points towards a triplet state $^3$**3**. $\Delta f$ scales from black to white, B [-6.2 Hz; -0.5 Hz], C [-6.7 Hz; +1.4 Hz], D [-6.5 Hz; +2.9 Hz]. All scale bars 5 Å.

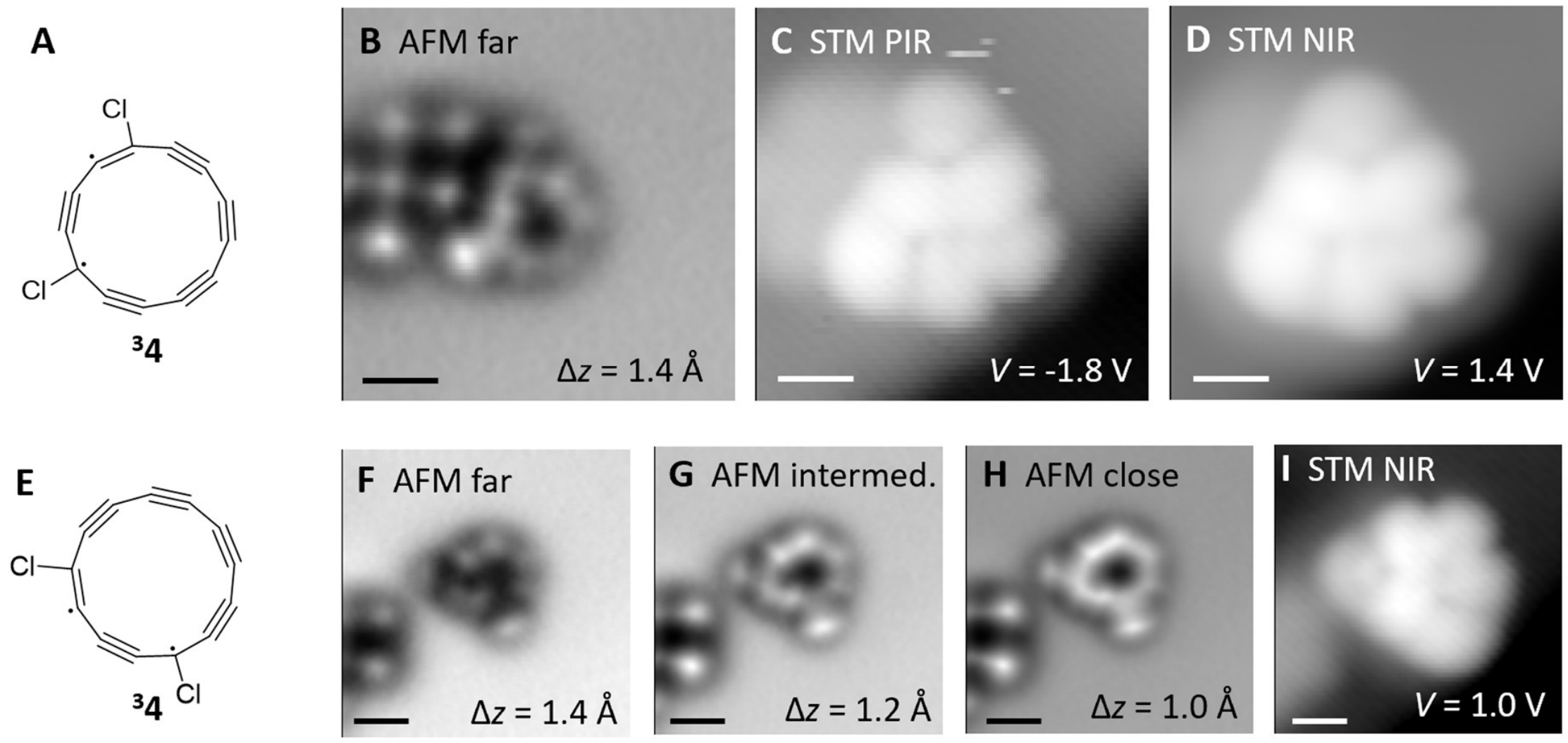

**Figure S12. $C_{13}Cl_2$ isomer 4**. (**A**) and (**E**) resonance structures of the planar, triplet $^3$**4**. (**B**) AFM data on bilayer NaCl near several Cl adatoms (setpoint $I$ = 0.6 pA, $V$ = 0.2 V; $\Delta z$ = 1.4 Å) revealing a planar geometry of the ring. (**C**) STM at the PIR at $V$ = -1.8 V, $I$ = 0.4 pA. (**D**) STM at the NIR at $V$ = 1.4 V, $I$ = 0.8 pA. (**F–H**) AFM data of $^3$**4** at an adsorption site further away from the Cl adatoms (setpoint $I$ = 0.6 pA, $V$ = 0.2 mV; $\Delta z$ as indicated). (**I**) STM at the NIR at $V$ = 1.4 V, $I$ = 0.8 pA. Scales from black to white, B [-3.9 Hz; +1.8 Hz], C [0 Å; +5.7 Å], D [0 Å; +6.0 Å], F [-3.3 Hz; +0.2 Hz], G [-4.2 Hz; +0.8 Hz], H [-4.6 Hz; +2.6 Hz], I [0 Å; +6.4 Å]. All scale bars 5 Å. The open-shell character of $^3$**4** is evidenced by the similarity of the PIR and NIR orbital-density images measured, see (C) and (D). PIR and NIR correspond to the singly occupied molecular orbital (SOMO) and singly unoccupied molecular orbital (SUMO) of the out-of-plane π-system, respectively (*77*). It is the same individual compound as in Fig. 6, A to D, however, after the bonding positions of the Cl atoms were manipulated (*22*) by applied voltage pulses. The molecule is adsorbed on bilayer NaCl, near a cluster of Cl adatoms.

*4.2. Adsorption geometries*

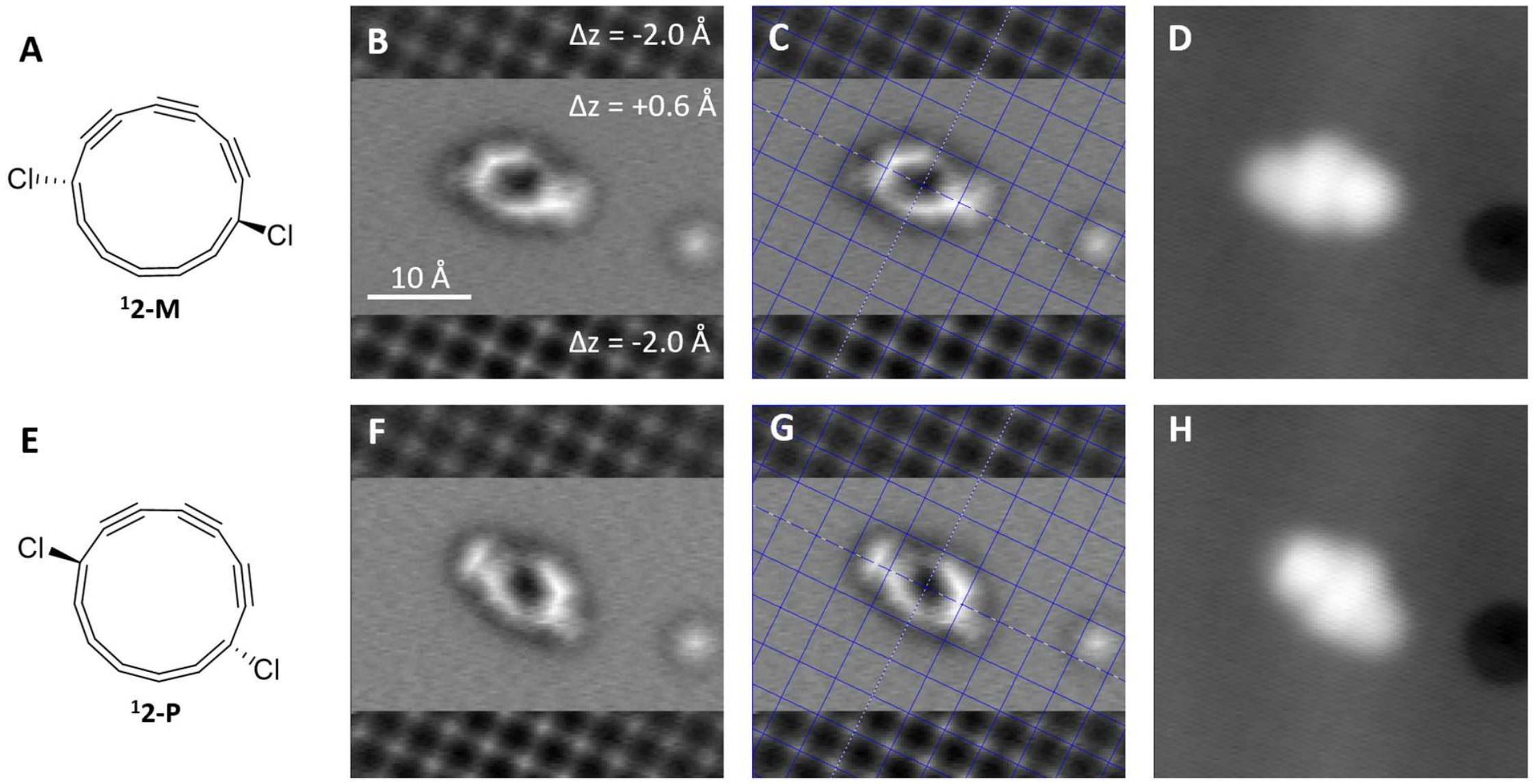

**Figure S13. Experimental adsorption-site determination of $^1$2-M and $^1$2-P.** (**A**) Lewis structure of **$^1$2-M**. (**B**) AFM data of **$^1$2-M** on bilayer NaCl on Au(111). The tip-height offset Δ*z* was different in the top (Δ*z* = -2.0 Å), center (Δ*z* = 0.6 Å) and bottom (Δ*z* = -2.0 Å) part of the image to resolve the substrate in top and bottom part and the molecule in the central part of the image (*41*). Setpoint $V$ = 0.15 V, $I$ = 0.4 pA. (**C**) Same data as in (B) with a grid overlaid, which vertices correspond to Cl sites of the NaCl surface. CO is adsorbed on a Na site. (**D**) STM of **$^1$2-M**, $V$ = 0.15 V, $I$ = 0.4 pA. (**E**-**H**) Corresponding data for **$^1$2-P**, after molecule had been switched from **$^1$2-M** to **$^1$2-P**. Scale bar 10 Å, applies to all measurements.

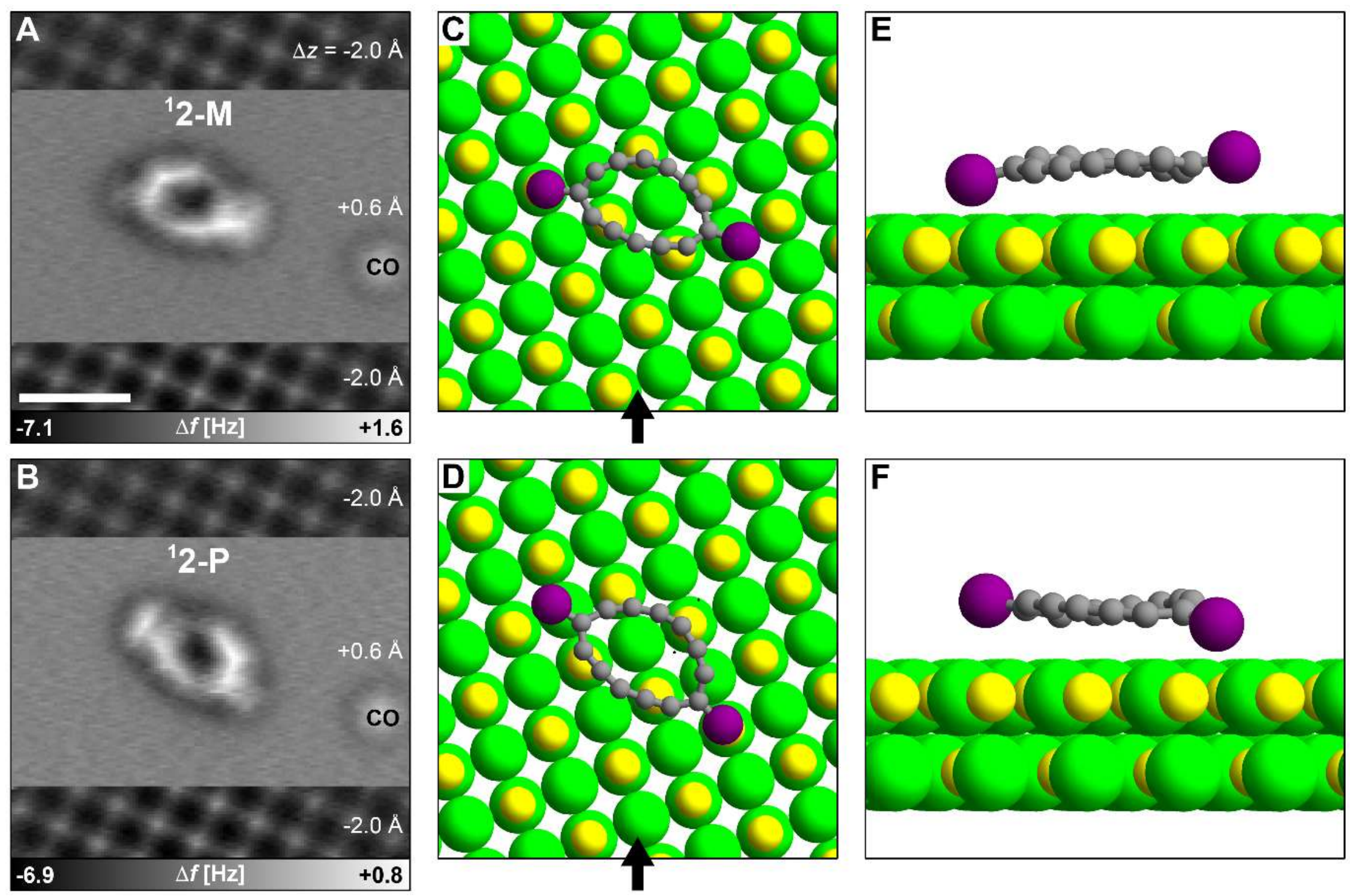

**Figure S14. Comparison of adsorption geometry of $^1$2-M and $^1$2-P, experiment and theory**. (**A, B**) AFM data to determine experimentally the adsorption site of **$^1$2-M** and **$^1$2-P**, respectively (see **fig. S13)**. Scale bar 10 Å. (**C, D**) Calculated relaxed adsorption geometries of **$^1$2-M** and **$^1$2-P**, respectively, on NaCl. Viewed from top, with the substrate oriented as in the experiment (A, B). Na cations in yellow and Cl anions of the NaCl surface in green; C atoms in gray and Cl atoms of the molecule in purple. (**E, F**) Side view along the direction of the arrows in (C, D), respectively. The theory reproduces the experiment very well, agreeing in adsorption site and adsorption orientation, comparing (A) with (C) and (B) with (D), and in the out-of-plane distortions of the molecule. Note that atoms that show a larger separation to the surface appear with brighter contrast in the AFM images (*41*), comparing (A) with (E) and (B) with (F). See also the corresponding AFM simulations using the Probe-Particle Model (*39*) in Fig. 4, D to F of the main text.

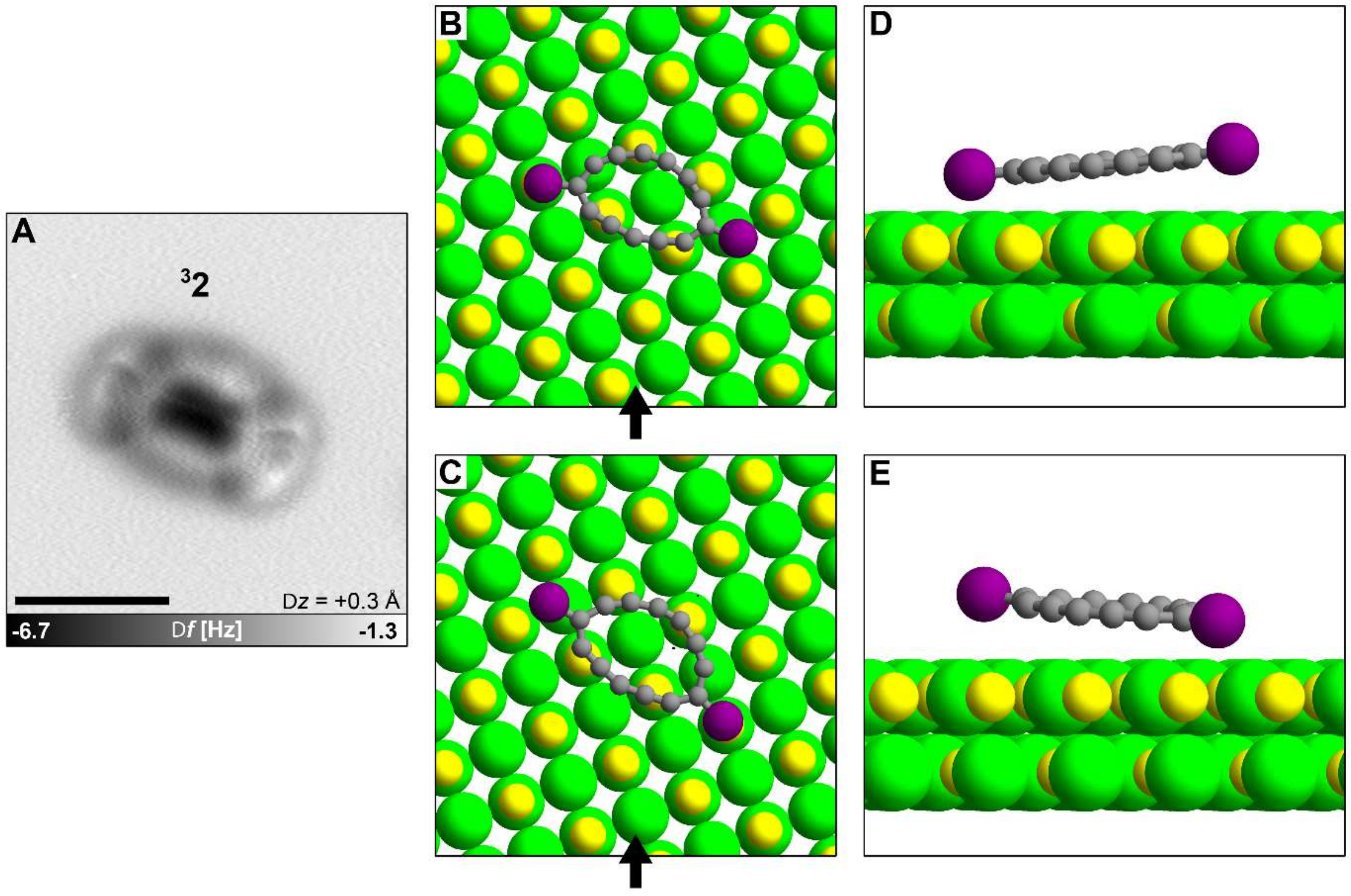

**Figure S15. Adsorption geometry of $^{3}$2**. (**A**) AFM data at $V$ = 0.34 V, setpoint $V$ = 150 mV, $I$ = 0.2 pA, $\Delta z$ = 0.3 Å. Scale bar 10 Å. At this elevated voltage, the molecule is switching between different states at a faster rate than the bandwidth of the AFM (see Fig. 6 for measurements at adsorption sites that stabilized $^{3}$**2**). In AFM images at $V$ = 0 V the molecule on defect free NaCl was not stabilized in the open-shell state and thus not observed in the $^{3}$**2** state and was always imaged in the states of either $^{1}$**2-M** or $^{1}$**2-P** (see **fig. S13**). The AFM image at 0.34 V (A) and the STM images at voltages >300 mV (**fig. S18**) show a twofold symmetry. The two-fold symmetry, the faint sharp lines and the multiple appearance of the Cl atoms at different locations in the AFM image (A) are indicative for a change of adsorption geometries, due to the interaction with the tip (*77*). (**B, C**) Calculated relaxed adsorption geometries of $^{3}$**2** on NaCl. Viewed from top, with the substrate oriented as in the experiment (A). Na cations in yellow and Cl anions of the NaCl surface in green; C atoms in gray and Cl atoms of **2** in purple. (**D, E**) Side view along the direction of the arrows in (B, C), respectively. For $^{3}$**2**, calculations reveal two prochiral adsorption sites (B, C). Note that even considering $^{3}$**2** being planar, the system becomes prochiral upon adsorption. The adsorption sites and orientations in the two prochiral adsorption geometries of $^{3}$**2** are similar to $^{3}$**2-P** and $^{3}$**2-M**, but the molecular geometry (out-of-plane distortions) is not. We assume that the AFM image (A) is affected by the molecule switching at a fast rate between four adsorption geometries: That of $^{1}$**2-P** and $^{1}$**2-M**, and of the two prochiral adsorption geometries of $^{3}$**2**. The $I(t)$ data (Fig. 5 and **figs. S21** and **S22**) suggest that at such conditions the $^{3}$**2** state has a longer lifetime than $^{3}$**2-P** and $^{3}$**2-M**. Therefore, we assume that under these conditions the molecule is most of the time in the $^{3}$**2** state. The AFM image in (A) seems in-line with a superposition of the two calculated adsorption geometries of $^{3}$**2**. However, because of the four states involved, and the switching being fast compared to the AFM and STM bandwidth, such images are challenging to simulate and interpret (*77*). Moreover, we assume that the relative lifetimes of all four states depend on the vertical and lateral tip position. Depending on the tip

position, one of the two adsorption geometries of $^{3}$**2** might be preferred and/or fast switching between the two adsorption geometries of $^{3}$**2** might be possible, which might be reasons for the observation of only one current plateau for the $^{3}$**2** state in $I(t)$ data (see **figs. S19**, **S20**, **S23** and **S24**). Stable AFM images of $^{3}$**2** were obtained near defects, see Fig. 6, A and D and **fig. S17**, agreeing well with the calculated adsorption geometry of $^{3}$**2** being planar, but adsorbed non-parallel with respect to the NaCl surface, see (B–E).

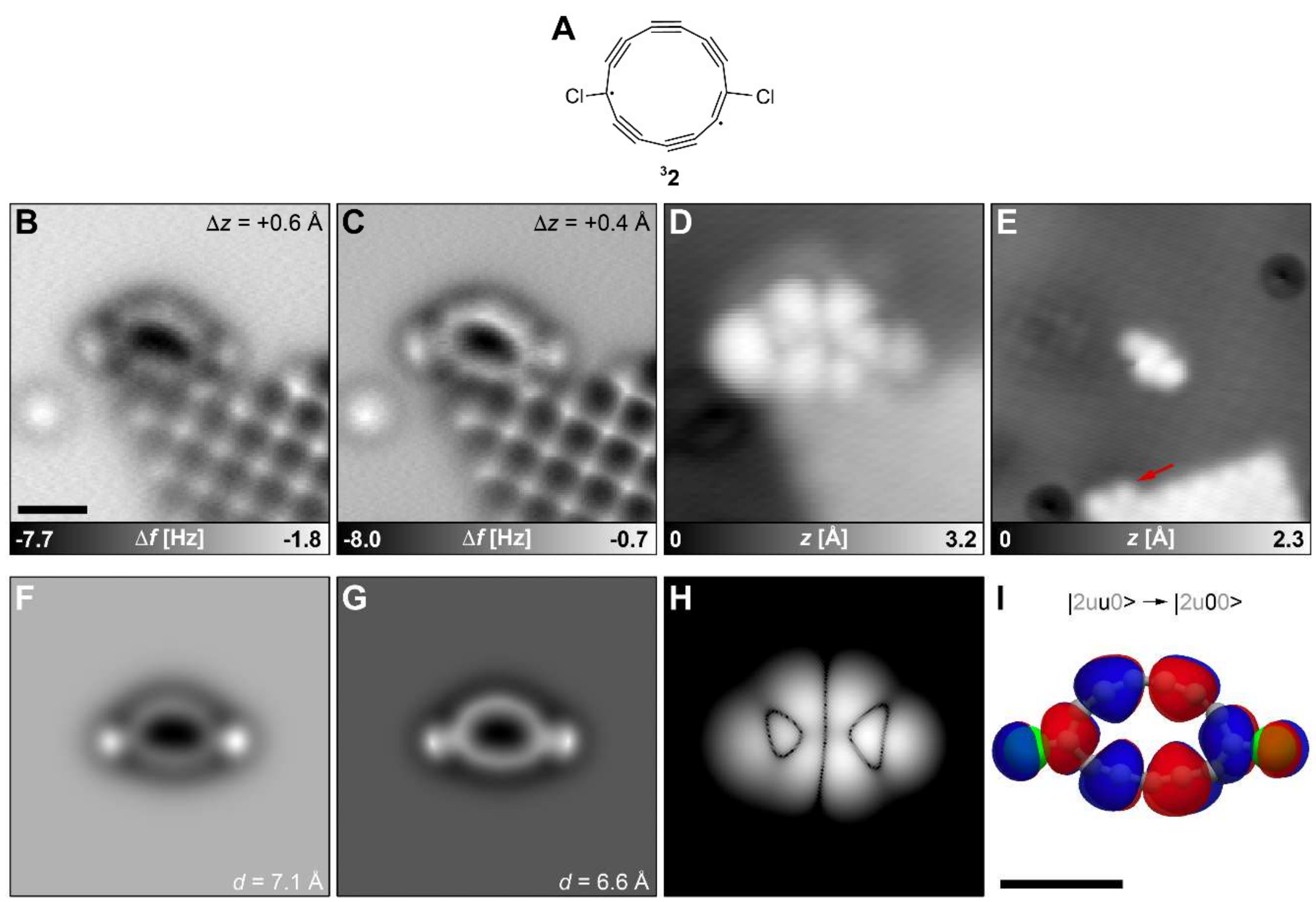

**Figure S16. AFM, STM and orbital density image of $^{3}$2.** (**A**) Resonance structure of $^{3}$**2**. (**B**, **C**) AFM data of $^{3}$**2** at $\Delta z$ = 0.6 Å (AFM-far) and $\Delta z$ = 0.4 Å (AFM-close), respectively. Setpoint: $V$ = 0.2 V, $I$ = 1.2 pA. (**D**) STM orbital density image of $^{3}$**2** at the PIR ($V$ = –0.54 V, $I$ = 0.5 pA). Image size (B–D) 28 x 28 Å$^{2}$, scale bar 5 Å in (B), applies also to (C, D, F, G, H). (**E**) STM data after the molecule had been manipulated away from the step edge and is in the $^{1}$**2-P** state. $V$ = –0.1 V, $I$ = 0.7 pA, image size 60 x 60 Å$^{2}$. The red arrow indicates a Cl atom of the 3$^{rd}$ layer NaCl island. The observation that this Cl remained at its position at the island, after lateral manipulation of the molecule off the island edge indicates that this Cl is part of the 3$^{rd}$ layer NaCl island and was not covalently bound to the molecule in (B–D). This is not obvious from the AFM images alone, because relaxations of the CO tip can give rise to apparent bonds at small tip-sample distances (*39, 42*), see (C). (**F**, **G**) AFM simulations (*39*) of $^{3}$**2** in its relaxed adsorption geometry on NaCl (see **fig. S15**), with the image plane parallel to the molecular plane to account for the adsorption parallel to the surface. The adsorption geometry parallel to the surface is likely related to the molecule being at a NaCl step edge in (B) and (C), in contrast to the tilted adsorption geometry on defect-free NaCl. Simulations of AFM images were performed based on CASPT2 calculated xyz geometries using the Probe-

Particle Model (*39*) with default CO-tip parameters (stiffness $k_x = k_y = 0.25$ N/m, $k_R = 30$ N/m) at an oscillation amplitude of 0.5 Å. The distance *d* denotes the distance between the CO-tip's oxygen atom and the topmost atom of the molecule. **(H)** Simulated STM orbital density image (*43*) based on (**I**) the Dyson orbital of the transition from the neutral $^3$**2** to the cation. Orbital densities of PIR and NIR of $^3$**2**, simulated for different tip-sample distances, are shown in **fig. S26**.

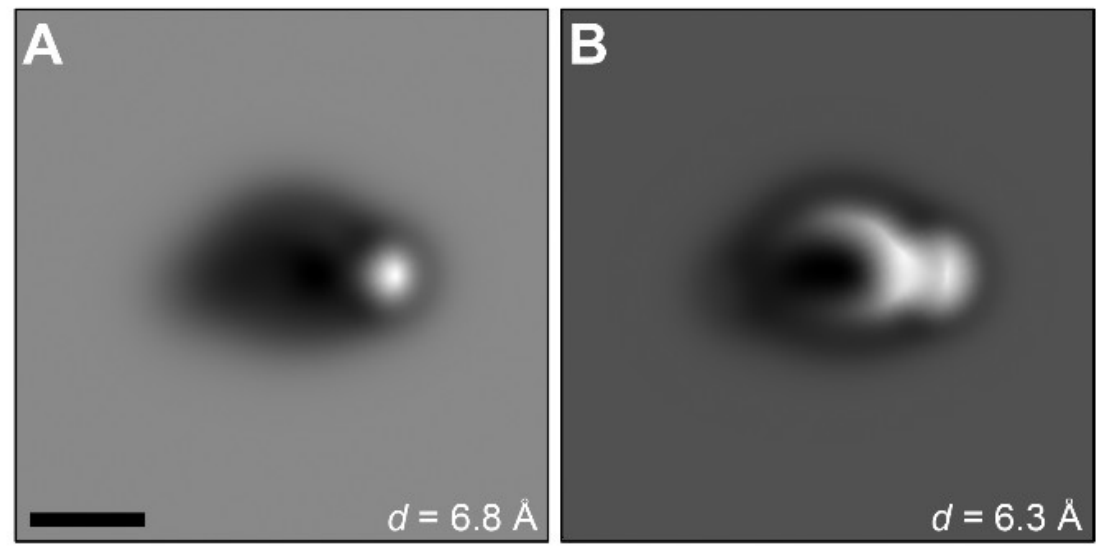

**Figure S17.** AFM simulations of $^3$**2** in its relaxed adsorption geometry on NaCl using the Probe-Particle Model (*39*), with the image plane parallel to the surface. The tilted adsorption of $^3$**2** on defect free NaCl (see **fig. S15**) is reflected in the image contrast, similar as in the AFM data shown in Fig. 6, A and D.

### *4.3. Transitions between configurations of* **2**

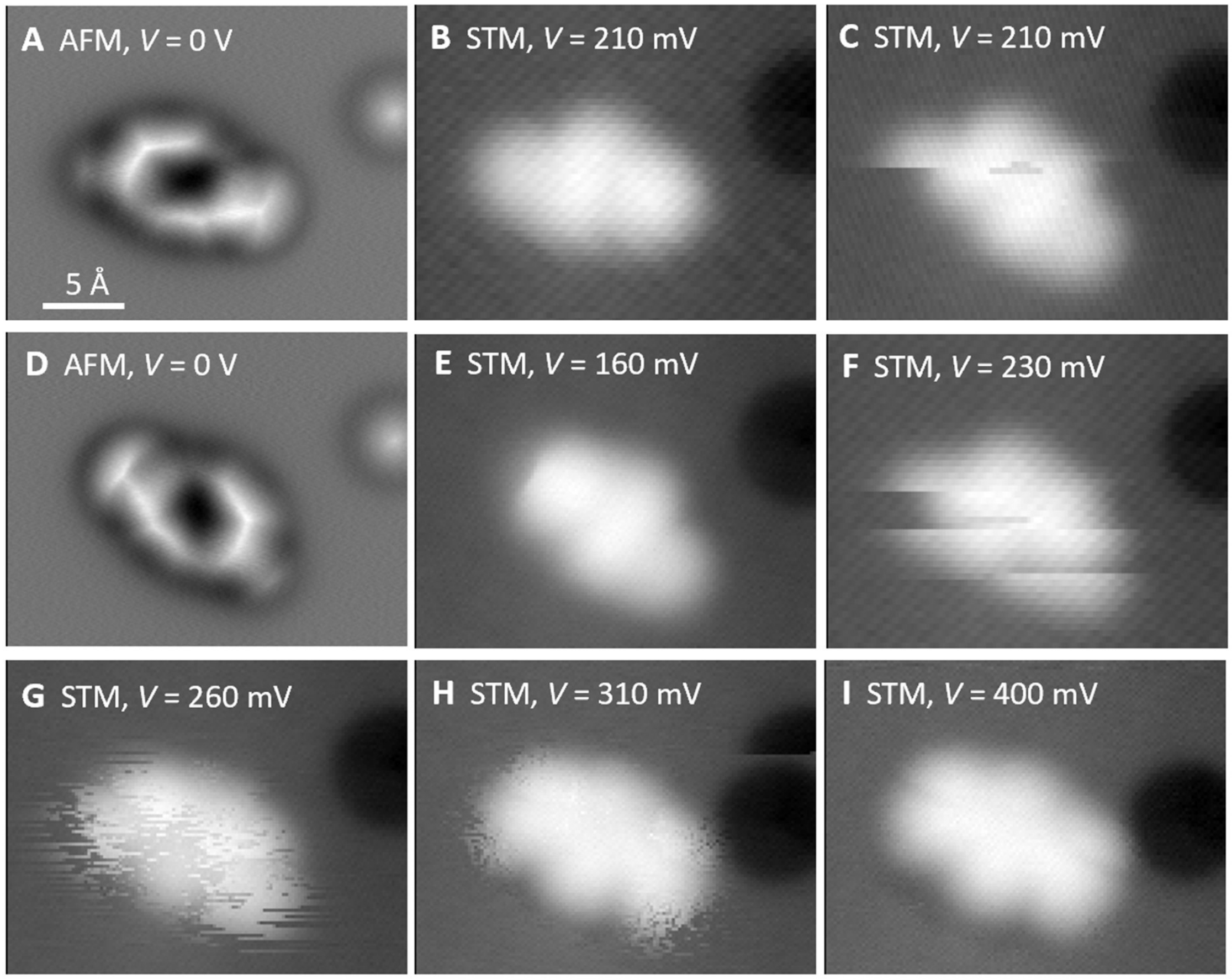

**Figure S18. Sequence of AFM and STM data at different voltages.** (**A**) Constant-height AFM data of $^{1}$**2-M**. (**B**, **C**) STM images at $V$ = 210 mV, that is, just at the threshold voltage for switching out of $^{1}$**2-M** and $^{1}$**2-P** states. Using these parameters, switching rates are on the order of minutes - about the timescale to acquire an STM image. We observed no switching of $^{1}$**2-M** during the acquisition of (B) and one switching event from $^{1}$**2-M** state to $^{1}$**2-P** during the acquisition of (C). The slow scan direction is from top to bottom. (**D**) Constant-height AFM data of $^{1}$**2-P**. (**E**) STM image of $^{1}$**2-P** at $V$ = 160 mV. For voltages below 200 mV we observed in general no switching events. (**F–I**) STM images at indicated voltages. By increasing $V$, the switching rate increased. At (F) $V$ = 230 mV, few switching events between $^{1}$**2-M** and $^{1}$**2-P** can be observed in one image. In (G) and (H), obtained at $V$ = 260 mV and $V$ = 310 mV, respectively, many switching events result in noisy features within the images and the switching rate is faster than the typical time per scan line (on the order of seconds). In (I), at $V$ = 400 mV, the switching rate is faster than the acquisition time per pixel (on the order of 10 ms), resulting in a smooth image. Note that the image shows two-fold symmetry. Next to **2** is a CO molecule, which moved by one lattice site during the acquisition of (H). All STM images were acquired at a constant current $I$ = 0.4 pA. Scales from black to white, A [-3.4 Hz; +3.8 Hz], B [0 Å; +2.0 Å], C [0 Å; +2.0 Å], D [-3.3 Hz; =3.9 Hz], E [0 Å; +2.1 Å], F [0 Å; +2.1 Å], G [0 Å; +2.1 Å], H [0 Å; +2.1 Å], I [0 Å; +1.9 Å]. Scale bar in (A) applies to all images.

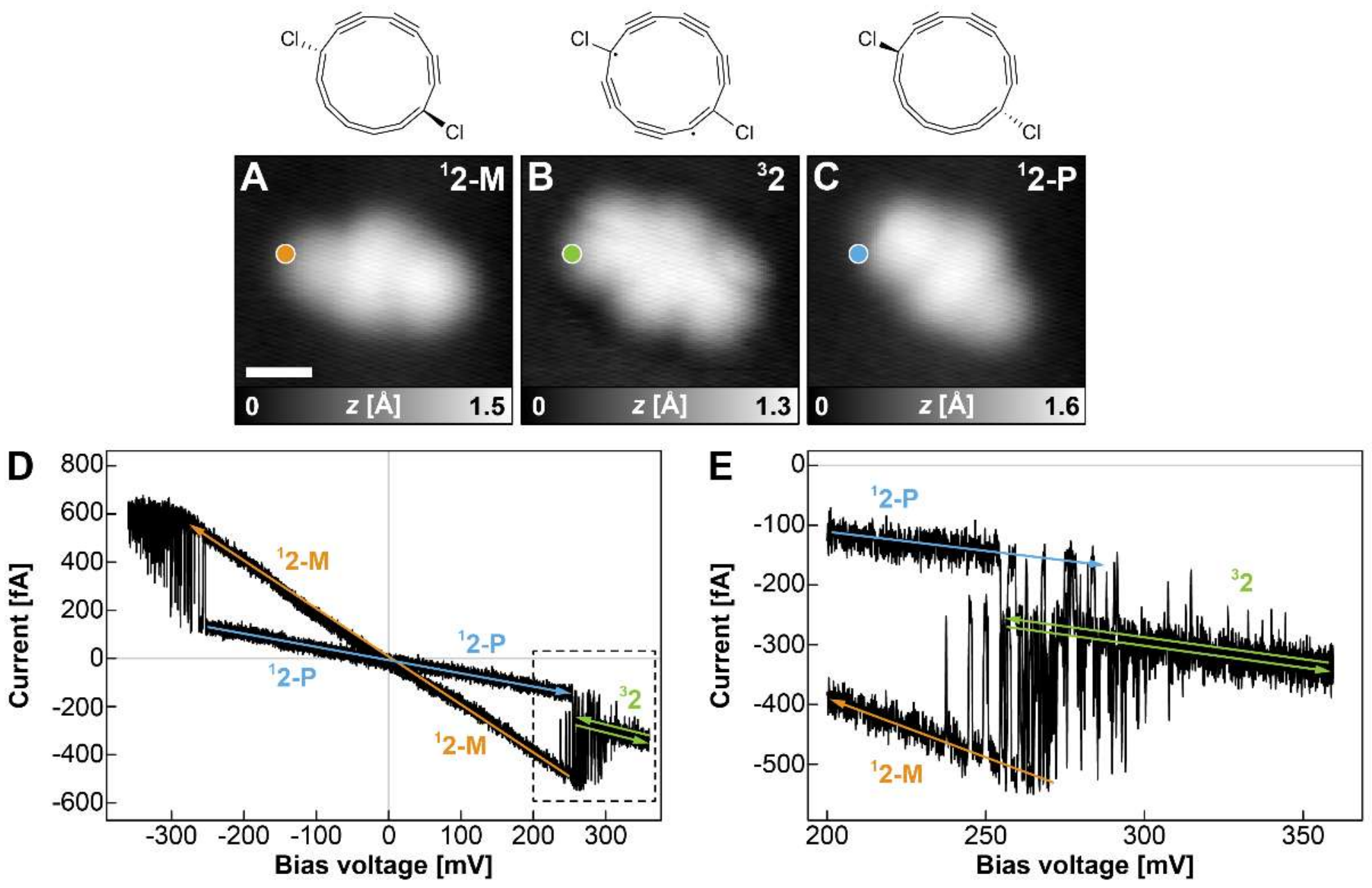

**Figure S19. *I*(*V*) spectroscopy of transitions.** (**A–C**) STM images and corresponding structures in the states of $^{1}$**2-M**, (predominantly in) $^{3}$**2**, and in $^{1}$**2-P**, respectively. The position at which the $I(V)$ spectra have been acquired is indicated. Scale bar in (A) is 5 Å and applies also to (B) and (C). Parameters for in (A) and (C) $I = 0.4$ pA, $V = 150$ mV that is, below the switching threshold. But for (B) $I = 0.4$ pA, $V = 400$ mV, that is, well above the switching threshold. As setpoint for the spectroscopy we chose $V = 400$ mV (STM parameters as in (B)), to be in the region of fast switching, with no measured telegraph noise, resulting in a well-defined tip height. (**D**) $I(V)$ spectrum on **2** at the position indicated in (A, B, C). Setpoint $V = 400$ mV, $I = 0.2$ pA, $\Delta z = -0.25$ Å. The sample voltage $V$ was ramped from 360 mV to -360 mV and back to 360 mV in 60 seconds. The occupation of either $^{1}$**2-P** or $^{1}$**2-M** in the region of small voltage (below the switching threshold) is stochastic. Here we show an example with different occupation during the down and up sweep of the voltage, to observe $I(V)$ traces in both states. The intermittent current state, however, is never observed in the voltage region between the threshold voltages [-210 mV; +210 mV]. (**E**) Zoom into the indicated voltage range (dashed box in D). At large voltages ($V > 320$ mV) individual switching events cannot be observed, because the switching rate is too fast. The three current plateaus assigned to the three states are indicated. The measured currents at $V > 320$ mV correspond to the extrapolated currents of the intermediate current plateau in the voltage region where switching between all three current plateaus can be observed (250 mV to 270 mV), see green line. This observation is in line with the assumption that for $V > 320$ mV (large voltages, fast switching rates) the molecule is switching at a rate too fast to be detected by our STM, but most of the time resides in the $^{3}$**2** state. This assumption is corroborated by the extracted lifetimes of $^{1}$**2-P**, $^{1}$**2-M** and $^{3}$**2** as a function of current and voltage (**figs. S21** and **S22**). Switching between states $^{1}$**2-P**, $^{1}$**2-M** and $^{3}$**2**, as observed for positive bias, was also observed at negative bias at similar absolute values of $V$ (see D).

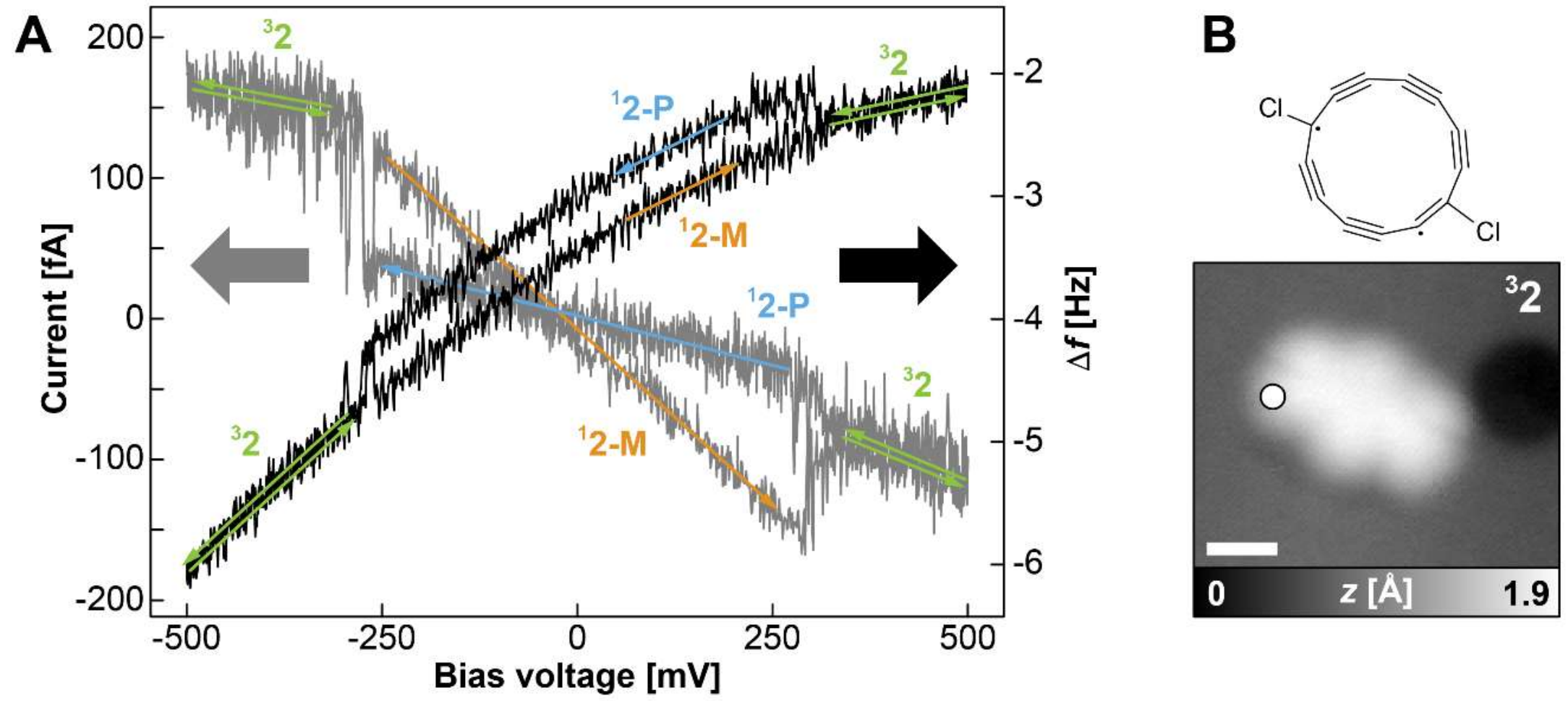

**Figure S20: Kelvin probe force spectroscopy.** (**A**) $I(V)$ spectrum (gray data, left y-axis) and simultaneously acquired $\Delta f(V)$ Kelvin probe force spectroscopy (KPFS, black data, right y-axis) on **2**. Setpoint $V$ = 400 mV, $I$ = 0.3 pA, $\Delta z$ = 0.5 Å. $V$ was ramped from 500 mV to -500 mV and back to 500 mV in 20 seconds. The configurations of **2** (for $^{3}$**2**, the predominant configuration) for respective regions of the spectra are indicated. The KPFS spectra corresponding to all three configurations show parabolas that are only shifted in the vertical direction, implying that the charge state of **2** is the same in all three configurations (*78, 79*). Note that different charge-states were transiently accessed at increased voltages, when measuring at the onset of the ion resonances, that is the NIR of $^{1}$**2** at $V$ = 1.0 V (see Fig. 6F) and the PIR of $^{3}$**2** at $V$ = -0.54 V (see Fig. 6J and **fig.** S16). The vertical shift of the KPFS parabola corresponding to the $^{1}$**2-P** configuration with respect to the other confirmations is explained by the different, non-voltage dependent background, mainly from van der Waals attraction and Pauli repulsion. The cantilever oscillation (amplitude $A$ = 0.5 Å) was only switched on for this measurement to obtain KPFS data, but all other $I(V)$ spectra shown were measured without cantilever oscillation. (**B**) STM image indicating the tip position at which spectra in (A) were acquired. STM parameters, $I$ = 0.3 pA, $V$ = 400 mV.

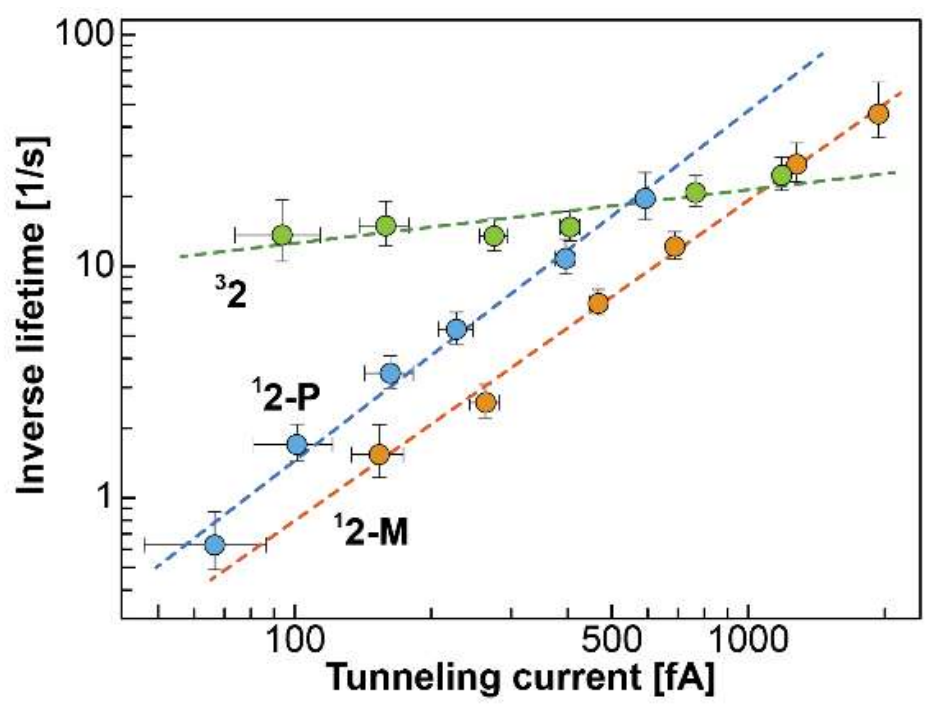

**Figure S21. Switching rate as a function of tunneling current**. Double logarithmic plot of the inverse lifetime, that is, the rate for switching out of the respective state, plotted against the current $I$ measured in that state. The data is the same as shown in Fig. 5F. The raw data, $I(t)$ traces, are shown in **fig. S23**. Data was obtained at $V$ = 250 mV at constant tip heights with $\Delta z$ in the interval of $\Delta z$ = -0.25 Å (largest current of each state) to $\Delta z$ = +1.0 Å (smallest current of each state). Linear fits (dashed lines) yield slopes $k$ of $k$($^{1}$**2-M**) = 1.4, $k$($^{1}$**2-P**) = 1.5, and $k$($^{3}$**2**) = 0.2. These slopes indicate that transitions out of $^{1}$**2-M** and $^{1}$**2-P** are triggered by tunneling electrons in one- and/or two-electron processes (*80*), whereas the slope of transitions from $^{3}$**2**, being much smaller than 1, indicates that transitions out of $^{3}$**2**, are not primarily induced by tunneling electrons.

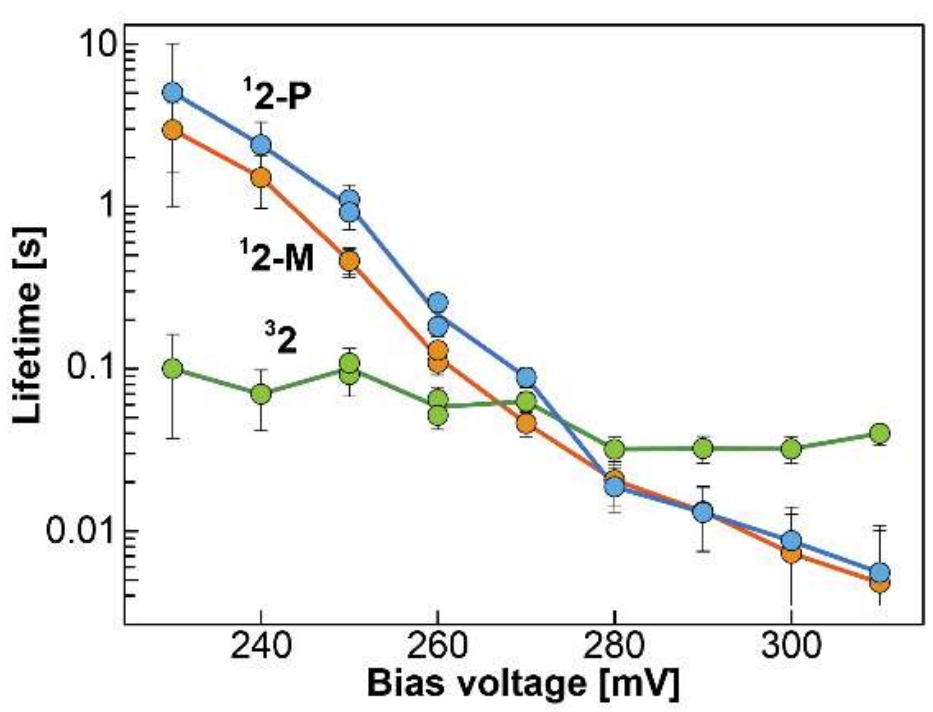

**Figure S22. Lifetime as a function of *V*.** Data was obtained at the same tip-height offset $\Delta z$ = 0.0 Å, but at different $V$ in the interval $V$ = 230 mV to $V$ = 310 mV. The raw data, $I(t)$ traces, are shown in **fig. S24**. For $V$ = 250 mV and $V$ = 260 mV, two measurements have been carried out, one before taking data at the other voltages and one after it, both are plotted to inquire reproducibility. We observe that for $V$ > 280 mV, state $^{3}$**2** has the longest lifetime and the molecule is most of the time in that state (see **fig. S24**). The small dependence of transitions out of $^{3}$**2** as a function of $I$ and $V$ indicates that these transitions are not predominantly triggered by tunneling electrons and is tentatively explained with a spontaneous decay of the triplet with a lifetime on the order of 0.1 seconds or longer, which is moderately reduced by the applied electric field in the presence of the tip and/or tunneling electrons. The steep decrease of the switching rate with applied voltage for transitions out of $^{1}$**2-M** and $^{1}$**2-P** is in line with a transition triggered by tunneling electrons, that might evolve from a multiple-electron process at small voltages to a one-electron process with increasing yield at larger voltages (*80*).

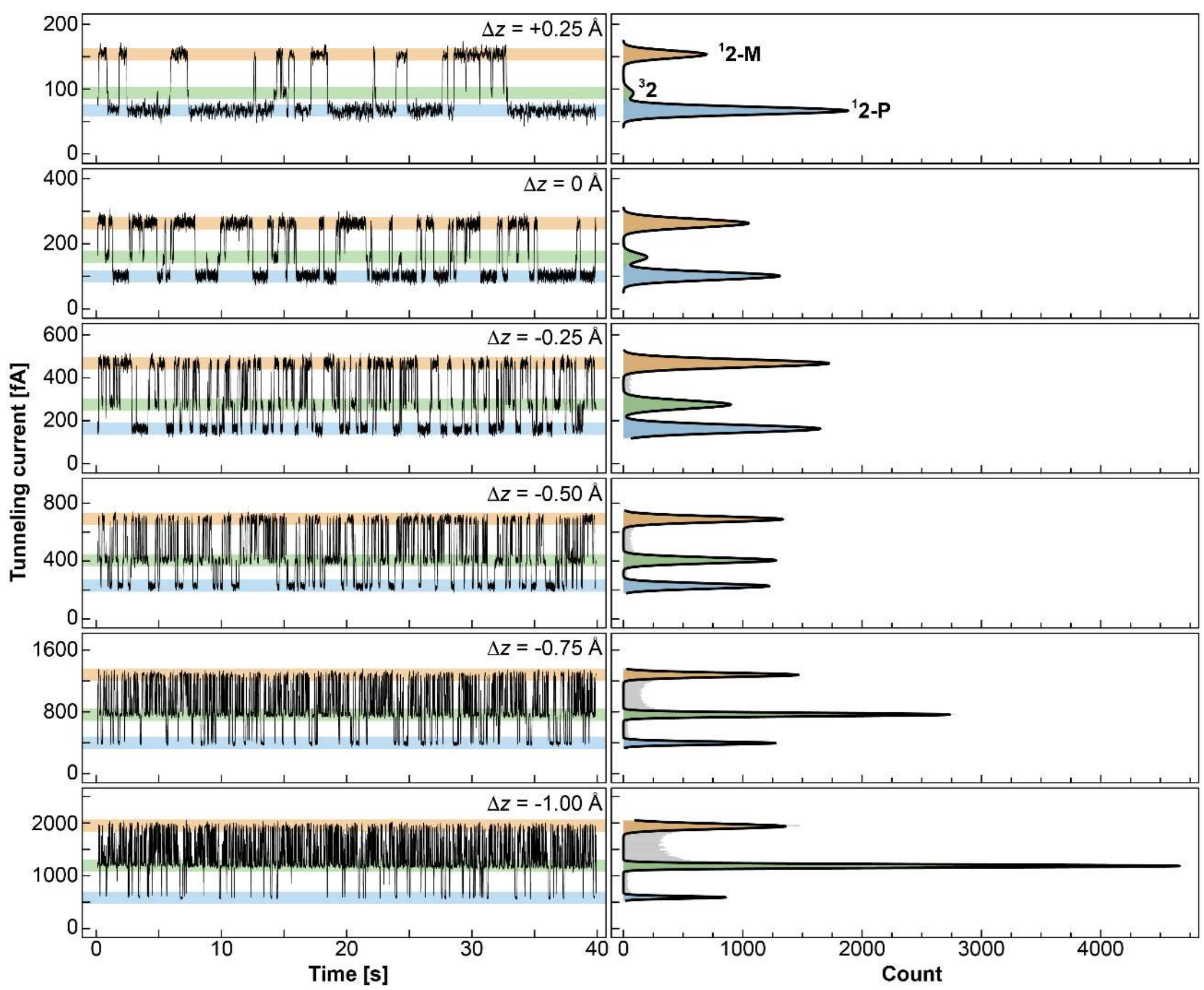

**Figure S23. Raw *I*(*t*) data of transitions at different tip heights**. Left-hand side panels: *I*(*t*) spectra at different tip-height offsets $\Delta z$ at constant $V = 250$ mV. The current plateaus assigned to the **$^{1}$2-M state** (orange), the **$^{3}$2** state (green) and the **$^{1}$2-P** state (blue) are indicated by colors. Right-hand side panels show histograms of the currents (100 equally sized current bins for each spectrum). Counts refer to occurrences of currents for 1 ms (time bins), during a *I*(*t*) trace of 40 s.

The molecule cannot be imaged in the **$^{3}$2** configuration at $V = 0$ V at this adsorption site, as it is not stable and decays with a lifetime on the order of 0.1 s. The assignment of **$^{3}$2** is based on STM and AFM images at increased bias of **$^{3}$2** at $V > 300$ mV, when the molecule is most of the time in that state, and its comparison with calculations (see **fig. S15**). These reveal that in the **$^{3}$2** state the molecule is in either of two adsorption orientations. Moreover, the **$^{3}$2** state is observed by STM and AFM on molecules at different adsorption sites at which the **$^{3}$2** state was stable (see Fig. 6, A and D and **fig. S16**).

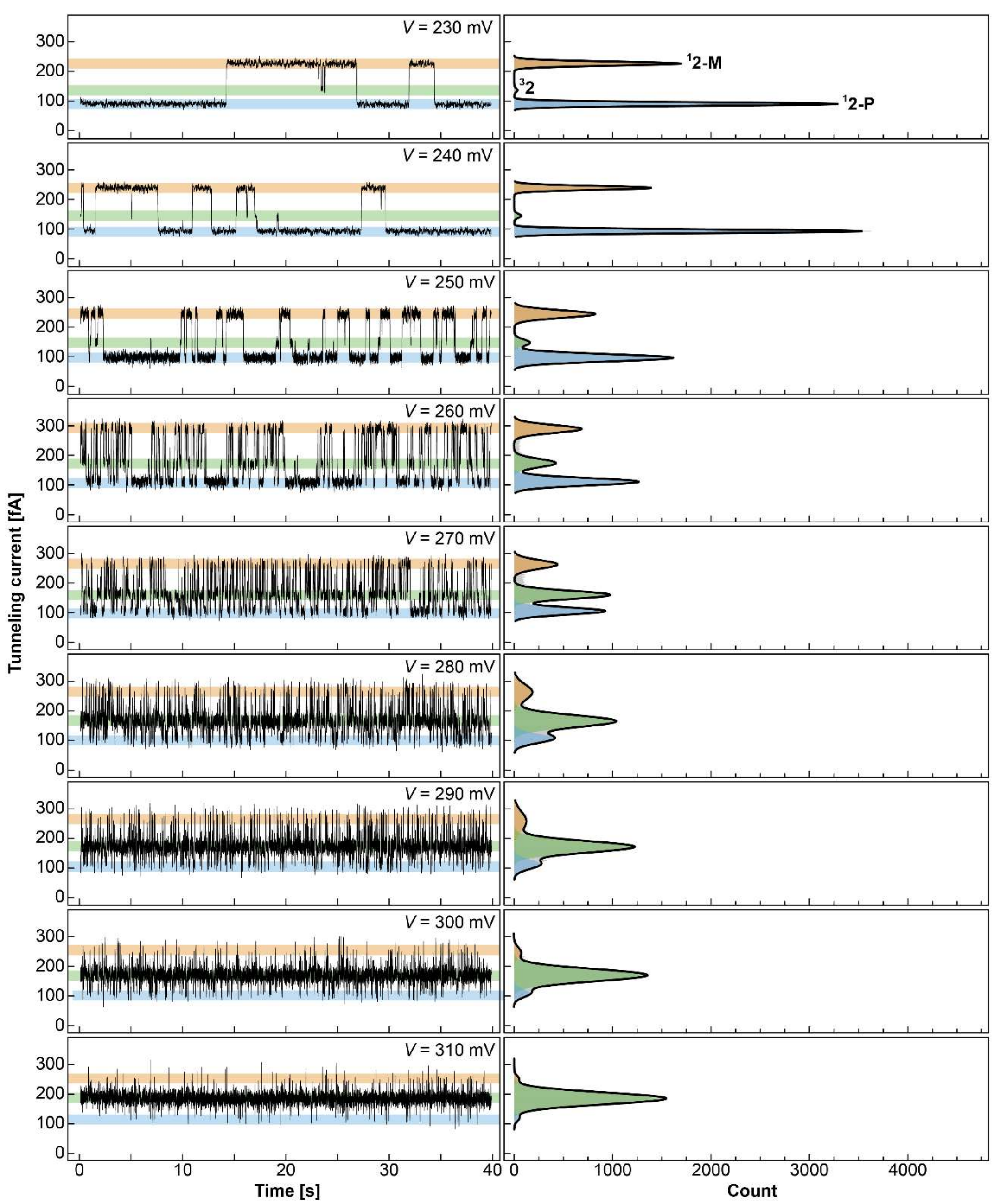

**Figure S24. Raw $I(t)$ data of transitions at different voltages**. Left-hand side panels: $I(t)$ spectra at different $V$, at constant tip-height offsets $\Delta z = 0$ Å. The current plateaus assigned to the **$^{1}$2-M** state (orange), the **$^{3}$2** state (green) and the **$^{1}$2-P** state (blue) are indicated by colors. Right-hand side panels show histograms of the currents (100 equally sized current bins for each spectrum). Counts refer to occurrences of currents for 1 ms (time bins), during a $I(t)$ trace of 40 s.

### *4.4. STM simulations*

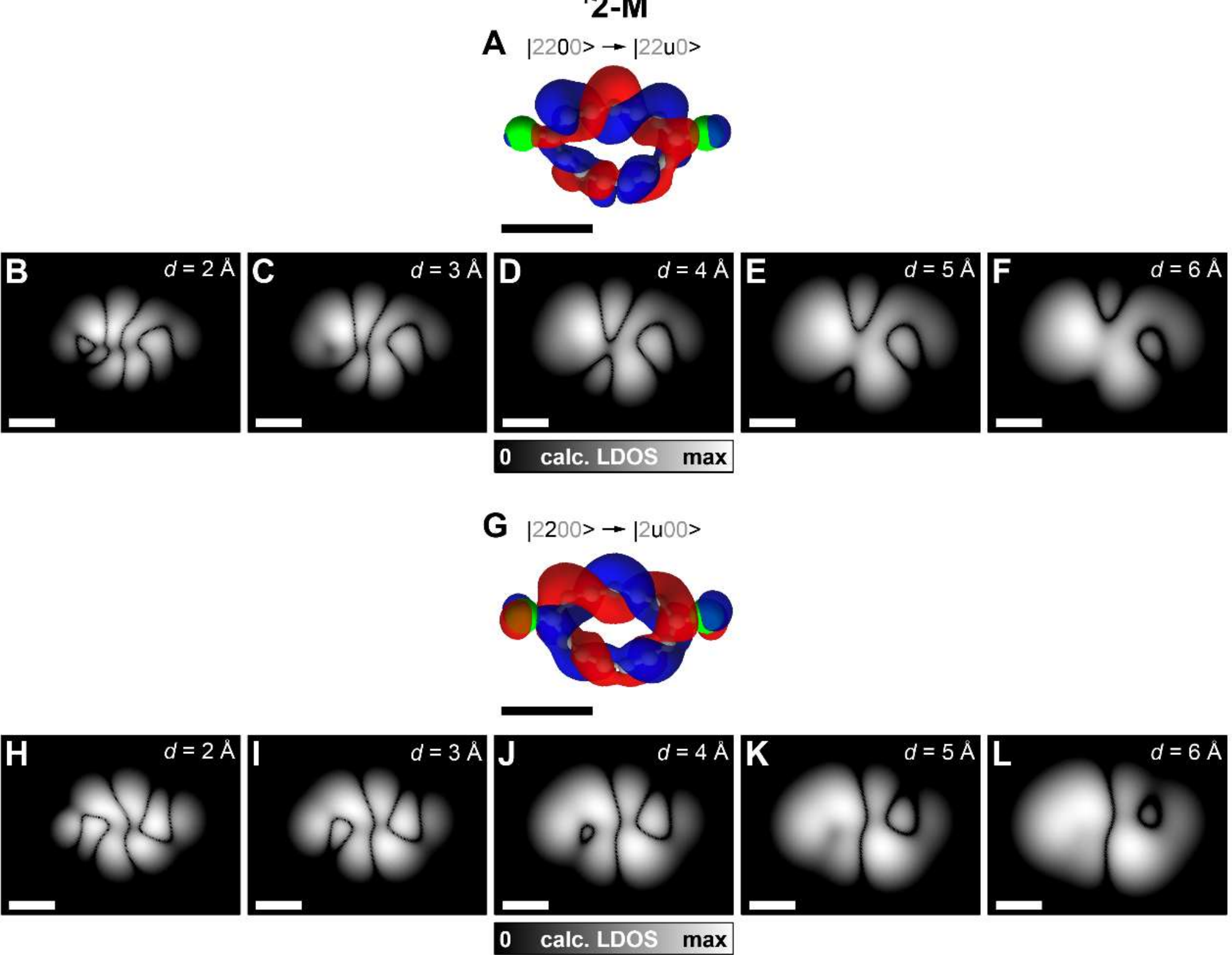

**Figure S25. Simulated STM images of ionic resonances of $^{1}$2-M**. Simulations of STM images of ion resonances of $^{1}$**2-M** based on Dyson orbitals for (**A**) the negative ion resonance (NIR) and (**G**) the positive ion resonance (PIR). The Dyson orbitals are used as inputs for simulating STM images. (**B**-**F**) Simulated STM images of the NIR based on the Dyson orbital for electron attachment to $^{1}$**2-M** (A). (**H**-**L**) Simulated STM images of the PIR based on the Dyson orbital for electron detachment from $^{1}$**2-M** (G). STM images have been simulated using Bardeen's tunneling theory, using an *s*-wave tip with a decay constant of 1 $\text{Å}^{-1}$ and distances *d* between the upmost atom of $C_{13}Cl_2$ and the center of the wavefunction of the tip (*43*). Scale bars 5 Å.

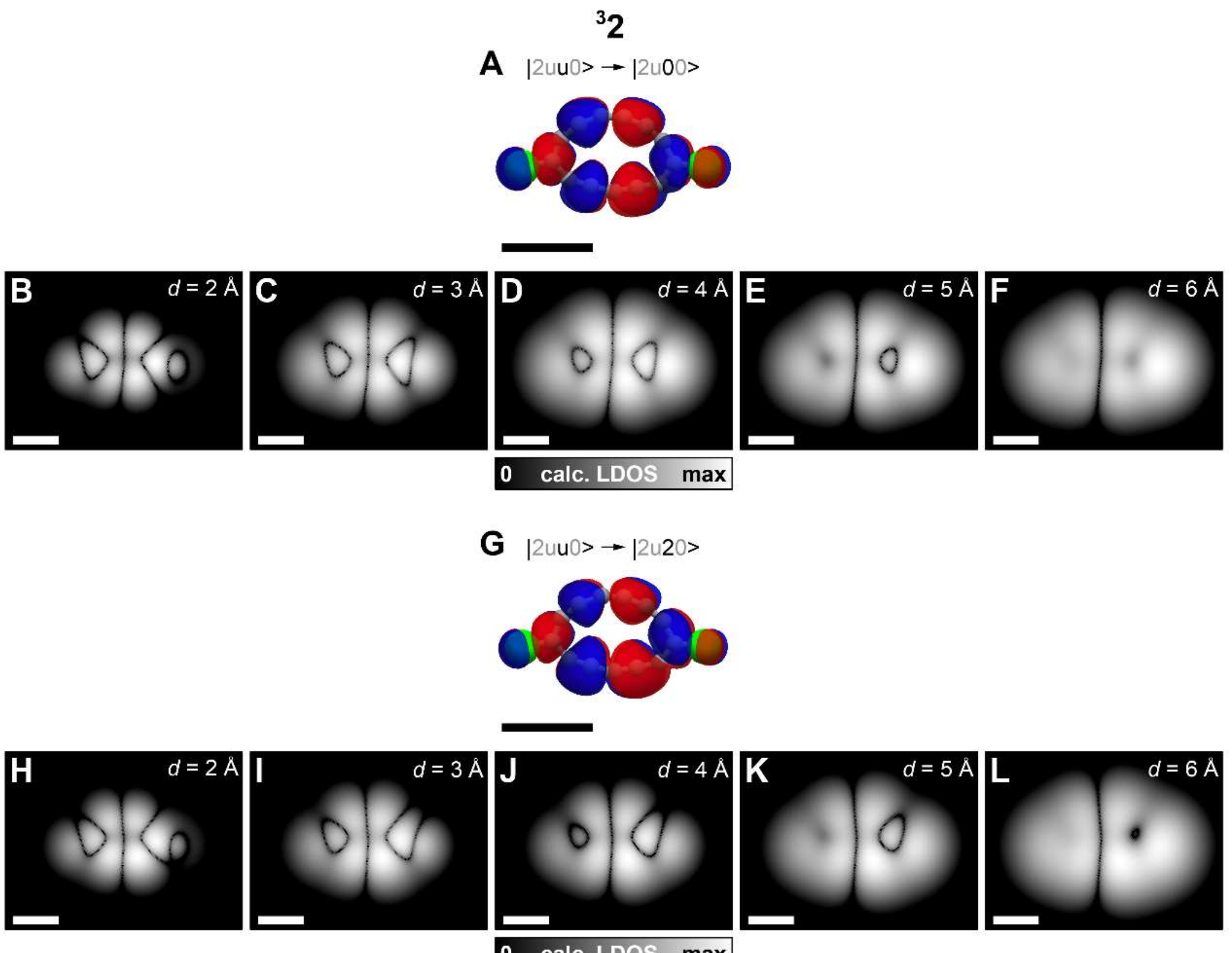

**Figure S26. Simulated STM images of ionic resonances of $^{3}$2**. Simulations of STM images of ion resonances of **$^{3}$2** based on Dyson orbitals for (**A**) the positive ion resonance (PIR) and (**G**) the negative ion resonance (NIR). The Dyson orbitals are used as inputs for simulating STM images. (**B**-**F**) Simulated STM images of the PIR based on the Dyson orbital for electron detachment from **$^{3}$2** (A); (**H**-**L**) of the NIR based on the Dyson orbital for electron attachment to the out-of-plane system of **$^{3}$2** (G) (*22*). STM images have been simulated using Bardeen's tunneling theory, using an *s*-wave tip with a decay constant of 1 $Å^{-1}$ and distances *d* between the upmost atom of $C_{13}Cl_2$ and the center of the wavefunction of the tip (*43*). Scale bars 5 Å.

### *4.5. Graphical Summary*

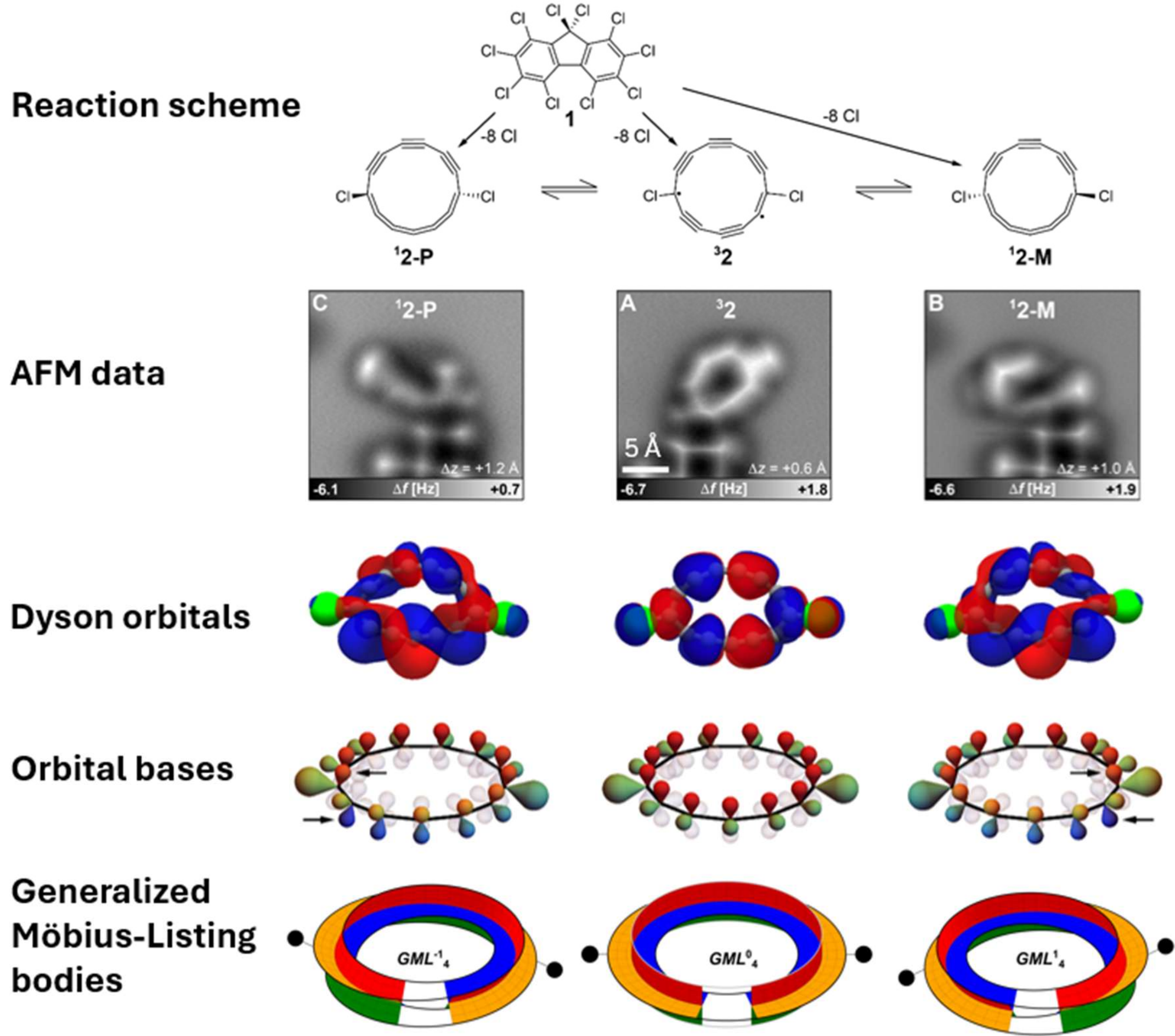

**Figure S27. Graphical summary: Correspondence between molecular configuration, geometry, Dyson orbitals, orbital bases and generalized Möbius-Listing bodies**. The first row shows the reaction scheme, see Fig. 2A. The second row shows AFM data, see Fig. 6, C, A and B. The third row shows Dyson orbitals obtained using CASPT2, see **figs. S9**. The shown Dyson orbitals, that is, electron attachment to the singlet and electron detachment from the triplet, correspond to the transitions accessed in STM measurements, see Fig. 6F and Fig. 6J. The fourth row shows the corresponding orbital bases, see Fig. 2, C, D and E. The fifth row shows the corresponding *GML* bodies, see Fig. 1, G, D, H in which schematic handles (black pins) have been added to highlight the correspondence to the Cl atoms and their out-of-plane distortions. The left column shows the singlet **$^{1}$2-P**, the central column shows the triplet **$^{3}$2** and the right column shows the singlet **$^{1}$2-M**.

## 5. Supplementary Notes

### *5.1. Supplementary Note 1*

The time-independent Schrödinger equation for a particle with mass $M$ constrained on a ring with radius $r$ can be written as:

$$\hat{H}\psi(\phi) = -\frac{\hbar^2}{2Mr^2}\frac{d^2\,\psi(\phi)}{d\phi^2} = E\psi(\phi) \tag{24}$$

where $\phi$ is the azimuthal angle. Solutions to Eq. 24 can be written as:

$$\psi_n(\phi) = \frac{1}{\sqrt{2\pi}}e^{\pm in\phi} \tag{25}$$

with $n$ as a quantum number. The associated energies are:

$$E_n = \frac{n^2\hbar^2}{2Mr^2} \tag{26}$$

and the z-component of the angular momentum is:

$$\hat{L}_z\psi_n(\phi) = -\mathrm{i}\hbar\frac{d\,\psi(\theta\phi)}{d\phi} = \pm n\hbar \tag{27}$$

where the subscript $z$ refers to the axis perpendicular to the ring, so that rotation about $z$ changes $\phi$.

If we assume that the wavefunction has the periodicity of 2π, i.e. it is single-valued for each angle $\phi$:

$$\psi_n(\phi + 2\pi) = \psi_n(\phi) \tag{28}$$

$$e^{\pm in(\phi+2\pi)} = e^{\pm in\phi} \tag{29}$$

$$e^{\pm \mathrm{i}n2\pi} = 1 \tag{30}$$

This equality is satisfied for $n \in \mathbb{Z}$:

$$n_{\mathrm{H}} = 0, \pm 1, \pm 2, \pm 3 \ldots \tag{31}$$

Where H in the subscript refers to the periodicity of 2π of a trivial (Hückel) system in Eq. 30. For a wavefunction with a periodicity of 8π, which corresponds to a quadruple-valued wavefunction in $\phi$, analogously to Eq. 30 we obtain:

$$e^{\pm in8\pi} = 1 \tag{32}$$

which is satisfied for:

$$n_{8\pi\text{ periodic}} = \frac{n_{\mathrm{H}}}{4} = 0, \pm\frac{1}{4}, \pm\frac{2}{4}, \pm\frac{3}{4} \ldots \tag{33}$$

While this consideration is correct for a particle on a ring with a boundary condition of 8π, it may be an (overly) simplistic representation of a half-Möbius topology. A more sophisticated approach is shown in *Supplelmentary Note 2*, below.

### *5.2. Supplementary Note 2*

In the cyclocarbons every carbon has two orthogonal *p*-orbitals contributing to the π system. This two-π-orbital-per-atom basis can be represented as a pseudospinor

$$|\psi\rangle = \begin{pmatrix} \psi_A \\ \psi_B \end{pmatrix}. \quad (34)$$

The boundary condition for making one circulation *C* in real space in the half-Möbius $GML^1{}_4$ topology is represented by

$$\boldsymbol{U_C} = \begin{pmatrix} 0 & 1 \\ -1 & 0 \end{pmatrix}. \quad (35)$$

Two full circumnavigations gives $\boldsymbol{U_C}^2 = -\boldsymbol{I}$, and hence a sign reversal of the spinor wavefunction, while the spinor wavefunction is periodic with respect to four full circumnavigations, $\boldsymbol{U_C}^4 = \boldsymbol{I}$. Using the Pauli matrix $\boldsymbol{\sigma_y}$, the boundary condition can be expressed as $\boldsymbol{U_C} = i\boldsymbol{\sigma_y}$, being a rotation on the Bloch sphere around the *y*-axis, accompanied by a global phase factor. Hence, the half-Möbius $GML^1{}_4$ topology entails a boundary condition with a wavefunction's sign change after *two* complete circumnavigations.

As shown in a prior work on Möbius graphene-based strips (*52*), boundary conditions resulting from a tight-binding Hamiltonian for a $GML^1{}_2$ body with *N* sites can be written as:

$$\widehat{H} = -t\sum_{j=1}^{N} e^{i\frac{\gamma}{N}}(c_j^\dagger c_{j+1}) \quad (36)$$

In Eq. 34, the orbital basis acquires a shift of $\frac{\gamma}{N}$ between each site, resulting in an overall shift of $\gamma$ over one circulation. In a Hückel topology ($GML^0{}_2$, Fig. 1, A and C), there is no phase shift and $\gamma = 0$, while in a Möbius topology ($GML^1{}_2$, Fig. 1, B and E) the wavefunction changes sign upon one circulation, meaning that $\gamma = \pi$.

The boundary conditions imposed in Eq. 36 are equivalent to an (effective) Aharonov-Bohm flux quantized in units of $\frac{\phi}{\phi_0}$ and resulting from a perpendicular magnetic field *B* applied to a particle on a ring with radius *R* (*53*),

$$\widehat{H} = -t\sum_{j=1}^{N} e^{i\frac{2\pi\Phi}{N\Phi_0}}\left(c_j^\dagger c_{j+1}\right);\ \gamma = \frac{2\pi\phi}{\phi_0} \quad (37)$$

where $\Phi = B\pi R^2$ is the magnetic flux relative to the Dirac flux quantum $\Phi_0 = h/e$. As shown by Berry, $\gamma$ is the geometric (Berry) phase accumulated upon one circumnavigation of the ring (*51*). Generalizing this analogy suggests that a $GML^n{}_m$ body may be characterized as having a Berry phase of $\gamma = \frac{2\pi\, n}{m}$. However, the Berry phase is only well defined, if interference effects become observable, that is, only after *two* full circumnavigations for a $GML^1{}_4$ body. The practical implications of such Aharonov-Bohm fluxes and Berry phases in real cyclocarbons with pseudospinor wavefunctions, where hopping terms between the $\psi_A$ and $\psi_B$ subsystems of nearest-neigbor atoms become nonzero, remain yet to be explored.